\documentclass{jfm}

\usepackage{graphicx}
\usepackage{array}
\usepackage{comment}
\usepackage{color}
\usepackage{multirow}
\usepackage{subfig}
\usepackage{natbib}
\usepackage{amsmath}
\usepackage{amssymb}
\usepackage{booktabs}
\usepackage{url}
\usepackage{enumitem}
\usepackage{lineno}
\definecolor{myred}{rgb}{0.8,0,0}
\definecolor{mygreen}{rgb}{0,1,0}

\newcommand{\nt}[1]{\textrm{#1}}

\renewcommand{\arraystretch}{1.2}
\newcommand{\ra}[1]{\renewcommand{\arraystretch}{#1}}

\title[Bifurcations in a Q2D Kolmogorov-Like Flow]{Bifurcations in a Quasi-Two-Dimensional Kolmogorov-Like Flow}
\author[J. Tithof, B. Suri, R. K. Pallantla, R. O. Grigoriev, M. F. Schatz]{Jeffrey Tithof, Balachandra Suri, Ravi Kumar Pallantla, Roman O. Grigoriev, Michael F. Schatz}
\affiliation{Center for Nonlinear Science and School of Physics, Georgia Institute of Technology, Atlanta, Georgia 30332-0430, USA}

\date{\today}

\begin{document}

\maketitle

\begin{abstract}
We present a combined experimental and theoretical study of the primary and secondary instabilities in a Kolmogorov-like flow. The experiment uses electromagnetic forcing with an approximately sinusoidal spatial profile to drive a quasi-two-dimensional (Q2D) shear flow in a thin layer of electrolyte suspended on a thin lubricating layer of a dielectric fluid. 
Theoretical analysis is based on a 2D model \citep{suri_2014}, derived from first principles by depth-averaging the full three-dimensional Navier-Stokes equations. 
As the strength of the forcing is increased, the Q2D flow in the experiment undergoes a series of bifurcations, which is compared with results from direct numerical simulations of the 2D model. 
The effects of confinement and the forcing profile are studied by performing simulations that assume spatial periodicity and strictly sinusoidal forcing, as well as simulations with realistic no-slip boundary conditions and an experimentally validated forcing profile. 
We find that only the simulation subject to physical no-slip boundary conditions and a realistic forcing profile provides close, quantitative agreement with the experiment. 
Our analysis offers additional validation of the 2D model as well as a demonstration of the importance of properly modelling the forcing and boundary conditions.

\end{abstract}

\section{Introduction} \label{sec:intro}

Fluid flows in two spatial dimensions have been the subject of substantial research efforts in recent decades. For the greater part of the twentieth century, it was generally considered that two-dimensional (2D) flows were merely a theoretical idealization with limited practical relevance. 
This conception has changed drastically since the 1980s, when experiments in thin electrolyte 
layers \citep{bondarenko_1979}, soap  films \citep{couder_1984}, and liquid metals \citep{sommeria_1982} demonstrated that nearly 2D flows can indeed be realized in the laboratory. 
Today, experimental approximations of 2D flows are widely employed as models of atmospheric and oceanic flows \citep{dolzhansky_2012, boffetta_2012}. 
Being theoretically and experimentally more amenable than their three-dimensional (3D) counterparts, 2D flows have also served as platforms for studying new phenomena such as turbulent cascades \citep{sommeria_1986,tabeling_1991}, coherent structures \citep{sommeria_1988}, and mixing \citep{haller_2000}. 

Perhaps one of the best known examples of 2D flows is the one introduced by Andrey Kolmogorov in 1959 as a mathematical problem for studying hydrodynamic stability \citep{arnold_1960}. 
The Kolmogorov flow represents the motion of a viscous fluid in two dimensions (we will refer to these as $x$ and $y$) driven by a forcing that points along the $x$-direction and varies sinusoidally in the $y$-direction. 
The fluid flow is considered incompressible, $\nabla \cdot {\bf u}=0$, and is governed by the 
2D Navier-Stokes equation,
\begin{equation}\label{eq:2dns}
\partial_t{\bf u} + {\bf u}\cdot{\bf{\nabla}}{\bf{u}}= -\frac{1}{\rho}\nabla p + \nu \nabla^2 \bf{u} + {\bf f}.
\end{equation}
Here, ${\bf u}=(u_x,u_y)$ is the velocity field, $p$ is the 2D pressure field, and ${\bf f} = A \sin(\kappa y) {\bf \hat x}$ represents the driving force with amplitude $A$ and wavenumber $\kappa$. 
The parameters $\rho$ and $\nu$ are the density and the kinematic viscosity of the fluid, respectively. 
Kolmogorov flow has served as a convenient model for understanding a wide variety of hydrodynamic phenomena in 2D, such as fluid instabilities \citep{meshalkin_1961, iudovich_1965, kliatskin_1972, nepomniashchii_1976}, 2D turbulence \citep{green_1974}, and coherent structures \citep{armbruster_1992, smaoui_2001, chandler_2013}.

Practically realizable flows, however, are never strictly 2D. Experimental approximations of Kolmogorov flow have often been carried out in either shallow layers of electrolytes \citep{bondarenko_1979} or in soap films \citep{burgess_1999}, wherein geometric confinement suppresses the component of velocity along one of the spatial directions ($z$).
The remaining two velocity components, however, generally depend on both extended and confined coordinates, making the flow ``quasi-two-dimensional'' (Q2D). 
To account for the dependence on the confined coordinate, Q2D flows in shallow layers have often been modelled by adding a linear friction term to the 2D Navier-Stokes equation  (\ref{eq:2dns}):
\begin{equation}\label{eq:2dns_wf}
\partial_t{\bf u} + {\bf u}\cdot{\bf{\nabla}}{\bf{u}}= -\frac{1}{\rho}\nabla p + \nu \nabla^2 \bf{u} - \alpha\bf{u} + {\bf f},
\end{equation}
where $\alpha$ is a constant. Here, ${\bf u}$ corresponds to the velocity field at the electrolyte-air interface. The addition of this term was first suggested by \citet{bondarenko_1979} to model a Q2D flow generated in a homogeneous shallow electrolyte layer. 
In such a flow, the bottom of the fluid layer is constrained to be at rest because it is in contact with the solid surface of the container holding the fluid. 
This no-slip constraint at the bottom of the fluid layer causes a gradient in the magnitude of the  horizontal velocity along the confined direction $z$. 
\citet{bondarenko_1979} rationalized that the dissipation due to this shear, for sufficiently shallow fluid layers, is captured by the linear friction term. In the context of Q2D flows in electrolyte layers, this term has come to be known as ``Rayleigh friction.'' 
Experimental flows in thin layers, and their 2D approximations employing equation (\ref{eq:2dns_wf}), are now commonly referred to as ``Kolmogorov-like'' when the forcing profile is nearly sinusoidal. 
Note that linear friction models, similar to that in equation (\ref{eq:2dns_wf}),  have also been employed to describe Q2D flows in liquid metals \citep{sommeria_1986} and soap films \citep{couder_1989,burgess_1999}. The motivation behind the addition of the friction term in these models is different from that in equation (\ref{eq:2dns_wf}). 
In this article we are only concerned with flows in shallow electrolyte layers.

Experimental realizations of Q2D flows in recent years have employed two-fluid-layer setups: either a  
setup with \emph{miscible} layers comprised of a heavy electrolyte fluid (salt water) beneath a lighter nonconducting fluid (pure water) \citep{marteau_1995, paret_1997a, kelley_2011a}, or a setup with \emph{immiscible} layers comprised of a heavy dielectric fluid beneath a lighter electrolyte \citep{rivera_2005, akkermans_2008a,akkermans_2010}. The rationale behind these modifications was that in addition to confinement, density stratification and immiscibility should enhance two-dimensionality in the top layer. Theoretical models of Q2D experimental flows realized in stratified layers of fluids, however, have not accurately modelled the effect of inhomogeneity in fluid properties as well as the gradient in the magnitude of horizontal velocity ${\bf u}(x,y)$ along the confined direction $z$. Consequently, experiments were compared with simulations based on the 2D model (\ref{eq:2dns_wf}) with empirically estimated parameters \citep{juttner_1997,boffetta_2012}.  

To address this deficiency, \citet{suri_2014} have investigated the variation in the horizontal velocity ${\bf v}(x,y,z,t)$ along the confined direction $z$ for a stratified two-immiscible-layer setup. 
Following \citet{dovzhenko_1981}, the Q2D velocity was approximated as 
\begin{align}\label{eq:profile}
{\bf v}(x,y,z,t)=P(z){\bf u}(x,y,t)=P(z)\left[u_x(x,y,t){\bf \hat{x}}+u_y(x,y,t){\bf \hat{y}}\right],
\end{align}
where ${\bf u}(x,y,t)$ corresponds to the 2D velocity field at the electrolyte-air interface and $P(z)$ models the variation of the horizontal velocity along $z$. 
By substituting the form of velocity in equation (\ref{eq:profile}) into the 3D Navier-Stokes equation and integrating along the $z$-direction, the following modified version of equation (\ref{eq:2dns_wf}) was derived:
\begin{equation}\label{eq:2dns_mod}
\partial_t{\bf u} + \beta{\bf u}\cdot{\bf{\nabla}}{\bf{u}}= -\frac{1}{\bar{\rho}}\nabla p + {\nu} \nabla^2 \bf{u} - \alpha\bf{u} + {\bf f}.
\end{equation}
In the above equation ${\bf f}$ is the depth-averaged force density and the parameters $\beta$, ${\bar \rho}$, ${ \nu}$, and $\alpha$ are given by:
\begin{equation}\label{eq:2dterms}
\begin{aligned}
\beta = \frac{\int_0^{h}\rho P^2 dz}{\int_0^{h}\rho P dz}, \quad \bar{\rho} = \frac{\int_0^h \rho P dz}{h}, \quad {{\nu}} = \frac{\int_0^{h}\mu P dz}{\int_0^{h}\rho P dz},  \quad \alpha = \frac{\mu\left(\frac{dP}{dz}\right)_{z=0}}{\int_0^{h}\rho P dz}, 
\end{aligned}
\end{equation}
where $h$ is the total thickness of the two fluid layers. The vertical profile $P(z)$ is very weakly dependent on the horizontal flow profile ${\bf u}(x,y,t)$. The profile $P(z)$ that corresponds to a sinusoidal horizontal flow (described as the straight flow below) was computed and validated against experimental measurements in \citet{suri_2014}.

The prefactor $\beta$ reflects the change in the mean inertia of the fluid layer due to the variation $P(z)$ of the horizontal velocity in the vertical direction. Since $P(z) \neq 1$ in experiments, $\beta \neq 1$, which distinguishes equation (\ref{eq:2dns_mod}) from all previous 2D models of flows in shallow electrolyte layers.   
For multi-layer setups, the coefficients $\beta$, ${\bar \rho}$, ${ \nu}$, and $\alpha$ account for both the inhomogeneity in fluid properties as well the vertical profile $P(z)$, as suggested by equation (\ref{eq:2dterms}). 
Equations (\ref{eq:2dns}) and (\ref{eq:2dns_wf}) can be treated as special cases of equation (\ref{eq:2dns_mod}) with suitable choices of the parameters $\alpha$ and $\beta$. Furthermore, equation (\ref{eq:2dns_wf}) can also be obtained from equation (\ref{eq:2dns_mod}) by rescaling the variables (cf. Appendix \ref{sec:nondim}).

In this article we study instabilities of a Q2D Kolmogorov-like flow realized in a setup with two immiscible fluid layers   
and compare experimental results with direct numerical simulations (DNS) of equation (\ref{eq:2dns_mod}).  
Most previous studies of Kolmogorov-like flow that compared experiments with theoretical predictions assumed a perfectly sinusoidal shear flow on an unbounded or periodic domain \citep{bondarenko_1979, dovzhenko_1984, batchaev_1989, krymov_1989, thess_1992a, dolzhanskii_1992}. 
While some of these studies reported quantitative agreement between theory and experiment in regards to the primary instability, none were able to match simultaneously both the critical Reynolds number and the critical wavenumber. Even matching one of these required treating the Rayleigh friction coefficient $\alpha$ as an adjustable parameter. 
To address these shortcomings, we have performed a systematic investigation of the effects of lateral confinement using numerical simulations with three different sets of boundary conditions. 
Furthermore, we investigate how the observed flow patterns and their stability are affected by the deviations in the forcing profile from perfect periodicity in the extended directions and by the variation of the forcing profile in the confined direction. 
Finally, we compare the results of numerical simulations with the experimental observations for the secondary instability, which introduces time-dependence into the flow.

Given that none of the previous models of Q2D flows were quantitatively accurate, 
the availability of an experimental setup and a matching 2D model that are in quantitative agreement is quite important for a number of reasons. In particular, this allows us to make substantial progress \citep{suri_2017} in understanding the role of coherent structures in turbulent flows \citep{hussain_1986, kawahara_2012,gallet_2013,chandler_2013,haller_2015}. Recent advances in transitional flows and weak turbulence rely on a deterministic, geometrical description where the evolution of the flow is guided by nonchaotic, unstable solutions of the Navier-Stokes equation, often referred to as exact coherent structures (ECS) \citep{nagata_1997, waleffe_1998, kerswell_2005, eckhardt_2007, gibson_2009}. 
The bulk of numerical studies have explored the role of ECS in 3D flows simulated on periodic domains with simple geometries, such as pipe flow, plane Couette flow, and plane Poiseuille flow. 

However, experimental evidence for the role of ECS in 3D flows has been scarce \citep{hof_2004,lozar_2012,dennis_2014}, in part due to technical limitations in obtaining spatially and temporally resolved 3D velocity fields. 
Q2D flows, on the other hand, can be quantified using 2D planar velocity fields which are relatively easy to measure. 
Recently, \citet{chandler_2013} and \citet{lucas_2014,lucas_2015} have identified dozens of ECS in numerical simulations of a weakly turbulent 2D Kolmogorov flow, governed by equation (\ref{eq:2dns}) with periodic boundary conditions which, however, do not describe flows that can be realized in experiments \citep{bondarenko_1979, dolzhanskii_1992, suri_2014}. Hence, the analysis presented herein should provide the much needed foundation for further studies of 2D turbulence which focus on experimental validation of theoretical predictions, building on the results of \citet{suri_2017}.

This article is organized as follows. 
In \S \ref{sec:exp_setup}, we describe the experimental setup employed to generate a Q2D Kolmogorov-like flow. 
In \S \ref{sec:model}, we introduce a realistic model of the forcing in the experiment and discuss different types of lateral boundary conditions which are used to study the effects of confinement theoretically. 
In \S \ref{sec:comparison}, we compare the flow fields obtained from experimental measurements with those from the numerical simulations for different flow regimes and characterize the bifurcations associated with increasing the forcing strength. 
In \S \ref{sec:pitchfork}, we discuss how the nature of the primary instability depends on the lateral boundary conditions. 
Conclusions are presented in \S \ref{sec:conclusion}.  

\section{Experimental Setup} \label{sec:exp_setup}

We generate a Q2D Kolmogorov-like flow in the experiment using a stratified setup with two immiscible fluid layers, first introduced by \citet{rivera_2005}. 
In this configuration, a lighter electrolyte is suspended on top of a denser dielectric, which serves as a lubricant between the electrolyte layer and the solid surface at the bottom of the container which holds the fluids. 
The fluid layers are set in motion using Lorentz forces resulting from the interaction of a direct current passing through the electrolyte and a spatially varying magnetic field.

\begin{figure}
\subfloat[]{%
\begin{minipage}[c][1.2\width]{0.5\textwidth}%
\includegraphics[clip,width=2.5in]{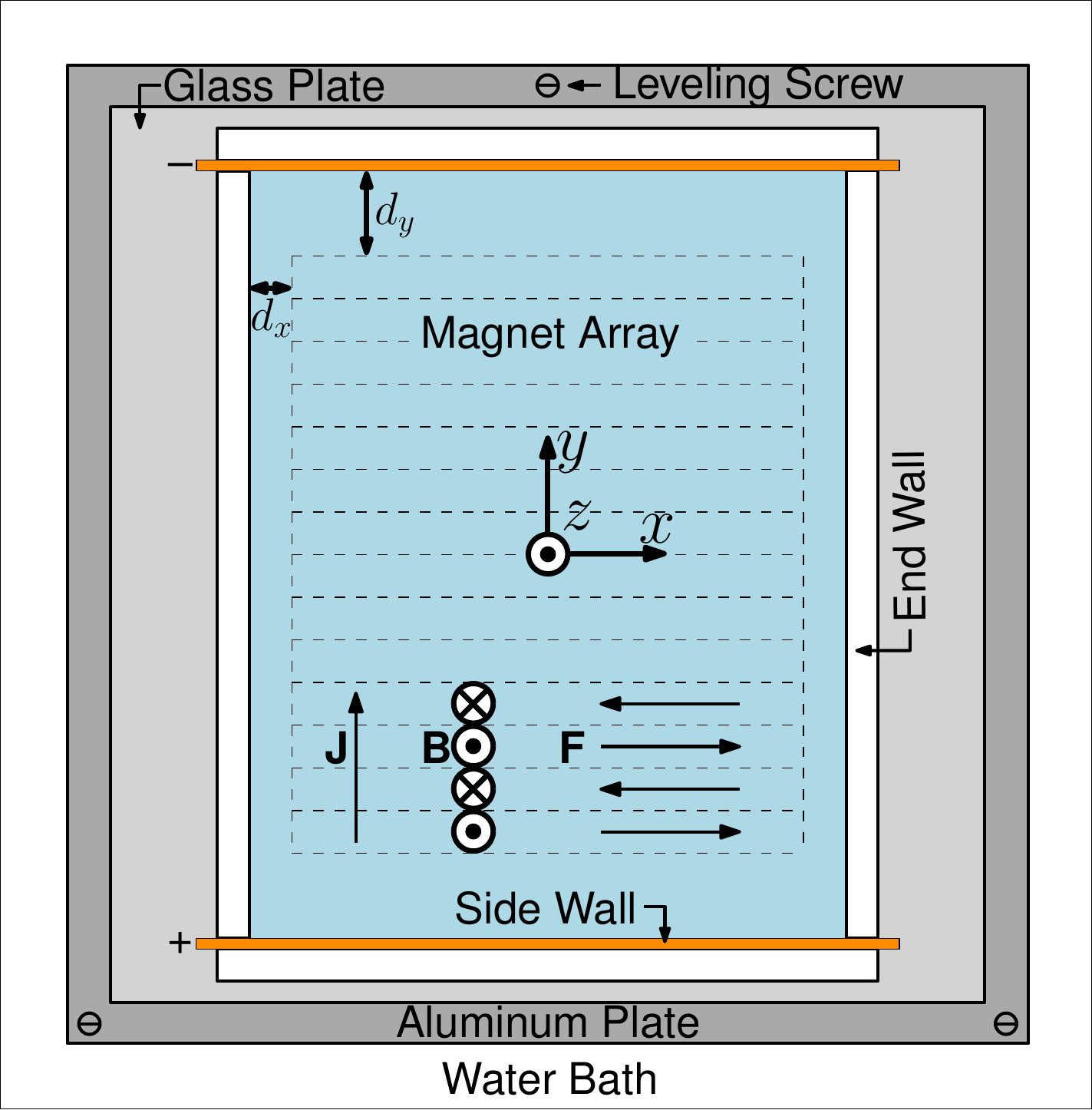}%
\end{minipage}}\subfloat[]{\centering{}%
\begin{minipage}[c][1.2\width]{0.5\textwidth}%
\begin{center}
\includegraphics[clip,width=2.8in]{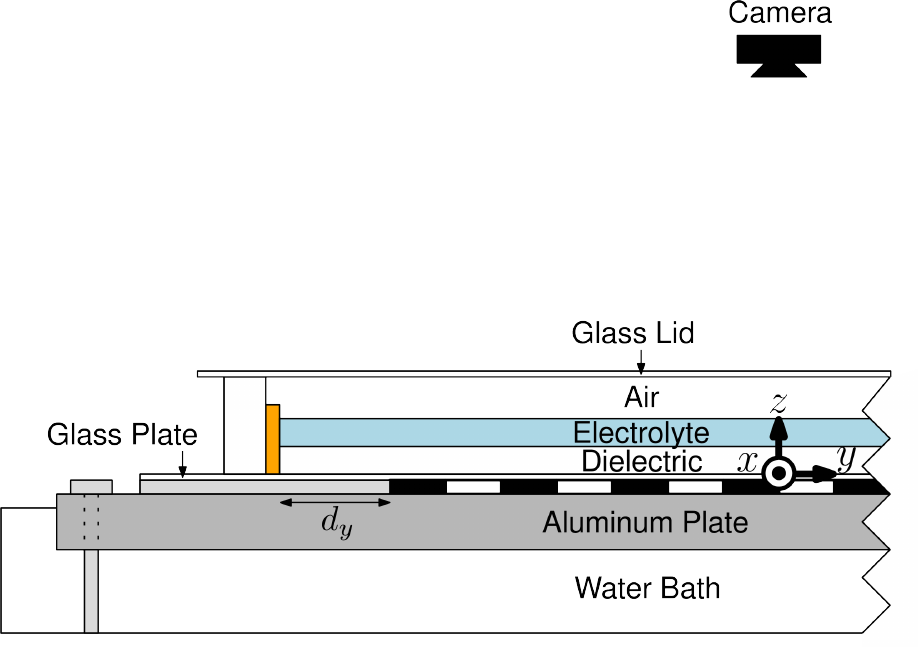}
\par\end{center}%
\end{minipage}}
\caption{\label{expsetup} A schematic diagram of the two-immiscible-layer experimental setup for generating Kolmogorov-like flow viewed (a) from above and (b) from the side. 
The vectors ${\bf J}$, ${\bf B}$, and ${\bf F}$ denote, respectively, the directions of the electric current, magnetic field, and the resulting Lorentz force. 
The flow is bounded by two end walls, two side walls (electrodes), and a no-slip bottom surface, while the top surface is a free electrolyte-air interface.  
This container is mounted on an aluminum plate which is levelled and submerged in a water bath that is temperature-regulated such that the electrolyte is maintained at 23.0 $\pm$ 0.2$^\circ$C.}
\end{figure}

We use a magnet array consisting of 14 NdFeB magnets (Grade N42) to generate a magnetic field that varies roughly sinusoidally along one direction, approximating the forcing in the Kolmogorov flow. 
Each magnet in the array is 15.24 cm long and 1.27 cm wide, with a thickness of $0.32 \pm 0.01$ cm. 
The magnetization is parallel to the thickness dimension, with a surface field strength of about 0.2 T. 
The magnets are positioned side-by-side along their width to form a 15.24 cm $\times$ ($14\times1.27$ cm) $\times$  0.32 cm array such that the adjacent magnets have fields pointing in opposite directions, normal to the plane of the array.  
This magnet array is placed on a flat aluminum plate of dimensions 30.5 cm $\times$ 30.5 cm $\times$  1.0 cm, and rectangular pieces of aluminum with the same thickness as the magnets ($0.32 \pm 0.02$ cm) are placed beside the magnet array to create a level surface. 
Manufacturing imperfections in the individual magnets and the aluminum siding result in a surface which is not adequately smooth. 
Hence, a thin glass plate measuring $25.4$ cm $\times$ $25.4$ cm in area with a thickness of $0.079 \pm 0.005$ cm is placed atop the magnets and siding to provide a uniform surface. 
A thin layer of black, adhesive contact paper (with approximate thickness 0.005 cm) is placed on top of the glass plate to serve as a dark background for imaging. 
The surface of the contact paper serves as the bottom boundary for the fluids.  
We place the origin of our coordinate system at this height and the lateral centre of the magnet array, with the $x$-coordinate aligned with the magnets' longest side, the $y$-coordinate pointing in the direction of the magnet array periodicity, and the $z$-coordinate in the vertical direction. 
A schematic diagram is shown in figure \ref{expsetup}.

Rectangular bars of acrylic are affixed directly onto the contact paper to create the lateral boundaries of the container that will hold the fluids. 
Parallel to the $y$-direction, two bars are placed at a distance of 17.8 cm apart, centred about the origin.  These solid boundaries for the fluid are henceforth referred to as the ``end walls.'' 
Similarly, running parallel to the $x$-direction, two electrodes mounted on rectangular bars of acrylic are placed at a distance of 22.9 cm, symmetrically relative to the origin. 
These boundaries are henceforth referred to as the ``side walls'' and are used to drive the current through the electrolyte. 
The placement of the end walls and side walls leaves a buffer region of $d_x=1.3$ cm and $d_y=2.5$ cm, respectively, between the edge of the magnet array and these solid boundaries. 

The aluminum plate upon which the magnets are mounted is supported by three screws, which are adjusted to level the system. 
The interior of the container is filled with $122\pm4$ mL of a dielectric fluid and $122\pm2$ mL of an electrolyte to form two immiscible layers that are $0.30 \pm 0.01$ cm and $0.30 \pm 0.005$ cm thick, respectively. 
The dielectric fluid used is perfluorooctane, which has a viscosity of $\mu_d = 1.30$ mPa$\cdot$s and a density of $\rho_d = 1769$ kg/m$^3$ at 23.0$^{\circ}$C.  
The electrolyte fluid is a solution consisting of 60\% 1 M copper sulfate solution and 40\% glycerol by weight.  
The electrolyte's viscosity is $\mu_c = 5.85$ mPa$\cdot$s and the density is $\rho_c = 1192$ kg/m$^3$ at 23.0$^{\circ}$C.  
Note that a large viscosity ratio $\mu_c/\mu_d=4.5$ has been chosen to enhance the two-dimensionality of the electrolyte, as described by \citet{suri_2014}. 
A small amount of viscosity-matched surfactant is added to the electrolyte to lower the surface tension, and a glass lid is placed on top of the container to limit evaporation.

A direct current, which serves as the control parameter, is then passed through the electrolyte; the resulting current density $J$ ranges from about 10 to 40 A/m$^2$ across the different runs. The interaction of this current with the spatially alternating magnetic field ${\bf B}$ results in a spatially alternating Lorentz force ${\bf F}$ which drives the electrolyte (cf. figure \ref{expsetup} (a)). 
The viscous coupling between the electrolyte and the dielectric fluids sets the dielectric fluid in motion as well. 
Since passing a current through a resistive conductor (the electrolyte) results in Joule heating, a calibrated thermistor is placed in the corner of the fluid domain to monitor the fluid temperature, and the aluminum plate is immersed in a temperature-controlled water bath. 
The water bath is regulated such that the temperature of the electrolyte is maintained to $23.0 \pm 0.2^{\circ}$C.  
By limiting the temperature fluctuations, the associated change in viscosity of the fluids is kept to a minimum.
 
For flow visualization, we add hollow glass microspheres (Glass Bubbles K15) manufactured by 3M, sieved to obtain particles with mean radius $24.5 \pm 2$ $\mu$m and mean density $150$ kg/m$^3$. Being lighter than the electrolyte, the microspheres stay afloat at the electrolyte-air interface for the duration of the experiment. The microspheres are illuminated with white light emitting diodes placed near the end walls, outside the container holding the fluids. The flow is  imaged at 15 Hz with a DMK 31BU03 camera manufactured by The Imaging Source, placed directly above the setup. This camera has a CCD sensor with a resolution of 1024 $\times$ 768 pixels, which results in an adequate resolution of about 53 pixels per magnet width.  
The flow velocities are calculated using the PRANA particle image velocimetry (PIV) package \citep{eckstein_2009, prana}. 
This software employs a multigrid PIV algorithm that deforms images to better resolve flows with high shear. 
The velocity field is resolved on a 169 $\times$ 126 grid, with about 9 points per magnet width. 

For the experimental measurements listed above, we obtain the following depth-averaged values for the parameters in equation (\ref{eq:2dns_mod}): $\alpha=0.064$ s$^{-1}$, $\beta=0.83$, ${\nu}=3.26 \times 10^{-6}$ m$^2$/s, and $\bar{\rho} = 976$ kg/m$^3$. These parameters were computed using the vertical profile $P(z)$ that corresponds to the strictly sinusoidal flow \citep{suri_2014}. 
The complexity of the flow in both the experiment and simulation is characterized by the Reynolds number, which we define as:
\begin{equation}\label{eq:re_def}
Re = \frac{Uw}{{\nu}}
\end{equation}
where $w=1.27$ cm is the width of one magnet and $U=\sqrt{\langle\bf{u}\cdot\bf{u}\rangle}$ is the measured root-mean-square (rms) velocity, where $\langle \cdot \rangle$ denotes spatial averaging over a sub-region with dimensions 10.16 cm $\times$ 10.16 cm at the centre of the magnet array. 
Note that from this point on, unless noted otherwise, the characteristic length scale $w$, velocity scale $U$, and time scale $w/U$ will be used for nondimensionalization.  

\section{Numerical Modelling} \label{sec:model}

In this section we present a model of the magnetic field generated by the finite array of permanent magnets in the experiment. 
We then introduce three types of boundary conditions used in our numerical simulations of the flow. 

\subsection{Modelling the Magnetic Field} \label{sec:dipole}

In the discussion so far, we have not addressed an important question of how the 2D forcing function ${\bf f}$ in equation (\ref{eq:2dns_mod}) relates to the 3D forcing ${\bf F}$ in the experiment. 
For a 2D Kolmogorov flow, the forcing ${\bf f}$ is sinusoidal, by definition. 
However, for Kolmogorov-like flows realized in electromagnetically driven shallow layers of electrolyte, ${\bf f}$ needs to be computed from the 3D Lorentz force {\bf F} arising from the interaction of the magnetic field ${\bf B}$ produced by a finite magnet array with a current density ${\bf J}$ \citep{suri_2014}. 
The current density is easily calculated from geometrical considerations, but the magnetic field generated by the array of permanent magnets is quite complicated. For ${\bf J} = J {\bf \hat y}$, the Lorentz force density at any location $(x,y,z)$ within the electrolyte layer is given by ${\bf F} = {\bf J}\times{\bf B} =  JB_z{\bf \hat x}-JB_x{\bf \hat z}$. 
Here, $B_x$ and $B_z$ are the $x$- and $z$-components of the magnetic field, respectively, which vary along all three coordinates $x$, $y$, and $z$. 
Experimental measurements show that the typical value of $B_x$ is less than 3\% of the value of $B_z$ at any given location within the electrolyte. 
Furthermore, the vertical component of the Lorentz force along with the gravitational force will be balanced by the vertical gradient of the pressure. 
Hence, the Lorenz force density for all practical purposes can be approximated as ${\bf F} \approx JB_z{\bf \hat x}$. One can then compute ${\bf f}$ using the expression:
\begin{equation}\label{eq:depth_averaging}
{\bf f} = \frac{1}{\bar \rho}\displaystyle\int\limits_{h_d}^{h_d+h_e}\frac{JB_z(x,y,z)\,dz}{h_d+h_e} {\bf{\hat x}},
\end{equation}
where $h_e$ and $h_d$ are the thicknesses of the electrolyte and dielectric layers, respectively.

\begin{figure}
\subfloat[]{\includegraphics[width=5.2in]{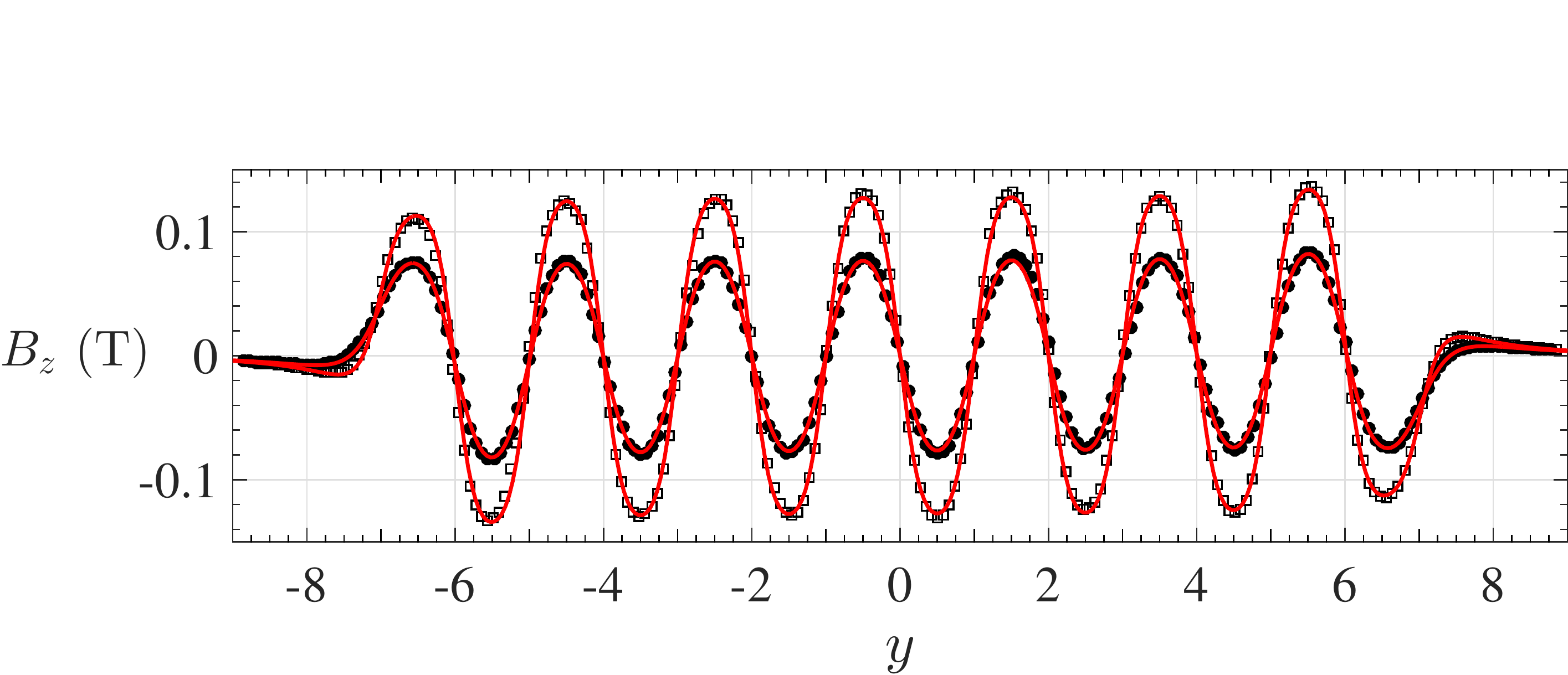}}\\
\centering\subfloat[]{\includegraphics[width=5.2in]{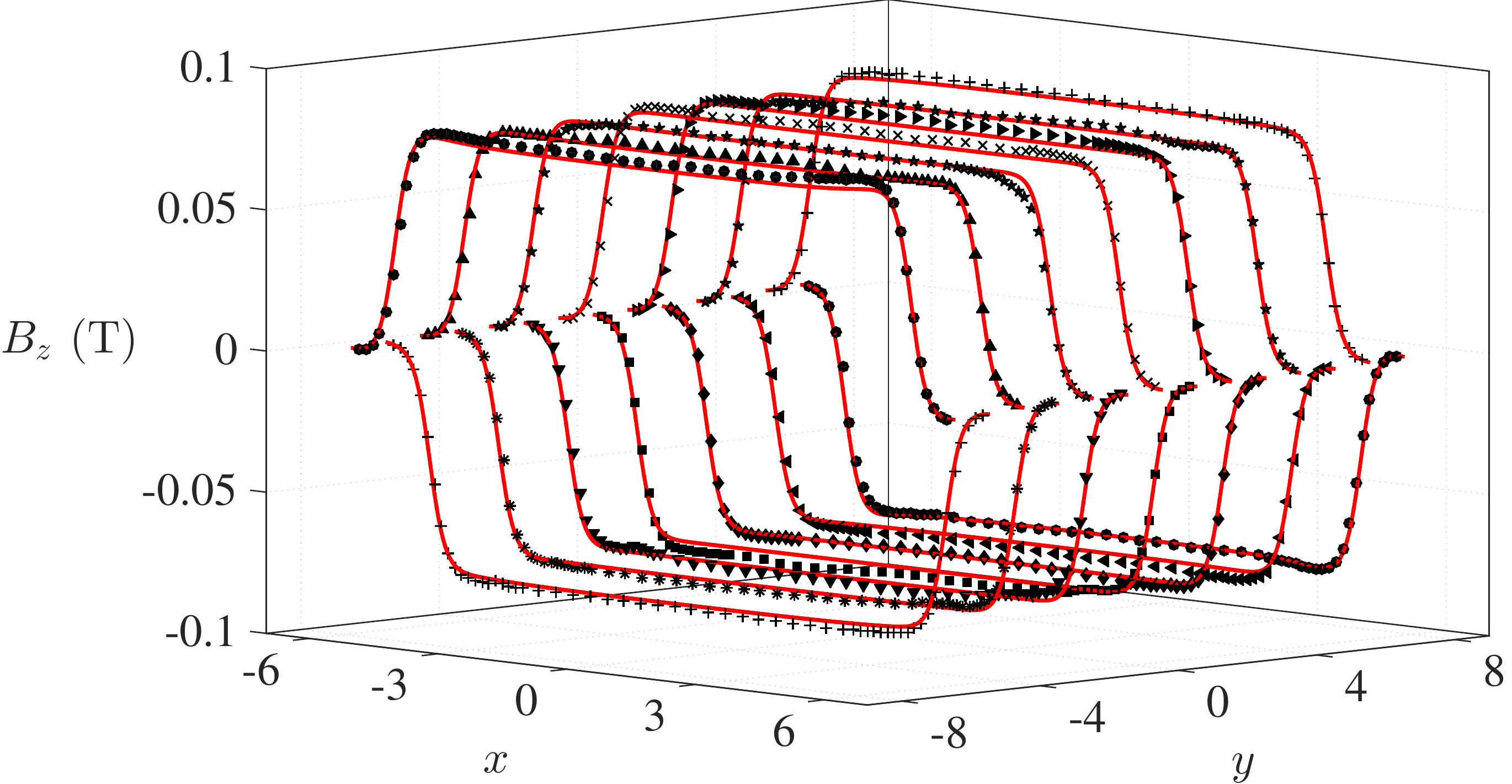}}
\caption{\label{fig:mag_prof} The $z$-component of the magnetic field, $B_z$, (a) at the longitudinal centre of the domain ($x=0$) and (b) along the magnet centrelines at $y=\pm\{ 0.5, 1.5, 2.5, 3.5, 4.5, 5.5, 6.5\}$. In (a), the experimental measurements at a height $z=0.265$ (just above the dielectric-electrolyte interface) and at $z=0.438$ (just below the electrolyte free surface) are shown, respectively, as open squares and filled circles. In (b), the black symbols indicate the experimental measurements at a height $z=0.438$ along 
the magnet centrelines. A least-squares fit has been performed using the data in (a) to determine the scaling factor for the dipole summation; the scaled dipole summation magnetic field is shown as the red lines. The experimental uncertainties are the size of the symbols or smaller.} 
\end{figure}

The black symbols in figure \ref{fig:mag_prof} (a) show the experimental measurements of $B_z$ along the line $x = 0$, passing above the centre of the magnet array at two different heights. Clearly, the magnetic field profile deviates significantly from that of a pure sinusoid. Furthermore, one cannot ignore the fringe fields near the edges of the array. To obtain a magnetic field profile that closely resembles the one in the experiment, one could measure the $z$-component of the magnetic field ($B_z$) across the entire flow domain at various heights above the magnet array. Using the measured field, one could then compute the depth-averaged forcing profile using equation (\ref{eq:depth_averaging}) \citep{suri_2014}. However, since measuring $B_z$ on a 3D grid is an extremely tedious process, we circumvent the labour by numerically modelling the magnet array as described below.

The magnets in the array are arranged such that adjacent ones have magnetization pointing in opposite directions, along $\pm {\bf \hat z}$. To obtain a magnetic field that closely resembles the one due to this array, we model each magnet 
as a uniformly magnetized medium, i.e., as a 3D cubic lattice of identical dipoles, each with a moment $m {\bf \hat z}$. Changing the sign of $m$ across adjacent magnets accounts for the alternating direction of magnetization. The magnetic field at any location $(x,y,z)$ above the array is then approximated using the linear superposition of the field contribution from all of the dipoles modelling the array. Hence, we refer to this model as the ``dipole summation.'' Since the strength of the dipole $m$ cannot be measured experimentally, a single scaling parameter is calculated from a least-squares fit with the experimental measurements, taken at two heights. The rescaled dipole summation magnetic field is shown in figure \ref{fig:mag_prof} (a) (red lines), along with the experimental measurements of $B_z$ (black symbols), corresponding to the line $x=0$ at heights $z = 0.265$ and $z = 0.438$. Figure \ref{fig:mag_prof} (b) shows the magnetic field comparison at $z = 0.438$ along the magnet centrelines. Note that the electrolyte layer in the experiment is bounded by the planes $z = 0.236$ and $z = 0.472$. Hence, we compute the magnetic field $B_z(x,y,z)$ using the dipole summation at various heights, in steps of $0.0197$, in the region $0.236<z<0.472$ and depth-average it using a discrete version of the expression (\ref{eq:depth_averaging}).

\subsection{Boundary Conditions for Direct Numerical Simulations}\label{sec:BCs}

In the experimental Kolmogorov-like flow, vertical solid walls serve as the lateral boundaries, resulting in a no-slip boundary condition for the velocity.  However, for reasons of analytical and computational feasibility, Kolmogorov flow has been studied almost exclusively using unbounded or periodic domains. 
Neither an infinite lateral extent nor periodicity offer a realistic representation of the effect of boundary conditions in the experiment, as far as the flow's structure and its stability are concerned. 
To explore the role of boundaries, we compare the experiment to numerical simulations using computational domains with increasing degrees of confinement. 
The three different computational domains we study are described below.

\begin{itemize}
\item {Doubly-Periodic Domain:} This computational domain is chosen to coincide with the central $8w\times8w$ region of the experimental domain ($|x|\le 4$ and $|y| \le 4$ in nondimensional units).
The simulated flow is constrained to be periodic in both the longitudinal and transverse directions, i.e., ${\bf u}(x=-4,y) = {\bf u}(x=4,y)$ and ${\bf u}(x,y=-4) = {\bf u}(x,y=4)$. 
Along the transverse direction it spans a width equaling that of $8$ magnets. 
The 2D forcing profile ${\bf f} = f_x(y){\bf \hat x}$  over this doubly-periodic domain is constructed from the depth-averaged magnetic field presented in \S \ref{sec:dipole} by retaining only the two dominant Fourier modes, $\sin(\kappa y)$ and $\sin(3\kappa y)$, along the $y$-direction, where $\kappa=\pi$ in dimensionless units.  Along the $x$-direction the profile is uniform: $f_x(y)=0.95\sin(\kappa y) + 0.05\sin(3\kappa y)$.

\item {Singly-Periodic Domain:} This computational domain coincides with the region $|x|\le 7$ and $|y| \le 4$. The longitudinal dimension is the same as that of the experiment, while the transverse one spans a width equaling that of 8 magnets, like in the doubly-periodic domain. No-slip boundary conditions are imposed at the end walls, i.e., ${\bf u}(x=\pm 7,y) = 0$, while periodic boundary conditions are imposed along the transverse direction, i.e., ${\bf u}(x,y=4) = {\bf u}(x,y=-4)$. 
The 2D forcing profile ${\bf f} = f_x(x,y){\bf \hat x}$ over this singly-periodic domain is constructed as a product of two one-dimensional profiles $f_x(x,y)=\chi(x)\psi(y)$. Along the $y$-direction the profile is once again constructed by retaining only two dominant Fourier modes of the depth-averaged magnetic field, $\psi(y)=0.95\sin(\kappa y) + 0.05\sin(3\kappa y)$. Along the $x$-direction the profile $\chi(x)$ is chosen to be the depth-averaged magnetic field profile from the dipole summation along the magnet centreline $y=0.5$.  We note that the effect of transverse confinement has been studied by \citet{thess_1992a}, and therefore is not investigated here separately.
	
\item {Non-Periodic Domain:} This computational domain is identical to the experimental one in both lateral dimensions, i.e., $|x|\le 7$ and $|y|\le 9$, with no-slip boundary conditions imposed at both the end walls and side walls, i.e., ${\bf u}(x=\pm 7,y) = 0$ and ${\bf u}(x,y=\pm 9) = 0$. As mentioned in \S \ref{sec:dipole}, the forcing over this domain is computed by depth-averaging the dipole summation.
\end{itemize}

To compare the experimental observations with those predicted by equation (\ref{eq:2dns_mod}) with the three types of boundary conditions described above, we have performed direct numerical simulations. The flow over the doubly-periodic domain is simulated using a pseudo-spectral method in the vorticity-stream function formulation, as described in \citet{mitchell_2013}.  
This simulation is henceforth referred to as the ``doubly-periodic simulation,'' abbreviated DPS.  For the singly-periodic and the non-periodic domains, numerical simulations have been performed using a finite-difference scheme, described in \citet{armfield_1999}. 
These simulations are hereafter referred to as the ``singly-periodic simulation'' (SPS) and the ``non-periodic simulation'' (NPS), respectively. 
Details of the spatiotemporal discretization and the integration schemes employed in all the numerical simulations can be found in Appendix \ref{sec:num_meth}.

\section{Comparison of Experiment and Simulations} \label{sec:comparison}

In this section we present the results of our comparison between the experiment and the numerical simulations on the three domains described above. 
First, we discuss the straight uniform flow found at lower Reynolds numbers, with a special emphasis on the effect of boundaries. 
We then perform linear stability analysis to demonstrate that equation (\ref{eq:2dns_mod}) describes the primary instability more accurately than equation (\ref{eq:2dns_wf}) on an unbounded domain, but still substantially underpredicts the experimentally observed critical Reynolds number. 
Next, we describe and compare the steady flow states found in the experiments and simulations above the primary instability. 
Finally, we discuss the secondary instability which gives rise to a time-periodic flow. 

\subsection{Straight Flow}\label{sec:str_flow_sec}

\begin{figure}
\quad\subfloat[]{\includegraphics[height=1.62in]{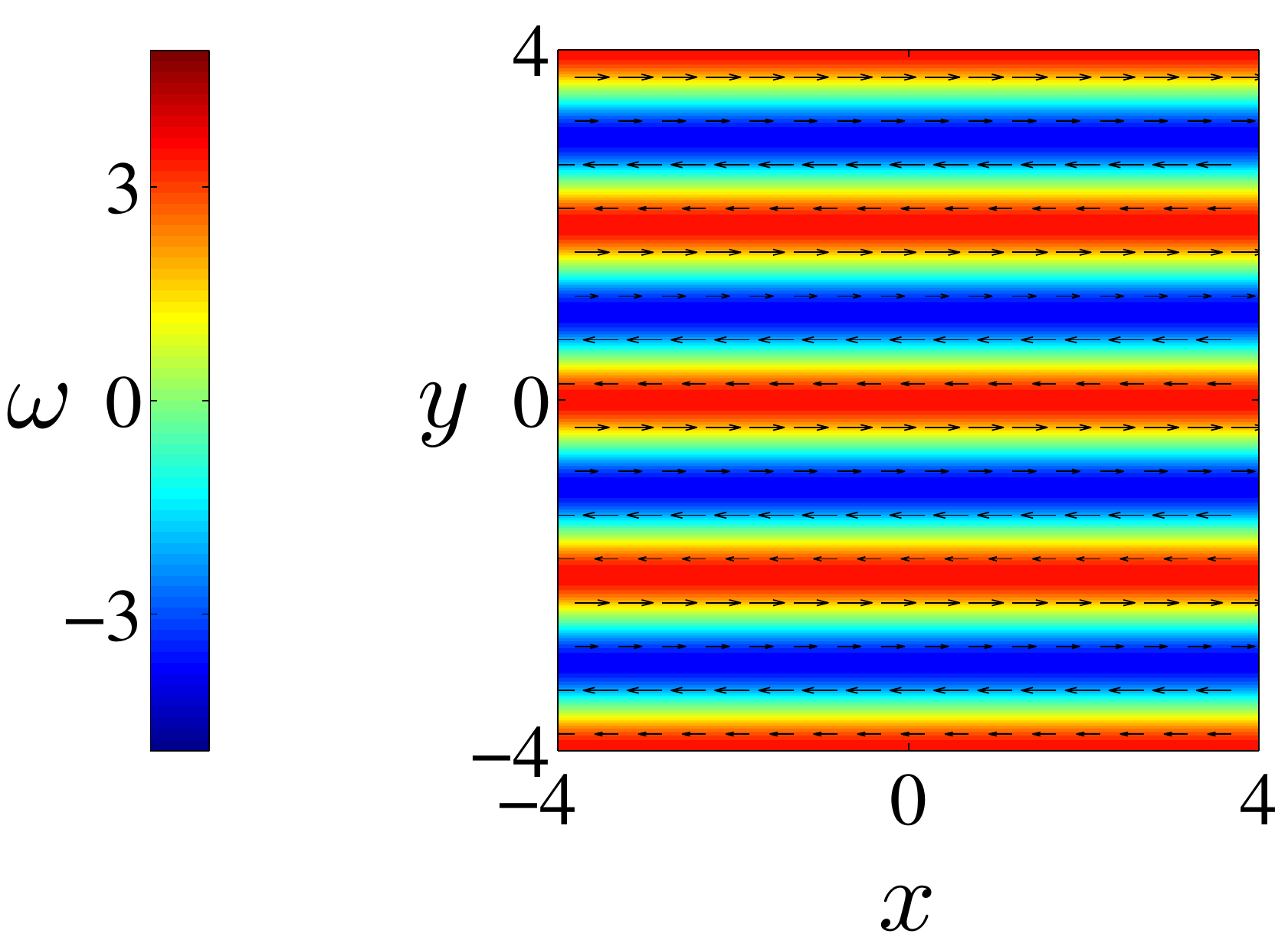}}\hspace{9mm}
\subfloat[]{\includegraphics[height=1.62in]{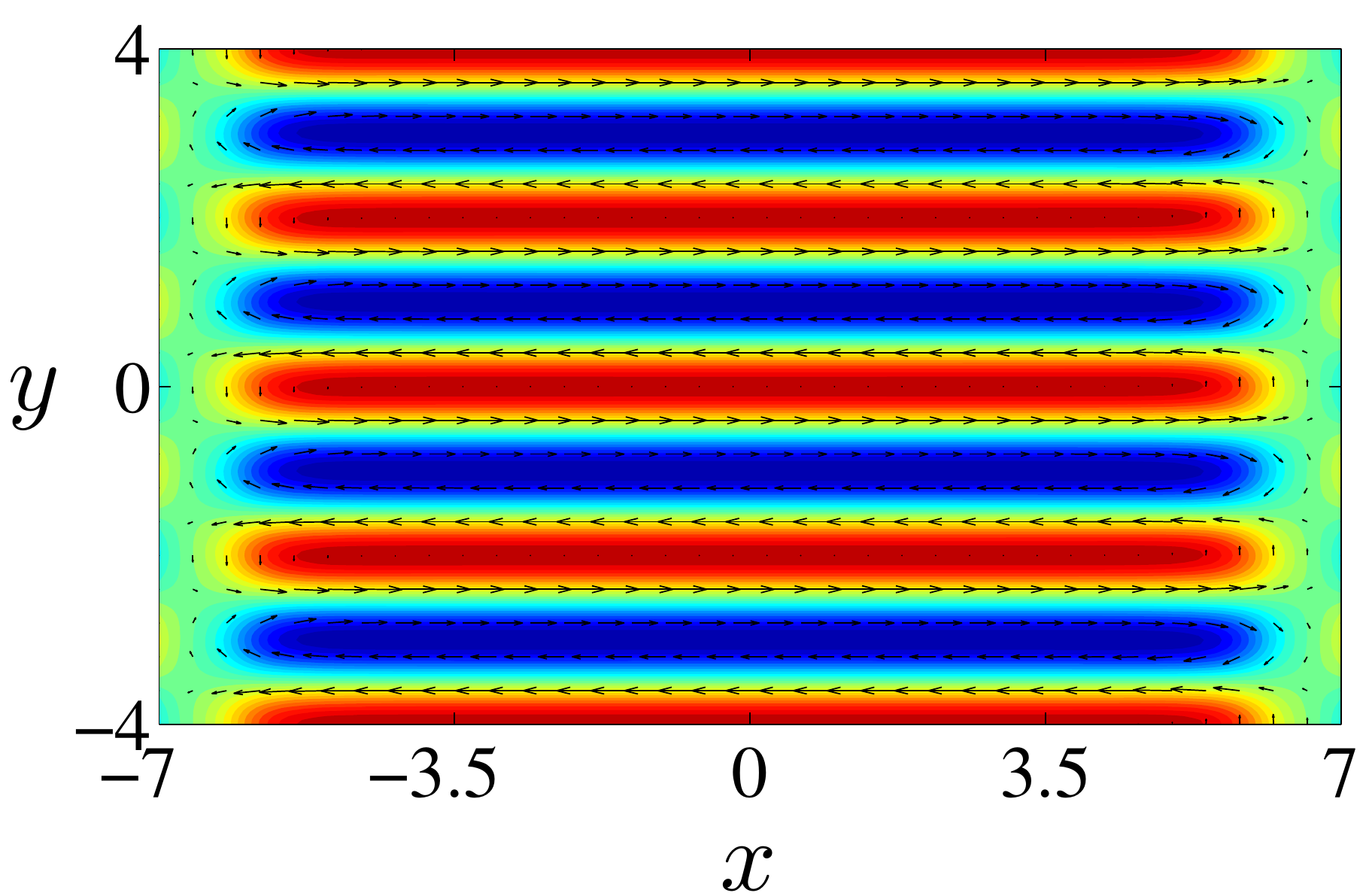}}\quad \\
\subfloat[]{\includegraphics[width=2.5in]{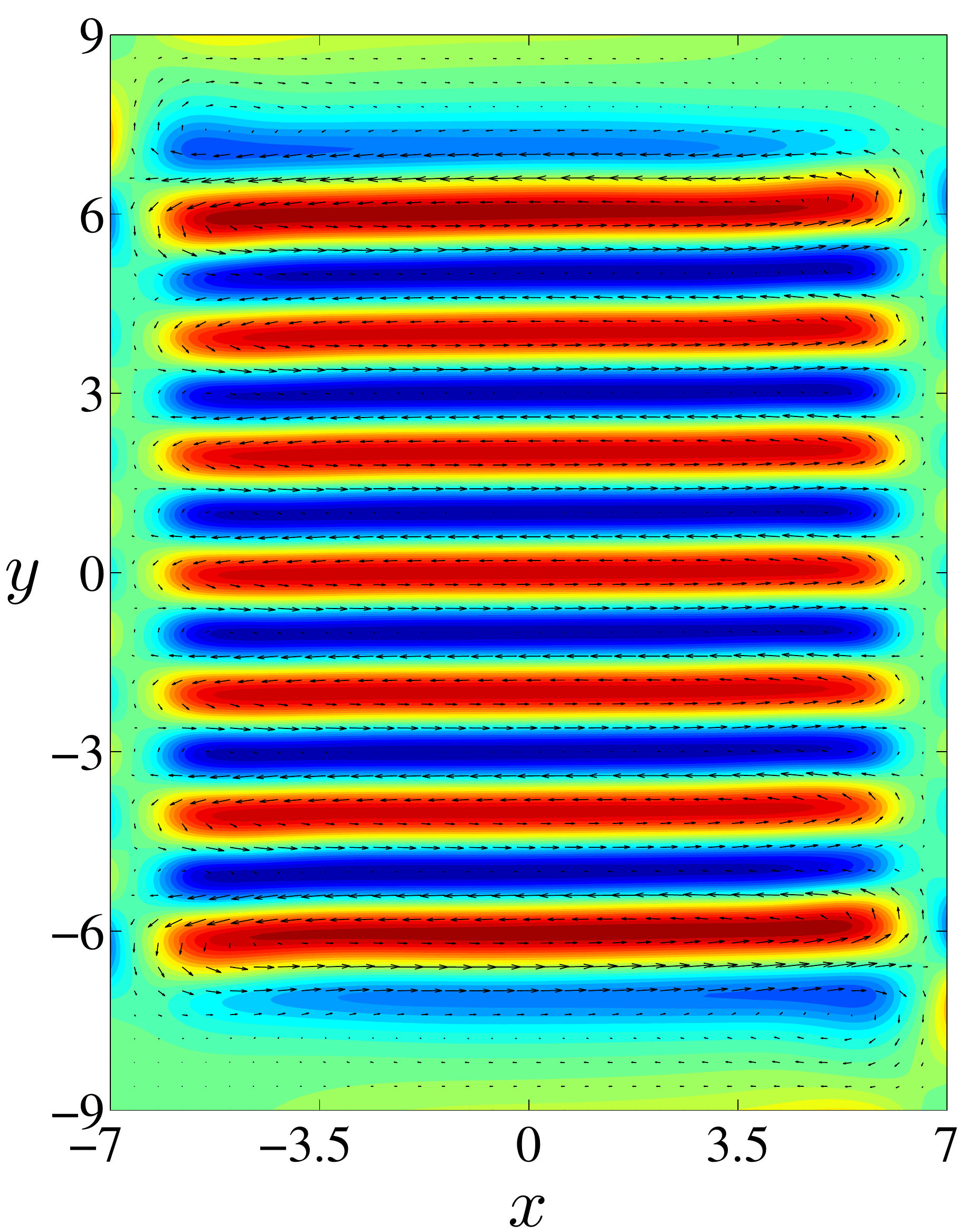}}\qquad
\subfloat[]{\includegraphics[width=2.5in]{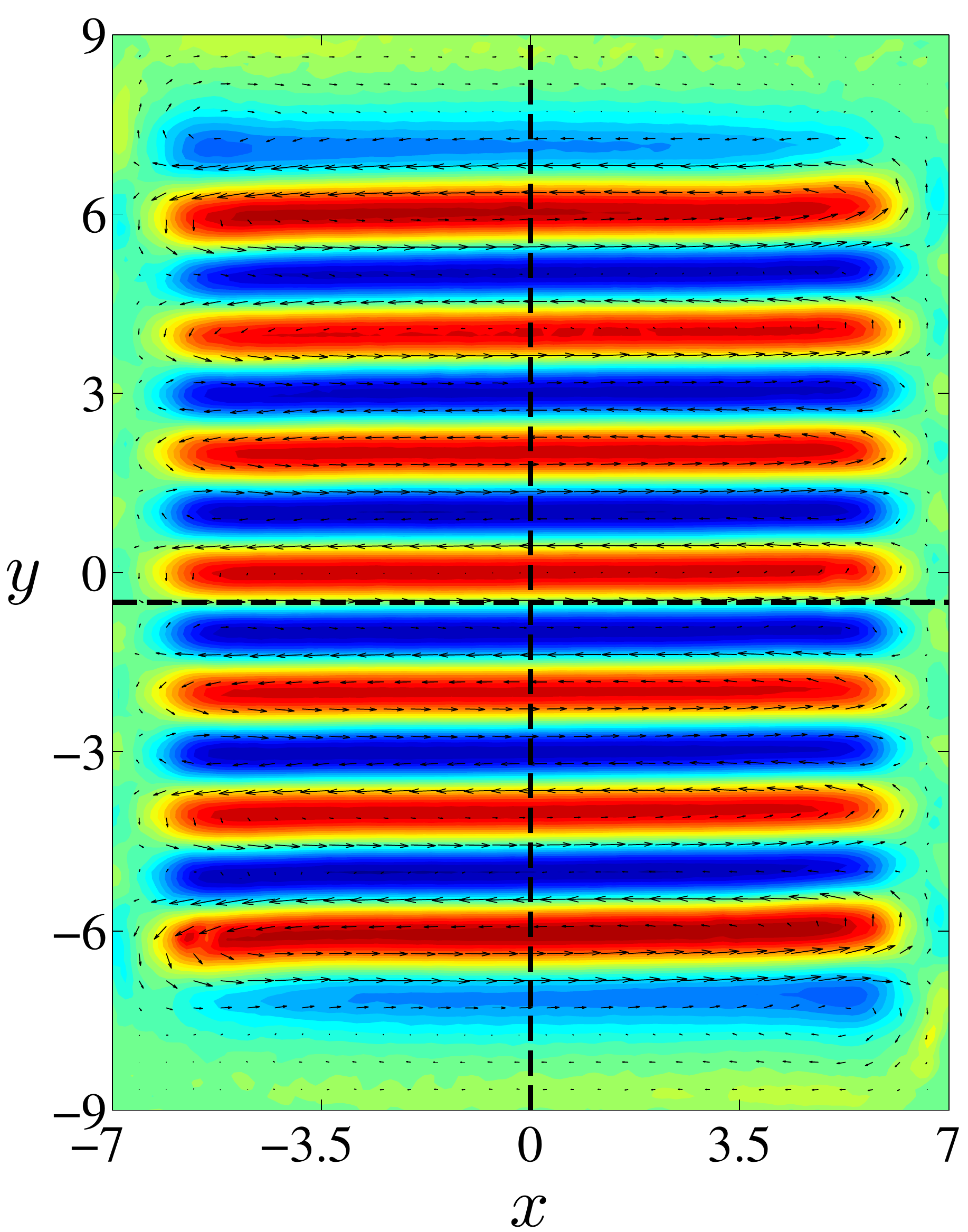}}\\
\caption{\label{fig:str_flow} Straight flow fields at $Re=8.1$ with $\alpha=0.064$ s$^{-1}$, $\beta=0.83$, and ${\nu}=3.26 \times 10^{-6}$ m$^2$/s for the (a) DPS, (b) SPS, (c) NPS, and (d) experiment. The dashed lines in (d) indicate the locations of velocity profiles in the experiment that are compared to the simulations. The vorticity colour scale plotted for (a) also applies to (b-d).  The velocity vectors are downsampled in each direction by a factor of 8 for the simulations and 4 for the experiment.}
\end{figure}

For weak driving, the flow mimics the forcing closely, with spatially alternating bands of fluid flow along the $\pm x$-directions, as can be seen in figure \ref{fig:str_flow} for $Re=8.1$. 
In this figure, black vectors represent the velocity field ${\bf u}$ and the colour indicates the vorticity $\omega = (\nabla\times{\bf u})\cdot{\bf {\hat z}}$. For the experiment (figure \ref{fig:str_flow} (d)), the $y$-component of the velocity measured in the central region of the domain is close to zero. 
However, there are regions of strong recirculation near the end walls, characterized by a nonzero $y$-component of velocity. 
A closer inspection of the flow shows a slight tilt in the alignment of the flow bands. 
This tilt is due to the global circulation, resulting from confinement and the fluid flowing in opposite directions over the end magnets at $y = \pm 6.5$. Figures \ref{fig:str_flow} (a) and (b) show the straight flows found in the DPS and SPS. 
It can be seen that flow fields in the DPS and SPS reproduce the experimental flow qualitatively away from the lateral walls. 
Futhermore, the SPS captures the turnaround flow near the end walls. 
However, neither the SPS nor the DPS displays the tilt of the flow bands observed in the experiment since the periodic flows are devoid of 
global circulation. 
In contrast, the NPS generates a flow field that looks indistinguishable from the experimental one (cf. figure \ref{fig:str_flow} (c)).

\begin{figure}
\centering
\subfloat[]{\qquad\quad\includegraphics[width=4.55in]{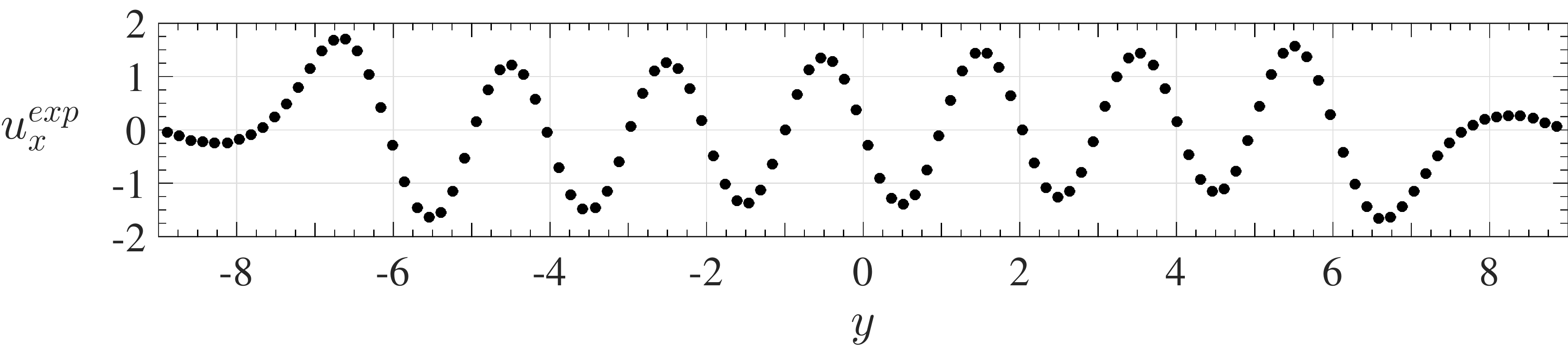}} \\
\subfloat[]{\includegraphics[width=4.95in]{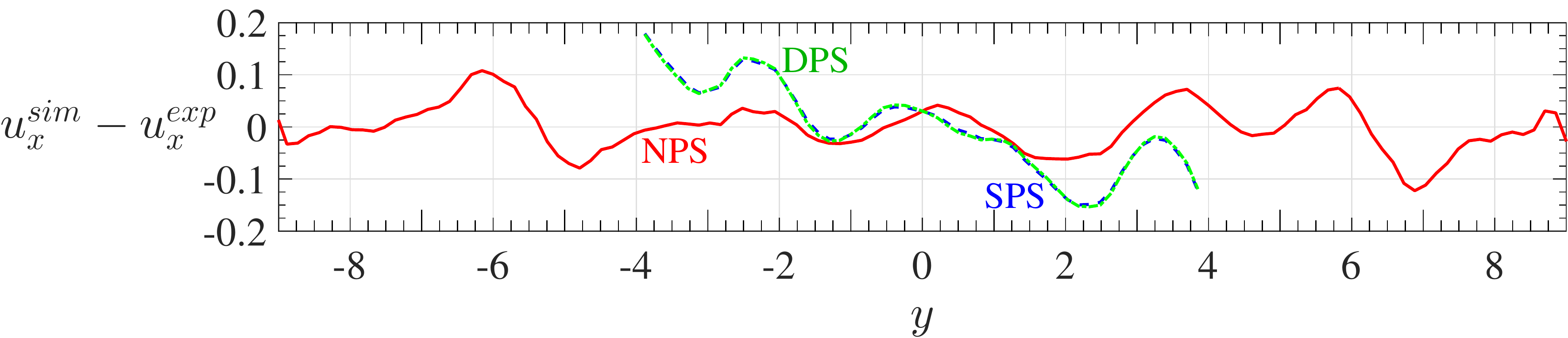}}\hfill  \\
\subfloat[]{\qquad\quad\includegraphics[width=4.6in]{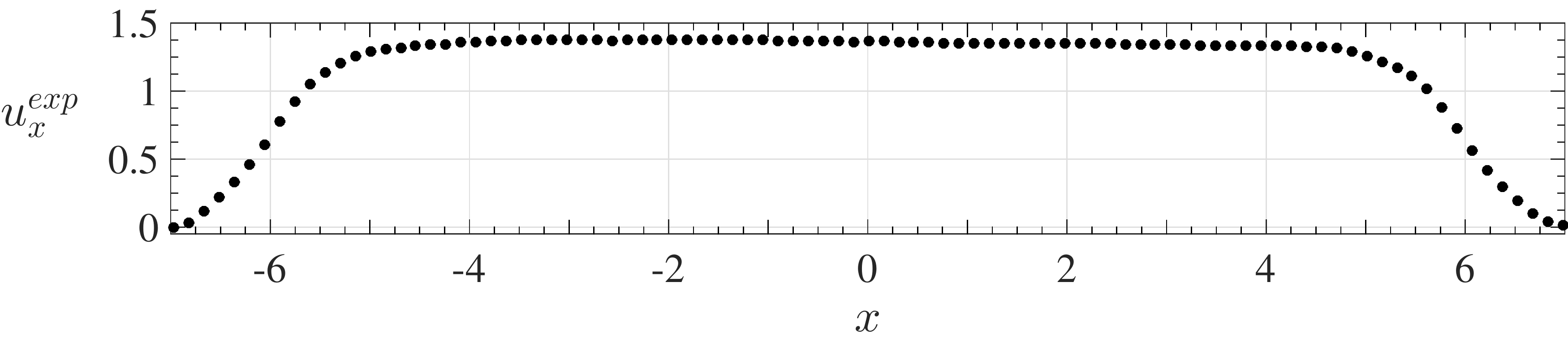}}\hfill \\
\subfloat[]{\includegraphics[width=5in]{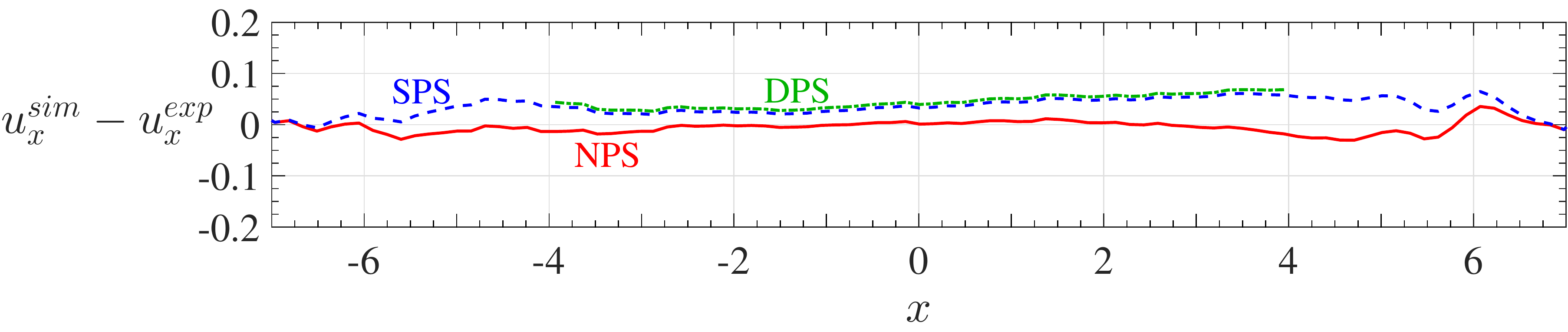}} \hfill 
\caption{\label{fig:vel_prof} Profiles of the longitudinal velocity and longitudinal velocity differences at $Re=8.1$ with $\alpha=0.064$ s$^{-1}$, $\beta=0.83$, and ${\nu}=3.26 \times 10^{-6}$ m$^2$/s. 
(a) $u_x^{exp}$ as a function of $y$ at the longitudinal centre ($x=0$), 
(b) the difference between the longitudinal velocity in the simulations and the experiment, $u_x^{sim}-u_x^{exp}$, as a function of $y$ at the longitudinal centre ($x=0$); note that the curves corresponding to the DPS and SPS are virtually indistinguishable, 
(c) $u_x^{exp}$ as a function of $x$ at the centreline of a middle magnet ($y=-0.5$), and 
(d) the difference between the longitudinal velocity of the simulations and the experiment, $u_x^{sim}-u_x^{exp}$, as a function of $x$ at the centreline of a middle magnet ($y=-0.5$); note that the curves corresponding to the DPS and SPS are virtually indistinguishable in the region $-4<x<4$, where the DPS is defined.
Experimental uncertainties are the size of the symbols or smaller.}
\end{figure}

For a quantitative description of the straight flow profile, we have plotted in figure \ref{fig:vel_prof} (a) the longitudinal component $u_x^{exp}$ of the velocity along the line $x=0$ in the experiment. 
The location of this cross section is indicated by the vertical dashed line in figure \ref{fig:str_flow} (d). The difference in $u_x$ between the experiment and the numerical simulations along this line is shown in figure \ref{fig:vel_prof} (b). 
As can be seen, the DPS and SPS, which are only defined for $|y|\le 4$, show systematic deviation from the experiment as high as 18\% since they do not capture global circulation. 
In comparison, the NPS agrees to within about 5\% over the same region, with no clear systematic deviation. 
The disagreement between the experiment and NPS in this region, we believe, is a result of the dipole summation not accounting for the variation in the strength of each individual magnet. 
Closer to the boundaries, at $y \approx 7$ and $y \approx -6$, the largest difference between the NPS and the experiment is around 12\%. The origin for this error is quite subtle and we shall defer its analysis to Appendix \ref{sec:sensitivity}.

The experimental longitudinal velocity component $u_x^{exp}$ at $y=-0.5$ (along a central magnet centreline) is shown in figure \ref{fig:vel_prof} (c). 
The very slight asymmetry in the longitudinal velocity is a result of the global circulation. 
In contrast, the flow in the DPS is perfectly uniform and thus does not capture this asymmetry, as can be seen from the plot of its difference with the experimental profile in figure \ref{fig:vel_prof} (d). 
The SPS, which is defined all the way to the end walls, also does not capture this asymmetry due to the lack of global circulation. 
The NPS produces the closest agreement: the corresponding flow displays the asymmetry observed in the experiment, with no significant systematic deviation.  In summary, the NPS succeeds in capturing the effects of confinement in the experiment with good accuracy, while the DPS and SPS show significant systematic deviations.

\subsection{Linear Stability Analysis of the Straight Flow}\label{sec:LSA}

As the strength of the forcing increases, the flow in the experiment undergoes a qualitative change at $Re_c = 11.07\pm0.05$, with uniform flow bands (cf. figure \ref{fig:vel_prof} (c)) developing 
modulation that eventually gives rise to distinct stationary vortices.
Hence, we shall refer to this flow as the ``modulated flow." 
Several previous experimental studies have reported this transition and have characterized it using the critical Reynolds number ($Re_c^{exp}$) and wavenumber ($k_{c}^{exp}$) of the 
modulation  \citep{bondarenko_1979, obukhov_1983, batchaev_1983}. 
In our experiments, the wavenumber just above this transition was measured to be $k_c^{exp} = 0.50\kappa$, where $\kappa$ is the wavenumber associated with the forcing.
In virtually all previous studies, theoretical estimates for these critical parameters have been obtained by using equation (\ref{eq:2dns_wf}) and modelling the straight flow in experiment as a strict sinusoid ${\bf u}_{s} \propto \sin(\kappa \it{y})$; the flow stability is then analyzed with respect to 
perturbations $\delta{\bf u}(y)e^{ikx}$ in the transverse component of the velocity.
In this section, we revisit this analytical approach for equation (\ref{eq:2dns_mod}) to provide estimates for the critical parameters.
Many previous studies used a different nondimensionalization, which corresponds to setting the nondimensional forcing wavenumber $\kappa$ to unity. To make comparison easier, we will introduce a scaled wavenumber $q=k/\kappa$ which corresponds to the convention used in those studies.

The strictly sinusoidal straight flow  governed by equation (\ref{eq:2dns_mod}) on an unbounded domain becomes unstable with respect to perturbations with wavenumber  $q$ above the Reynolds number $Re=Re_n(q)$, which to a very good accuracy is given by:
\begin{equation}\label{eq:Re_0}
Re_n(q) = \frac{\pi}{\beta}\frac{1}{q}\sqrt{\frac{(1+q^2)}{(1-q^2)}\left(q^2+\frac{\alpha}{{\nu}\kappa^2}\right)\left(1+q^2+\frac{\alpha}{{\nu}\kappa^2}\right)}.
\end{equation}
This expression was computed by linearizing equation (\ref{eq:2dns_mod}) around ${\bf u}_{{s}}$ and calculating its stability with respect to perturbations including three dominant modes, 
\begin{equation}
\delta{\bf u}(y)e^{ikx}=\sum_{n=-1,0,1}\epsilon_ne^{i\kappa (ny+qx)}.
\end{equation}
A detailed discussion of the stability analysis and the analytical expression for the neutral stability curve, similar in form to that in equation  (\ref{eq:Re_0}), can be found in Appendix B of \citet{dolzhansky_2012}. The critical Reynolds number $Re_{c} = \min_{q}Re_n(q)$ and the corresponding critical wavenumber $k_{c} = \kappa q_{c}$ computed using the expression (\ref{eq:Re_0}) can be compared with experimental observations.

\begin{figure}
\centering
\includegraphics[width=3in]{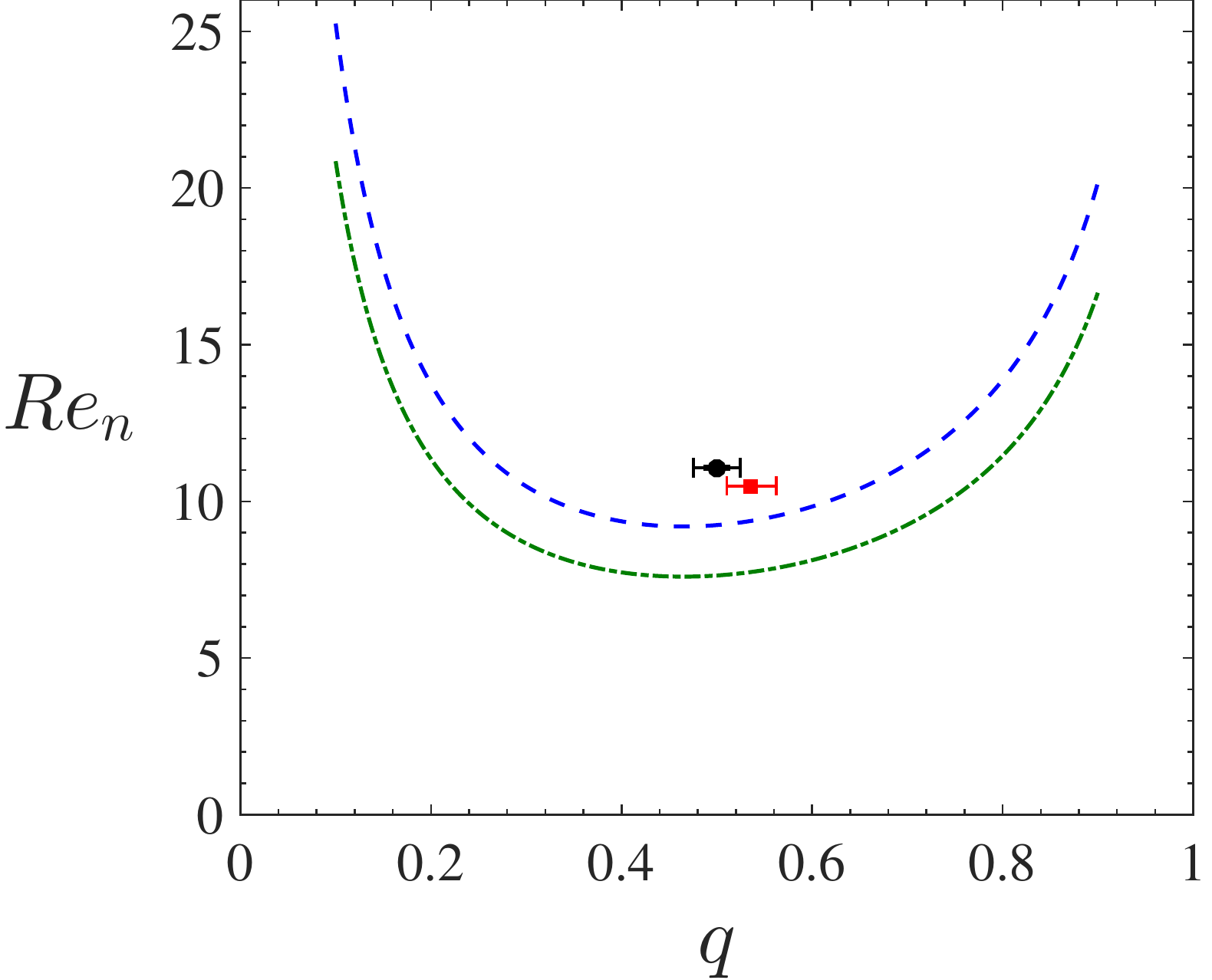}
\caption{\label{fig:nsc} Neutral stability curves (\ref{eq:Re_0}) describing the primary instability. The blue dashed line corresponds to $\alpha=0.064$ s$^{-1}$ and $\beta=0.83$, while the green dot-dashed line corresponds to $\alpha=0.064$ s$^{-1}$ and $\beta=1.0$.  The measurement from the experiment (NPS) is plotted as a black dot (red square); note that the uncertainties in $Re_c$ are smaller than the size of the symbols. 
In all the cases, ${\nu}=3.26 \times 10^{-6}$ m$^2$/s is held constant. }
\end{figure}

The neutral stability curve (blue dashed line) which corresponds to the experimental values of parameters $\alpha$, $\beta$, and ${\nu}$ is shown in figure \ref{fig:nsc}. 
The minimum of this neutral stability curve yields a critical Reynolds number $Re_c=9.16$ and an associated critical wavenumber $q_c = 0.465$.  
The black dot on the plot indicates the critical values $Re_c^{exp} = 11.07$ and $q_{c}^{exp} = 0.50$, corresponding to the instability we observe in the experiment. 
The relative difference $(Re_c^{exp}-Re_c)/Re_c^{exp}$ between the theoretical estimate for the critical Reynolds number and that measured in experiment is about 17\%. The critical wavenumber, however, is in better agreement with the experimentally measured one, with a 7\% relative error. 

While the critical Reynolds number obtained from the linear stability analysis clearly disagrees with the experimentally observed one, it is still a significant improvement over analytical estimates for a flow modelled using equation (\ref{eq:2dns_wf}), which corresponds to setting $\beta = 1$ in equation (\ref{eq:2dns_mod}). 
The corresponding neutral stability curve is indicated by the green dot-dashed line in figure \ref{fig:nsc}. 
From equation (\ref{eq:Re_0}) it can be seen that the entire neutral stability curve scales as $1/\beta$. This implies that the critical wavenumber ($q_c = 0.465$) is independent of $\beta$, while the predicted critical Reynolds number for $\beta = 1$ is $Re_c = 7.60$. 
This is a 31\% discrepancy with the experimental value, which is comparable to the 30\% discrepancy reported by \citet{bondarenko_1979} in a study based on equation (\ref{eq:2dns_wf}). 

As we discussed previously, the parameter $\beta$ describes the effect of the vertical variation in the magnitude of the horizontal velocity on the effective inertia and nonlinearity of the flow. Equation \eqref{eq:2dns_wf} does not account for this effect, so it is natural that its predictions are substantially less accurate.

\subsection{Modulated Flow}\label{sec:mod_flow_sec}

\begin{figure}
\quad\subfloat[]{\includegraphics[height=1.62in]{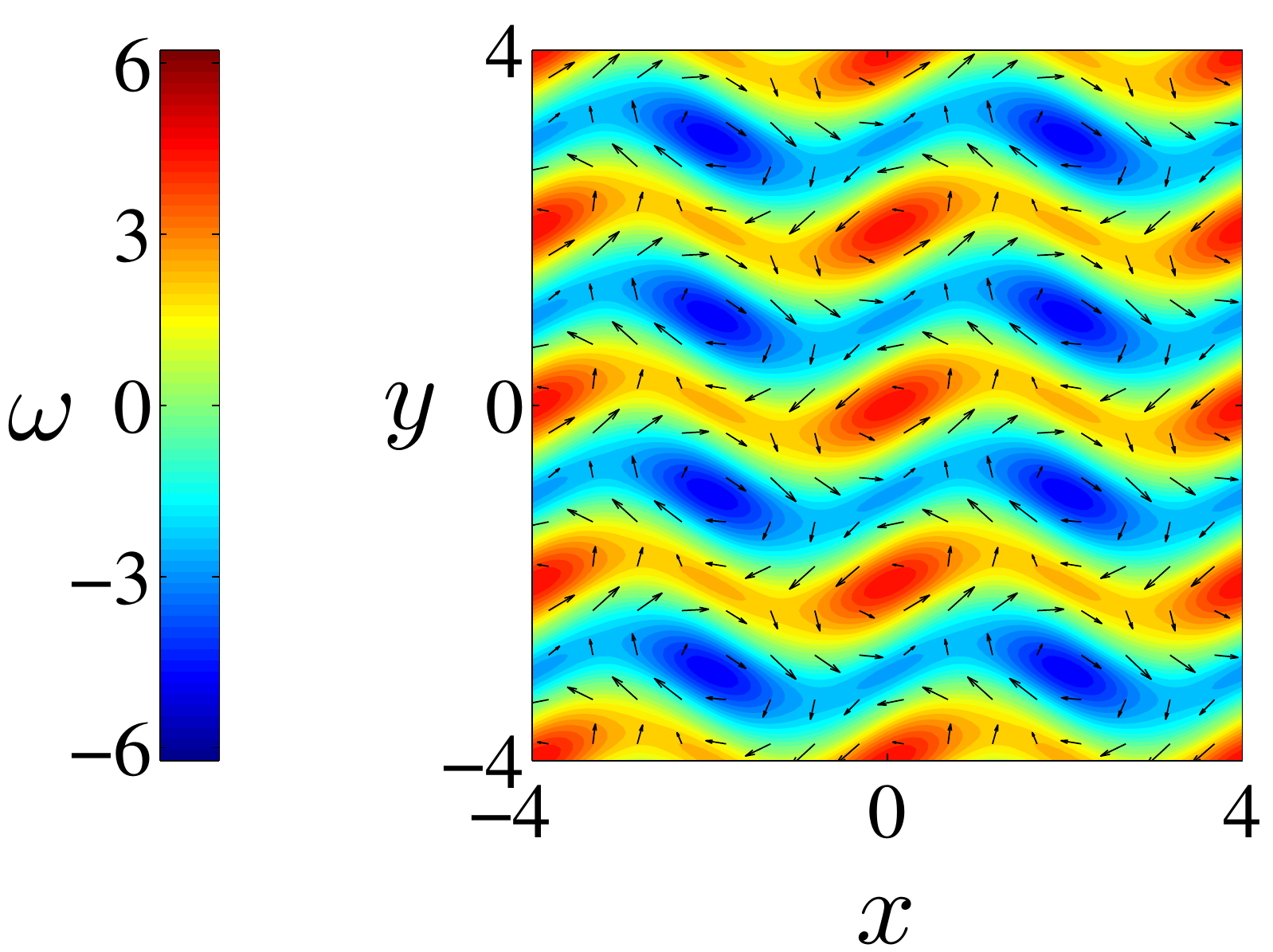}}\hspace{11mm}
\subfloat[]{\includegraphics[height=1.62in]{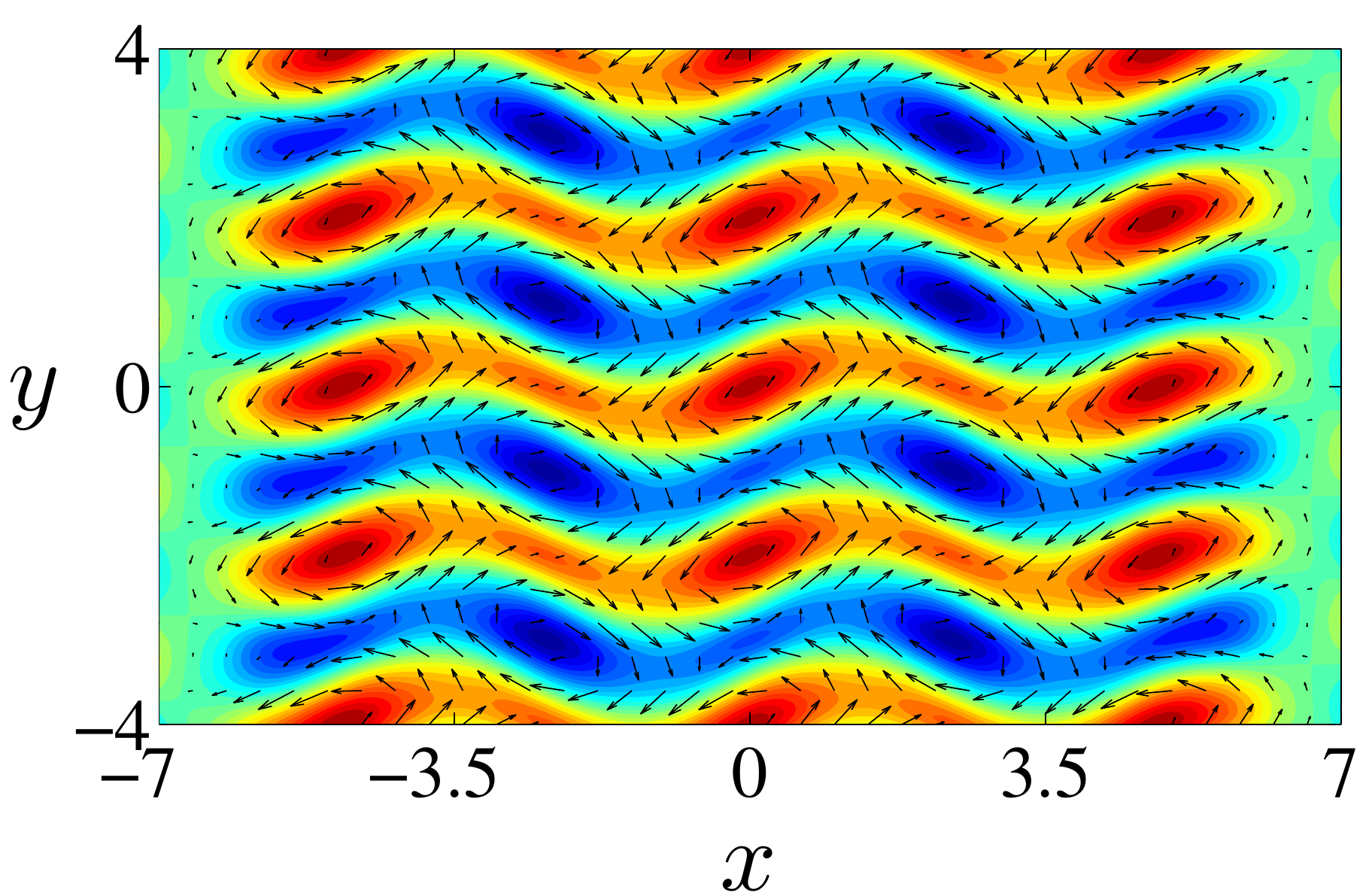}}\quad \\
\subfloat[]{\includegraphics[width=2.5in]{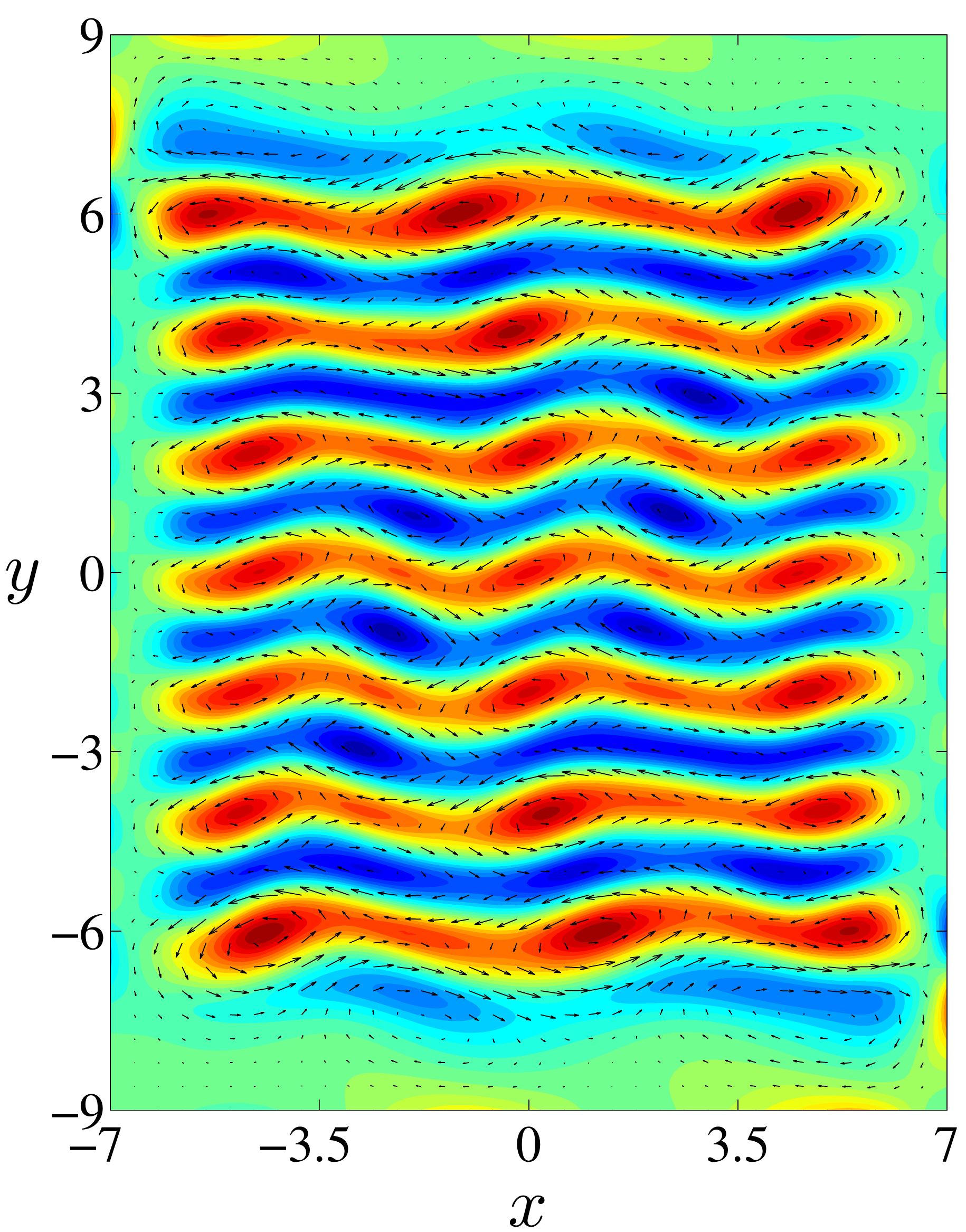}}\qquad
\subfloat[]{\includegraphics[width=2.5in]{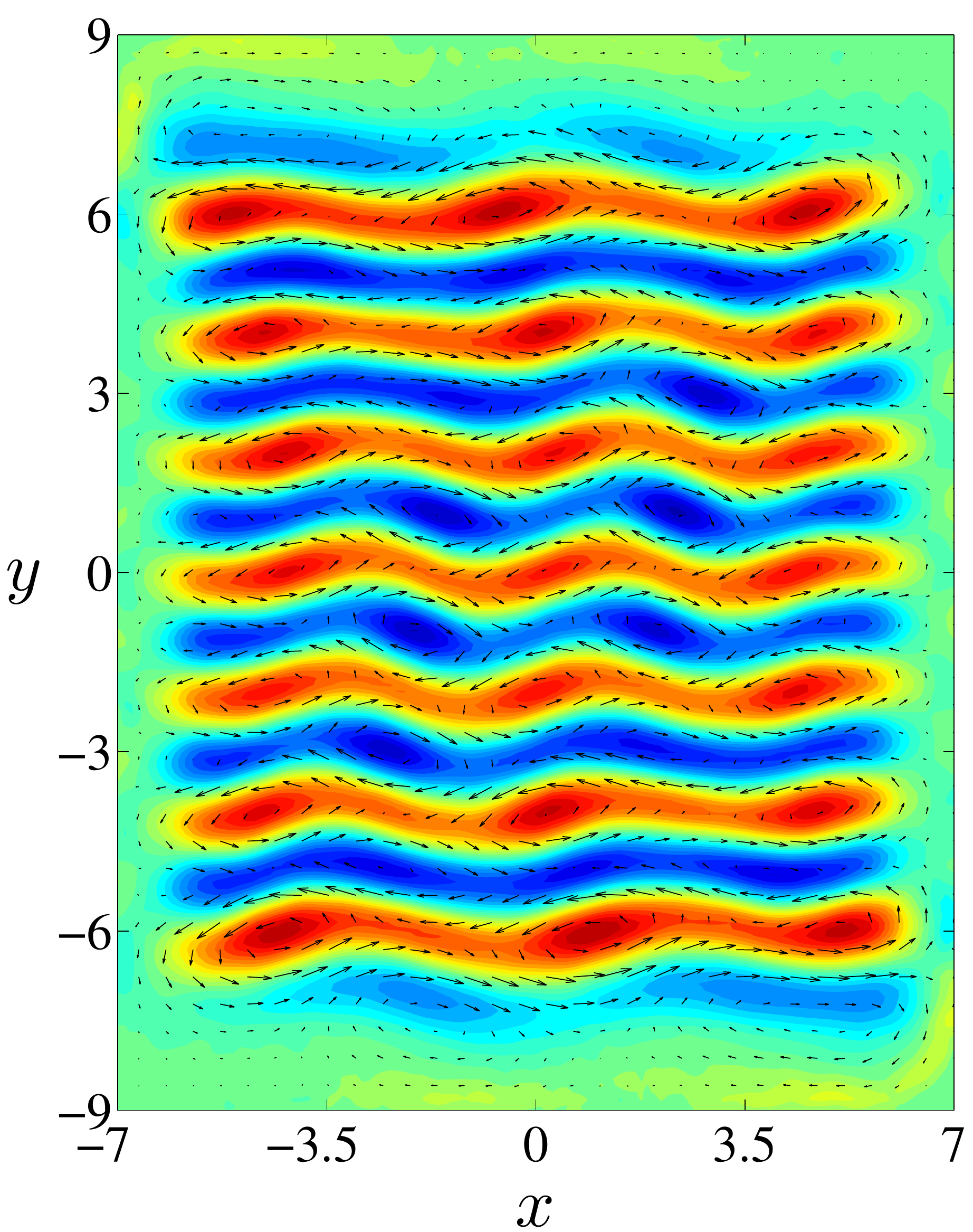}}\\
\caption{\label{fig:mod_state} Modulated flow fields at $Re=14$ with $\alpha=0.064$ s$^{-1}$, $\beta=0.83$, and ${\nu}=3.26 \times 10^{-6}$ m$^2$/s for the (a) DPS, (b) SPS, (c) NPS, and (d) experiment.  The vorticity colour scale plotted for (a) also applies to (b-d).  The velocity vectors are downsampled in each direction by a factor of 8 for the simulations and a factor of 4 for the experiment.}
\end{figure}

Figure \ref{fig:mod_state} (a-d) shows the modulated flow fields corresponding to the DPS, SPS, NPS, and experiment, respectively, at $Re = 14$. 
At this Reynolds number, the modulated flow is well developed and is visually quite distinct from the straight flow. 
The counterclockwise global circulation in the experiment strongly affects the alignment of the vortices (see figure \ref{fig:mod_state} (d)) as can be seen by comparing the modulated flows in the DPS and SPS with the relevant regions of the experimental flow. 
Unlike the DPS and SPS, the flow field in the NPS captures the features observed in the experiment remarkably well. 
This unambiguously demonstrates the importance of properly modelling the confinement effects in both the longitudinal and the transverse direction to reproduce the features of the flow in the experiment.

\begin{figure}
\subfloat[]{\includegraphics[width=2.55in]{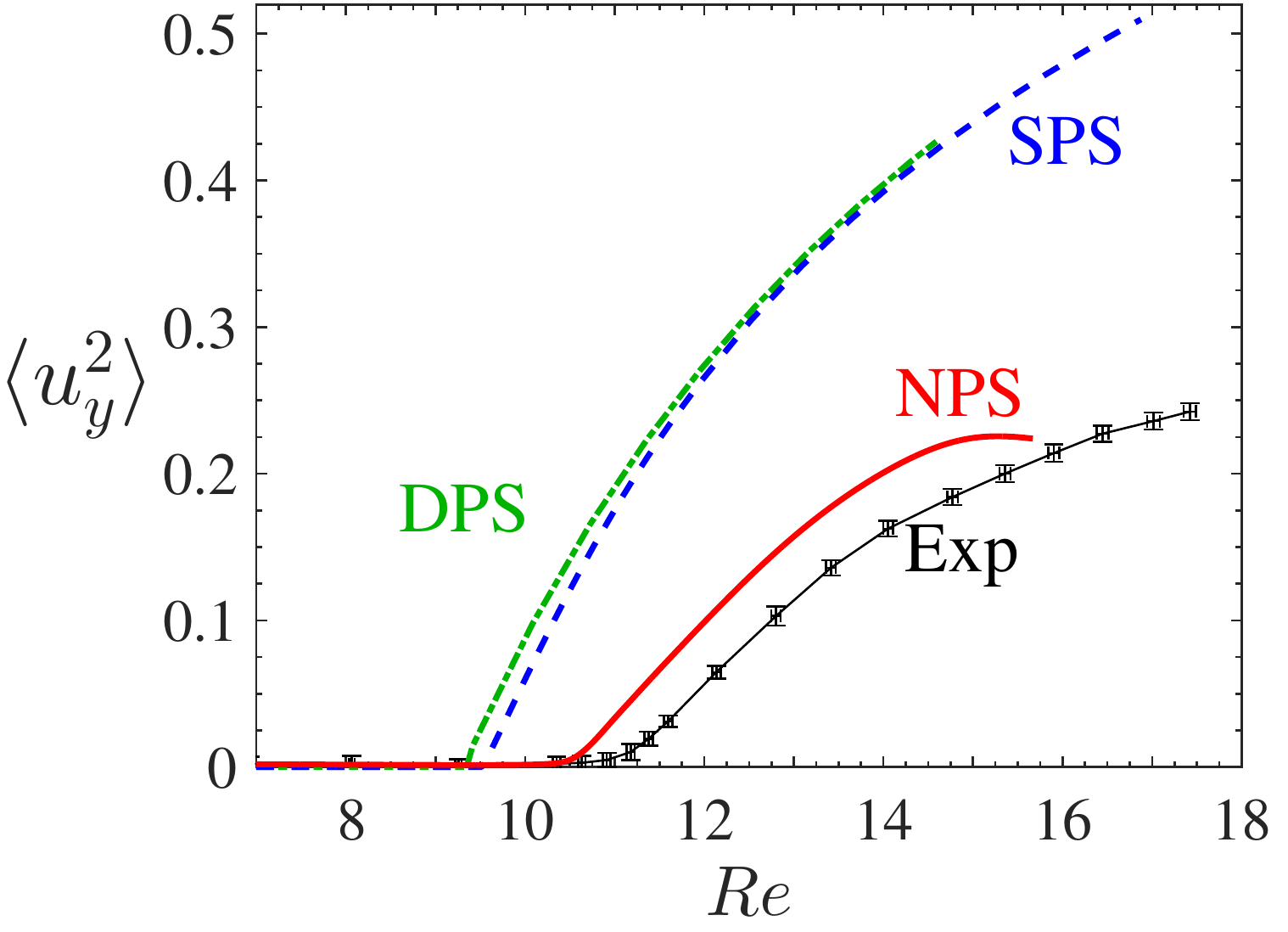}}\qquad
\subfloat[]{\includegraphics[width=2.34in]{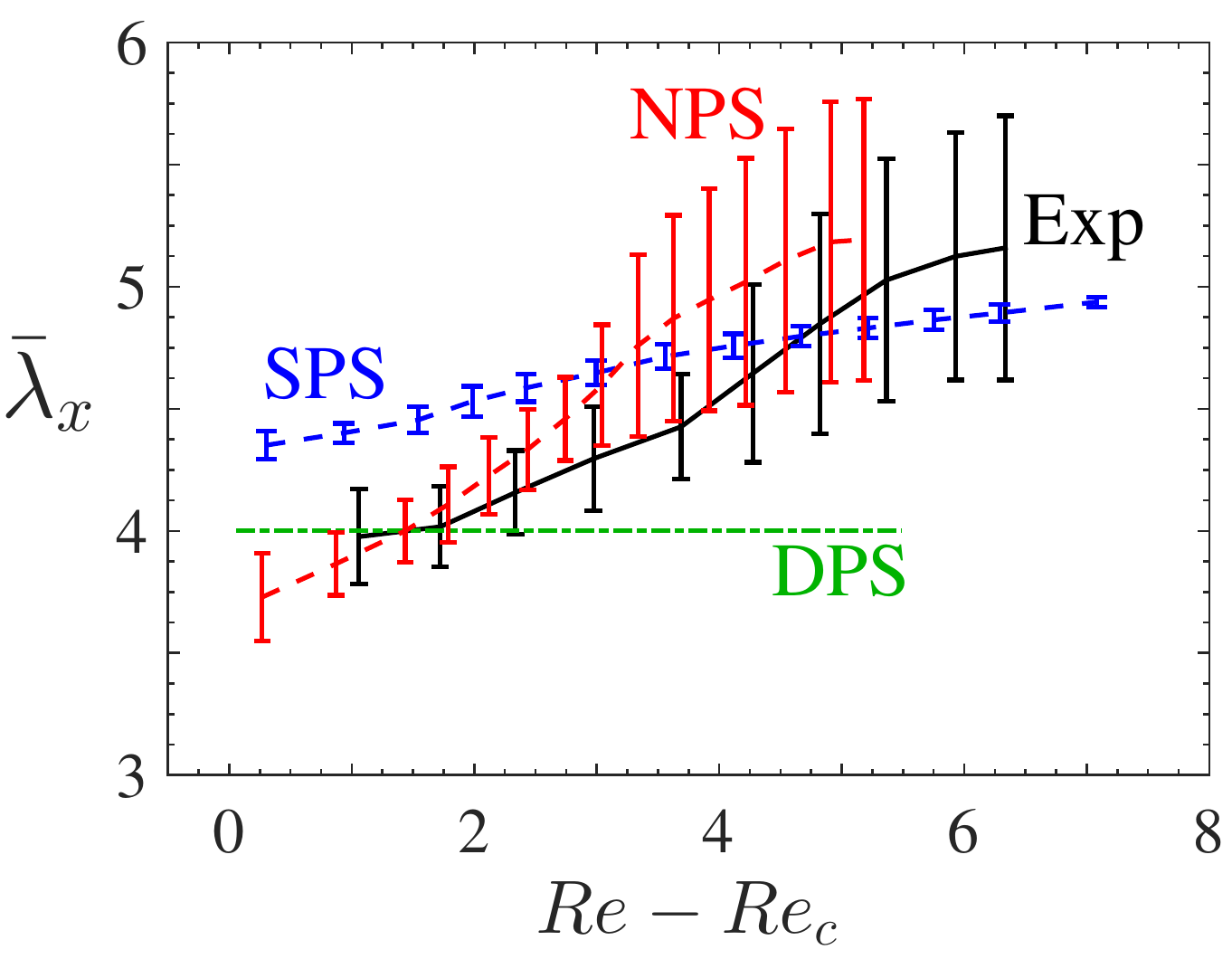}}
\caption{\label{fig:bif1} Primary instability for $\alpha=0.064$ s$^{-1}$, $\beta=0.83$, and ${\nu}=3.26 \times 10^{-6}$ m$^2$/s. (a) A bifurcation diagram and (b) the average wavelength of the pattern, $\bar{\lambda}_x$, as a function of $Re$ for the modulated flow regime.  
At each $Re$, wavelength measurements are made in the central region $|y|\le 4$ then averaged; the uncertainty bars indicate one standard deviation in the spatial measurements.}
\end{figure}

The onset of the modulated flow is characterized by the appearance of the transverse component $u_y$ of the velocity throughout the flow domain. 
As the driving is increased, the magnitude of $u_y$ also increases. 
A bifurcation diagram characterizing the transition from the straight to the modulated flow is shown in figure \ref{fig:bif1} (a).  
We use the spatial mean square transverse velocity, $\langle u_y^2\rangle$, 
as the order parameter and plot it as a function of $Re$.  
The spatial average is computed over the central region $|x|\le 4$ and $|y|\le 4$ for all simulations and experiment.  
In comparison to the experimental value of $Re_c^{exp} = 11.07$, the primary instability in the DPS and SPS occurs at much lower Reynolds numbers $Re_c=9.39$ and $Re_c=9.53$, respectively. 
In contrast, by imposing the correct (no-slip) boundary conditions in both the longitudinal and transverse directions, in addition to using a realistic model of the magnetic field, the transition can be predicted quite accurately. The straight to modulated transition in the NPS occurs at $Re_c = 10.49$ (red square in figure \ref{fig:nsc}.), which is within 5.2\% of $Re_c^{exp}$. 
Finally, we note that setting $\beta=1$ results in a poor prediction $Re_c=8.71$ even in the NPS, which corresponds to a 21\% error.

Given that the pattern of vortices observed in the experiment lacks perfect periodicity, we compute the average longitudinal wavelength $\bar{\lambda}_x$ using the spatial average of the separation between adjacent vortex centres in the central region $|y|\le 4$; the vortex centres are identified by locating local minima in the velocity magnitude. 
Just above onset, the vortices in the experiment form a lattice with a fairly uniform separation, $\bar{\lambda}_{x}^{exp} \approx 4.0$ ($q_c^{exp}=0.50$).
As the forcing is increased, the mean separation between the vortices increases, as can be seen from the plot of $\bar{\lambda}_x$ versus $Re-Re_c$ shown in figure \ref{fig:bif1} (b). 
Additionally, the vortex lattice becomes spatially irregular, as can be seen in figure \ref{fig:mod_state} (d). This spatial variation is quantified in the plot in figure \ref{fig:bif1} (b) wherein the uncertainty bars indicate one standard deviation in the spatial variation of the separation between adjacent vortices. 
Note that immediately above onset, accurate identification of the vortex centres in the experiment is not possible because of the very weak modulation; hence, experimental measurements are only plotted for $Re - Re_c \geq 1$.

For comparison, figure \ref{fig:bif1} (b) also shows the average wavelength of the flow pattern in the DPS, SPS, and NPS. 
Finer spatial resolution in the simulation, compared to that in experiment, facilitates measuring $\bar{\lambda}_x$ closer to onset with greater accuracy. 
In the DPS, the size of the domain along $x$ was chosen {\it a posteriori} to be commensurate with the critical wavelength at onset in the experiment. Despite this, neither the spatial variation of the wavelength nor its variation with $Re - Re_c$ observed in the experiment are captured.  
The SPS, however, shows a qualitatively similar trend for the dependence of $\bar{\lambda}_x$ on $Re - Re_c$.
The periodicity in the transverse direction results in a uniform vortex pattern with smaller spatial variation in the separation between vortices compared to the experiment. 
In contrast, the NPS captures both the spatial variation of the wavelength and the distortion of the lattice with increasing forcing quite satisfactorily. 

At $Re-Re_c\approx1$, the discrepancy is much smaller than the uncertainty bars, 
but for $Re-Re_c \gtrsim 1.5$, the NPS overestimates the wavelength compared to what is observed in the experiment. 
The largest discrepancy, which is 0.46 (a 10\% relative error), occurs around $Re-Re_c= 3.7$. 
The difference in the flow patterns in the NPS and the experiment is due to the deviation of the latter from being perfectly Q2D. The analysis in Appendix \ref{sec:sensitivity} shows that the wavelength of the pattern sensitively depends on relatively minor changes in the forcing profile, which is responsible for the observed discrepancy between the numerics and experiment.

\begin{figure}
\centering
\subfloat[]{\includegraphics[height=1.9in]{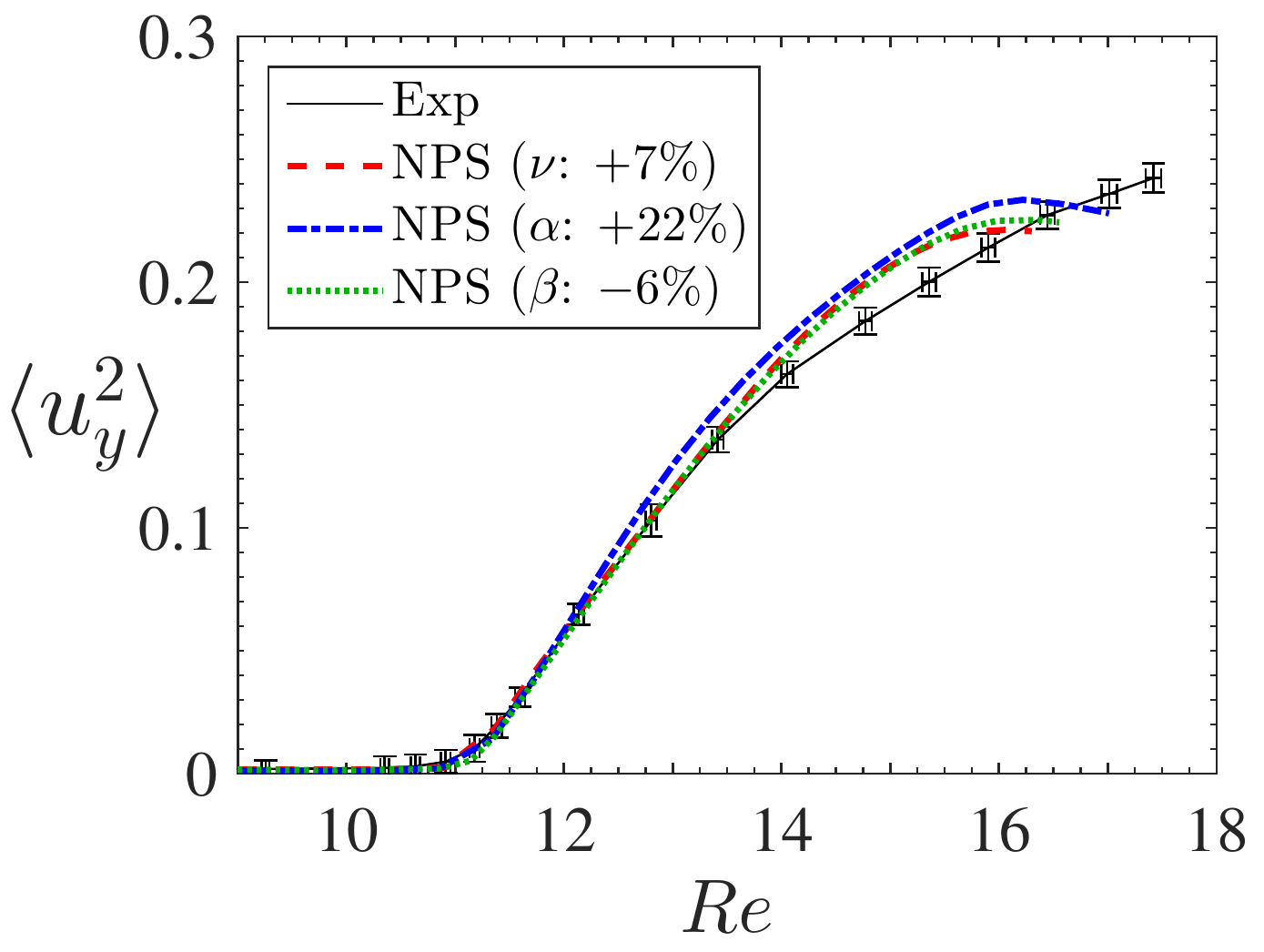}} \quad
\subfloat[]{\includegraphics[height=1.9in]{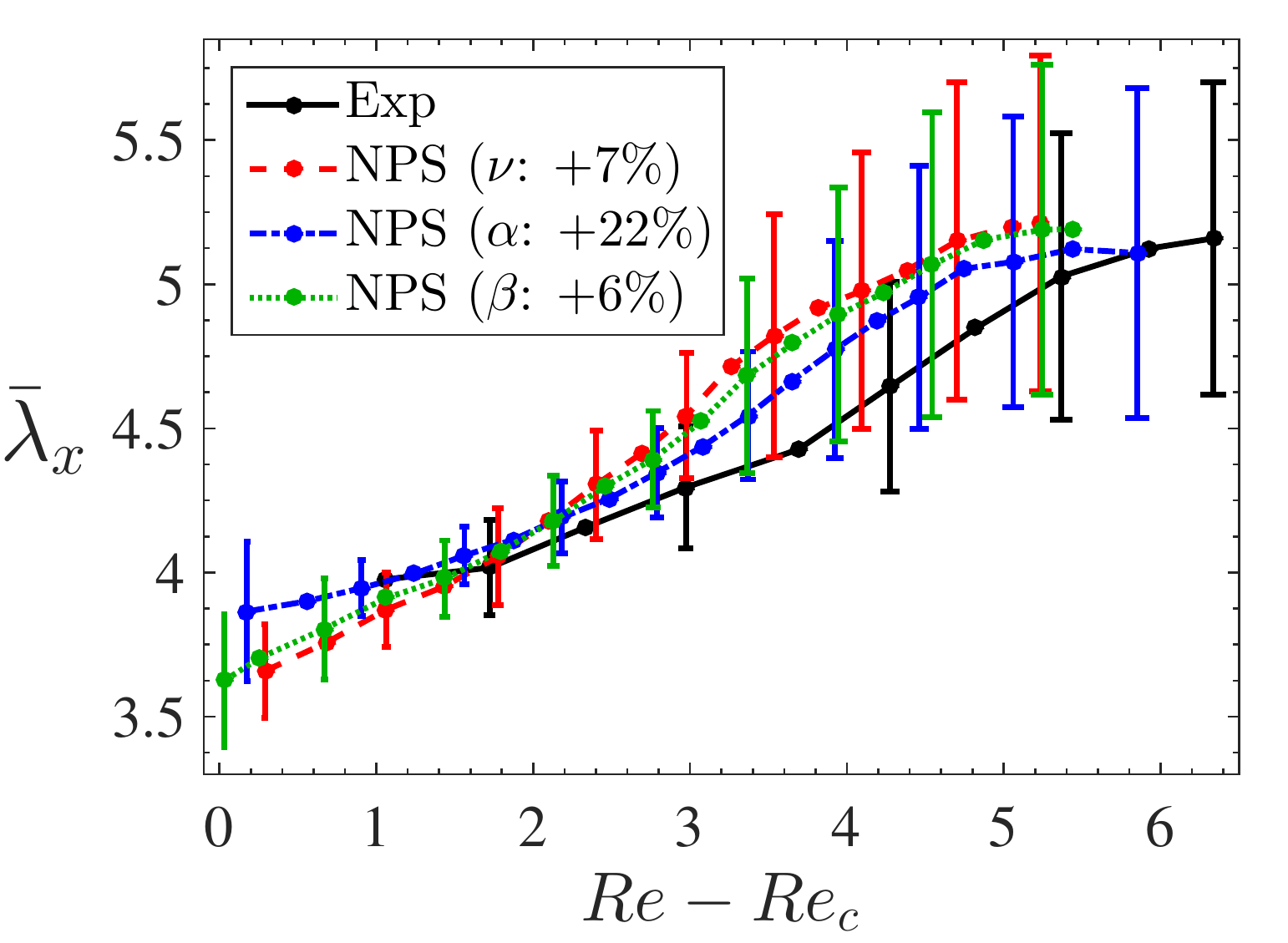}}
\caption{\label{fig:bif1_vary_params} The effect of the variation in model parameters. (a) A bifurcation diagram and (b) the average wavelength of the pattern in the modulated flow regime. 
The numerical results correspond to either a 7\% increase in ${\nu}$, a 22\% increase in $\alpha$, or a 6\% decrease in $\beta$ compared with the depth-averaged values for the straight flow ($\alpha=0.064$ s$^{-1}$, $\beta=0.83$, and ${\nu}=3.26 \times 10^{-6}$ m$^2$/s). The uncertainty bars in (b) are only shown for every other data point for clarity.}
\end{figure}

While the NPS provides a reasonably accurate description of the transition from the straight to the modulated flow in the experiment, as we mentioned previously, it somewhat underestimates the critical Reynolds number.
To resolve this discrepancy, we tested the sensitivity of the transition in the NPS to changes in the forcing profile as well as variations in parameters $\beta$, ${\nu}$, and $\alpha$. 
In particular, we found that $Re_c$ is fairly insensitive to spatial variation in the strength of the magnets in the array. Consequently, we turned our attention to studying the sensitivity of $Re_c$ to the values of the parameters $\beta$, ${\nu}$, and $\alpha$.
In order to match $Re_c$ in the NPS and experiment, we had to either decrease $\beta$ by 6\%, increase ${\nu}$ by 7\%, or increase $\alpha$ by 22\%.
Figures \ref{fig:bif1_vary_params} (a) and (b) show that the variation of parameters has a fairly weak effect on both the amplitude of the modulation and the wavelength of the pattern, 
 which suggests that the disagreement between the simulation and experiment is primarily due to the deviation of the flow and/or forcing from quasi-two-dimensionality, which is discussed in Appendix \ref{sec:sensitivity}. 

Note that the Reynolds number $Re = Uw/{\nu}$ is defined using the measured rms velocity $U$ and parameter ${\nu}$ which cannot be measured, but has to be computed. While in the simulation the value of ${\nu}$ is well-defined (it is one of the parameters of the model), in the experiment it is not, so the corresponding $Re$ depends on the choice of ${\nu}$. Hence, to enable a proper comparison of experiment with numerics, we defined $Re$ in both cases using the analytically computed depth-averaged value ${\nu} = 3.26\times 10^{-6}$ m$^2$/s \citep{suri_2014}, regardless of the actual value of ${\nu}$ used in the simulation. Matching $Re$ using this convention is effectively equivalent to matching the rms velocity $U$. 

\subsection{Secondary Instability}\label{sec:sec_instablity}

As we increase the forcing further, the modulated flow in the experiment becomes unstable giving way to a time-periodic flow with a period $T_p=42.8\pm0.4$ ($120\pm1$ s in dimensional units) at onset, which corresponds to $Re_p = 17.6\pm0.1$. The modulated state in the NPS, with the depth-averaged ($\alpha=0.064$ s$^{-1}$, $\beta=0.83$, and ${\nu}=3.26 \times 10^{-6}$ m$^2$/s) as well as the adjusted parameters, undergoes a Hopf bifurcation as the forcing is increased. Table \ref{table:hopf} compares $Re_p$ and $T_p$ in the experiment with those from the NPS for the different parameter sets. A video comparing the time-periodic flow in the NPS (with depth-averaged parameters) and experiment is included as online supplementary material. 

\setlength{\tabcolsep}{1em}
\begin{table*}\centering
\ra{1.3}
\begin{tabular}{@{}lccc@{}}
\toprule
{} & $Re_p$ & $T_p$ \\  
\midrule
Experiment & $17.6\pm0.1$ & 42.8 ($120\pm1$ s) \\
NPS (Depth-Averaged) & 15.6 & 43.2 (137 s)  \\ 
NPS (${\nu}$: +7\%) & 16.4 & 43.2 (130 s)  \\ 
NPS ($\alpha$: +22\%)  & 17.1 & 48.1 (139 s) \\ 
NPS ($\beta$: $-6$\%)  & 16.5 & 45.8 (137 s) \\
\bottomrule
\end{tabular}
\caption{\label{table:hopf}Critical transition parameters characterizing the stable periodic regime for the experiment and the NPS with different sets of parameters.}
\end{table*}

$Re_p$ and $T_p$ in the simulation are within $15$\% of the experimental measurements for the depth-averaged values of parameters computed using the vertical profile $P(z)$ that corresponds to the straight flow. 
However, the values of ${\nu}$, $\alpha$, and $\beta$ should vary slowly with $Re$, since $P(z)$ is weakly dependent on the horizontal flow profile. 
Hence, a different set of parameters is required to describe the two instabilities and, more generally, there is no universal set of parameters $\beta$, ${\nu}$, and $\alpha$ that correctly describes the experimental flow at all $Re$. 
From table \ref{table:hopf} we see that $Re_p$ and $T_p$ show very different sensitivity to changes in each of the parameters.
Hence, while separately modifying ${\nu}$, $\alpha$, and $\beta$ shows some improvement in matching either $Re_p$ or $T_p$, it should be possible to obtain even better agreement by modifying \textit{all} the model parameters simultaneously, each by only a few percent.

The necessity for modifying parameters across different dynamical regimes also raises the question of how robust $\alpha$, $\beta$, and $\nu$ are to changes in the (local) wavenumber of the flow. To test this, we have recomputed the parameters using the wavenumber $k \approx \sqrt{5/4}\kappa$ associated with the modulated flow. We found that $\beta$ and $\nu$ change by less than 1\%, and  $\alpha$ by about 3.5\%, compared to those computed using $k = \kappa$. This robustness suggests that, once adjusted to match the experiment, the 2D model should provide a reasonably accurate description of the dynamics even in the weakly turbulent regime where the wavenumber may vary in space and time \citep{suri_2017}.

\section{Nature of the Primary Instability}\label{sec:pitchfork}

An important consequence of confining the flow in the longitudinal or transverse directions is that we restrict the set of coordinate transformations (symmetries) that leave the governing equation (\ref{eq:2dns_mod}) equivariant. 
The symmetries of the governing equation, in turn, determine the number of, and the relation between, distinct modulated flow solutions created as a result of the primary bifurcation.  Below we discuss each of the different flow domains, in the order of decreasing symmetry.

\subsection{DPS}

On an unbounded or a doubly-periodic domain, equation (\ref{eq:2dns_mod}) is equivariant under the following symmetry operations \citep{chandler_2013}: 
\begin{enumerate}[leftmargin=!,labelindent=20pt,itemindent=-5pt]
\item \ Continuous shift by $\delta x$ in $x$: $\mathcal{T}^{\delta x}_x(x,y)\rightarrow(x+\delta x,y)$. 
\item \ Reflection in $x$ combined with a discrete shift of half a period in $y$: $\mathcal{R}_x\mathcal{T}^w_y(x,y)\rightarrow(-x,y+w)$.
\item \ Reflections in both $x$ and $y$: $\mathcal{R}_x\mathcal{R}_y(x,y) \rightarrow (-x,-y)$.
\end{enumerate}
Note that the double reflection $\mathcal{R}_x\mathcal{R}_y$ is equivalent to a rotation by angle $\pi$ about the $z$-axis, while the square of the symmetry operation $\mathcal{R}_x\mathcal{T}^w_y$ corresponds to a discrete  shift $\mathcal{T}^{2w}_y$ in the $y$-direction,  i.e., $(\mathcal{R}_x\mathcal{T}^w_y)^2=\mathcal{T}^{2w}_y$. 
For the DPS, $0\le\delta x<L_x$, so the corresponding symmetry group is $\mathcal{G}=SO(2)\times Z_2\times Z_n$, where $n$ is the number of magnets (here, $n=8$). It is a subgroup of $E(2)$ isomorphic to ${O}(2)\times{Z}_n$.

\begin{figure}
\centering
\subfloat[]{\includegraphics[height=1.3in]{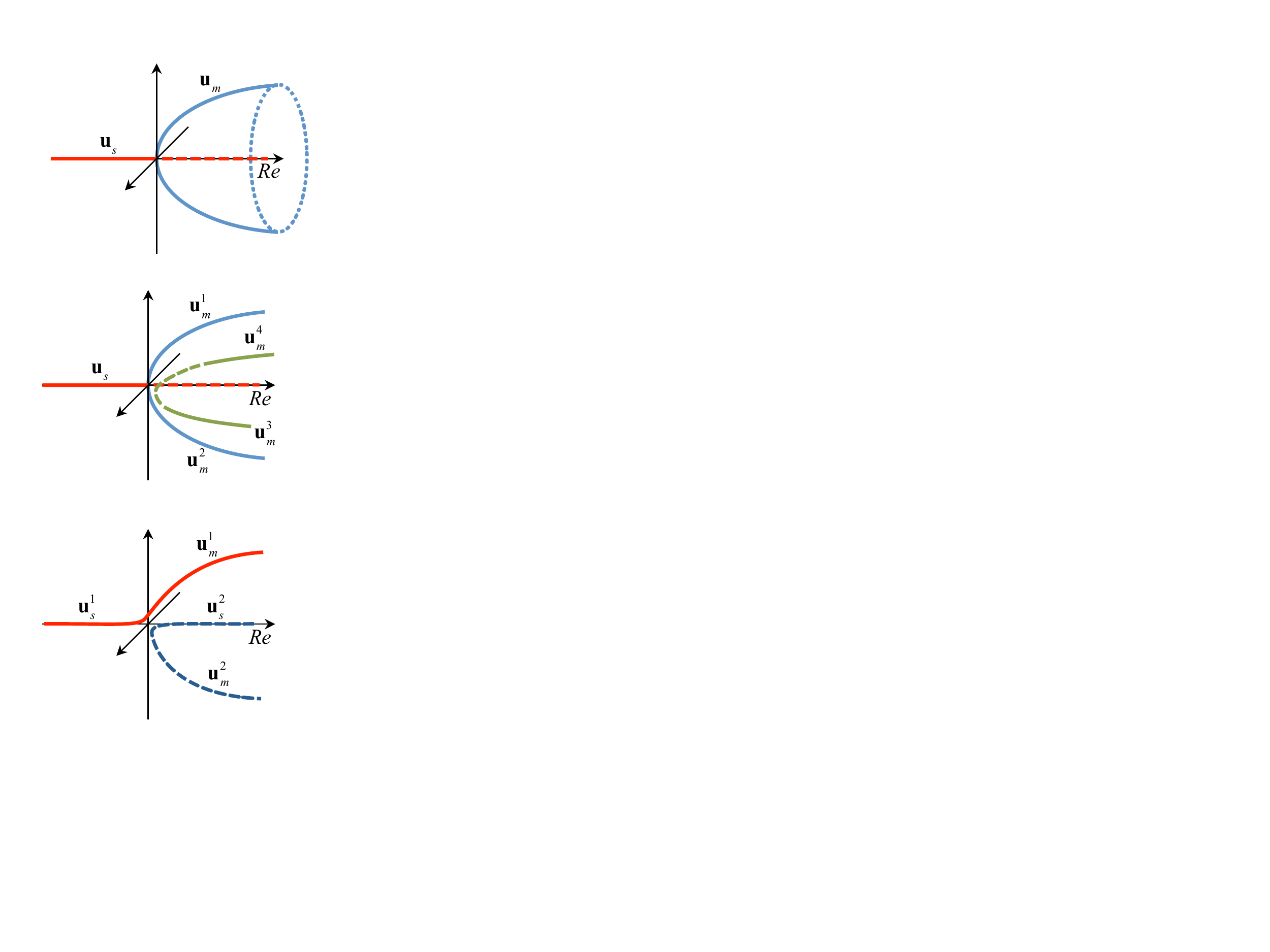}}\quad
\subfloat[]{\includegraphics[height=1.3in]{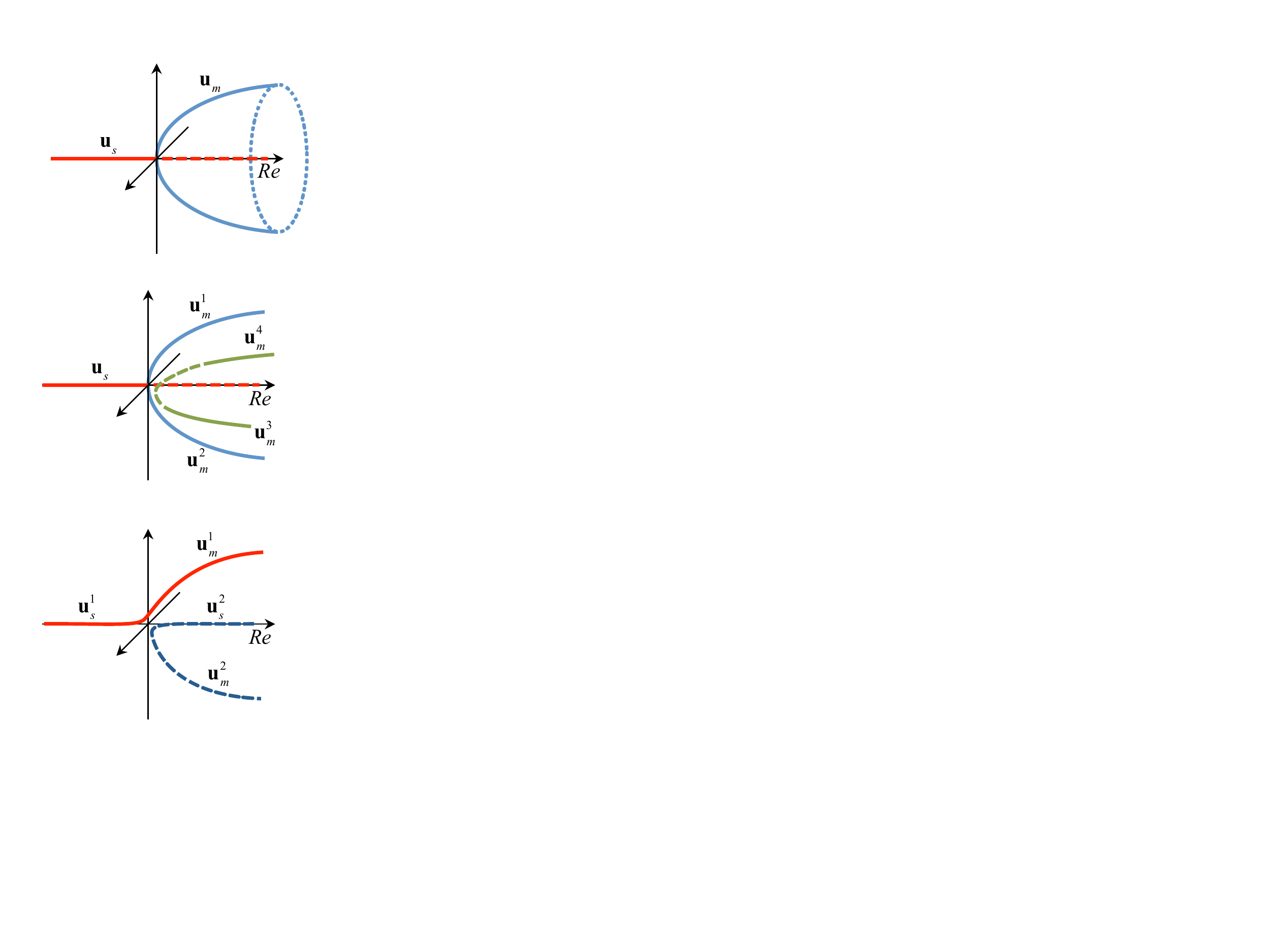}}\quad
\subfloat[]{\includegraphics[height=1.3in]{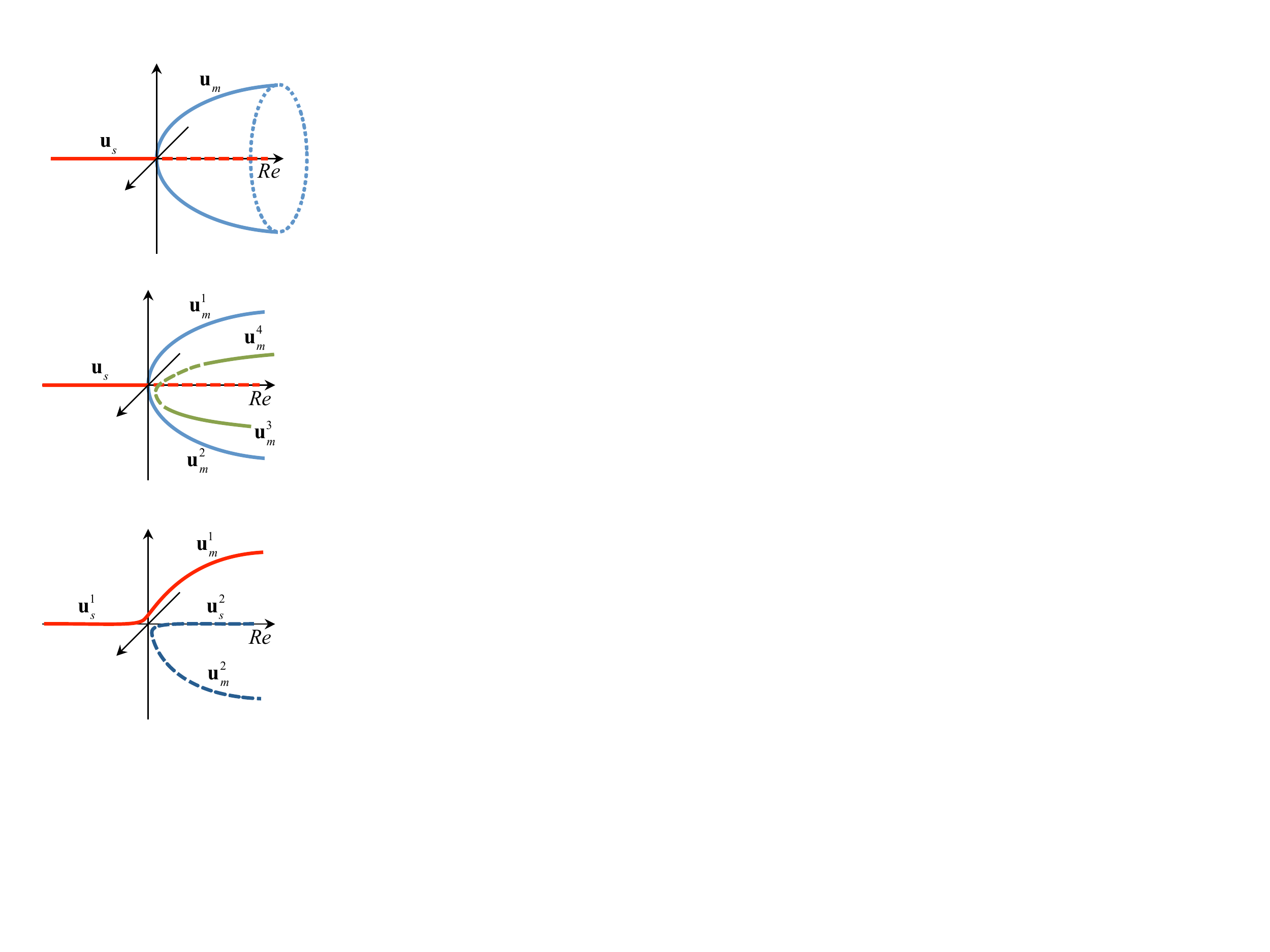}}
\caption{\label{fig:pitchfork} A schematic showing the bifurcations corresponding to the primary instability: (a) circle pitchfork in the DPS, (b) sequence of pitchfork bifurcations in the SPS, (c) imperfect pitchfork bifurcation in the NPS. Solid (dashed) lines indicate stable (unstable) solution branches. The vertical and out-of-plane axes correspond to deviations of the flow from straight that are invariant under $\mathcal{R}_x\mathcal{R}_y$ and $\mathcal{R}_{x}\mathcal{T}^w_y$, respectively. }
\end{figure}

The above transformations that leave the governing equation equivariant, however, need not leave the flow fields invariant. 
When a flow field does not share a certain symmetry of the governing equation, one can generate -- by applying the corresponding coordinate transformation -- a dynamically equivalent symmetry-related copy of the flow. 
The straight flow in figure \ref{fig:str_flow} (a) remains unchanged when an arbitrary translation $\delta x \in [0,L_x]$ is applied along the $x$-direction, so there is a unique solution ${\bf u}_s$.
However, since the primary instability breaks the translational symmetry, there is a continuum of distinct modulated flow solutions ${\bf u}_m$ related by translations in the $x$ direction. 
This instability therefore corresponds to a circle pitchfork bifurcation (cf. figure \ref{fig:pitchfork} (a)). 

\begin{figure}
\centering
\subfloat[]{\includegraphics[height=1.6in]{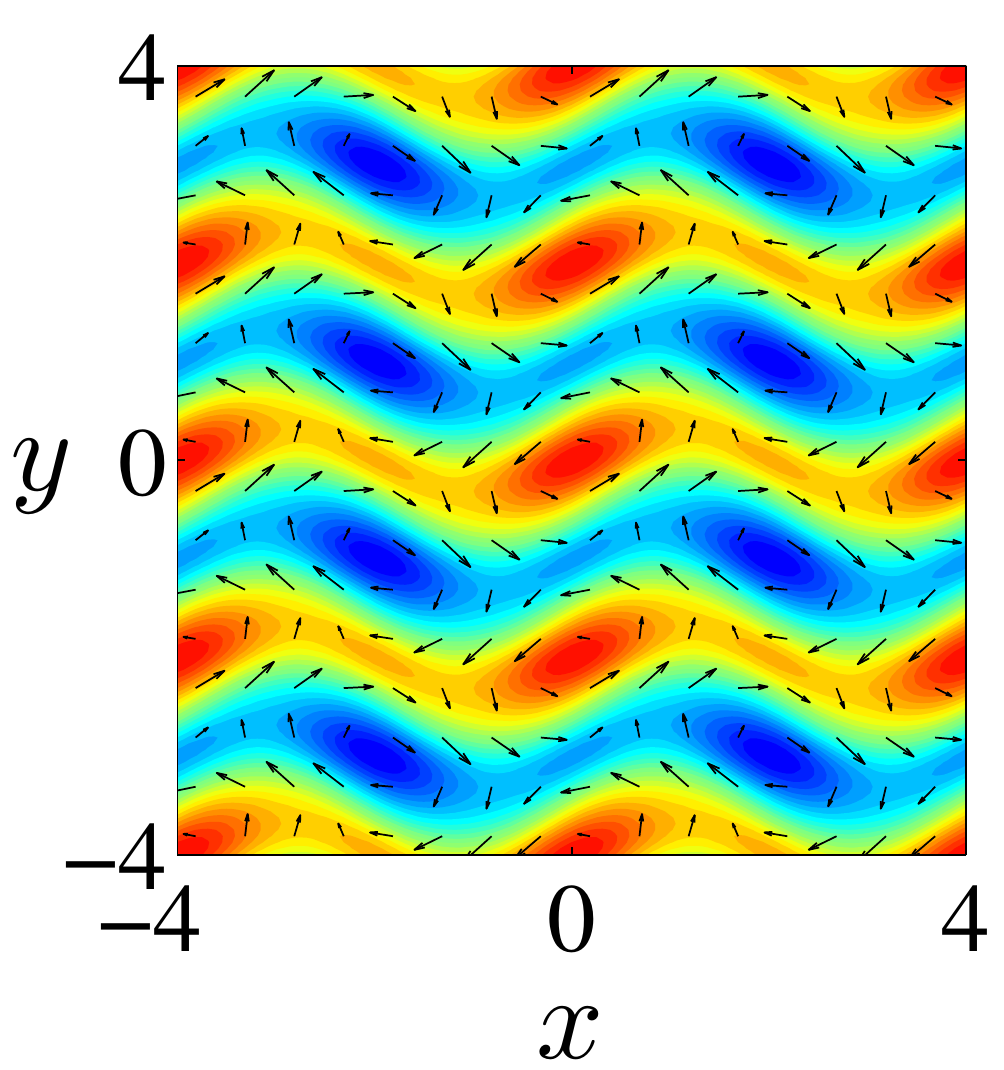}} \qquad
\subfloat[]{\includegraphics[height=1.6in]{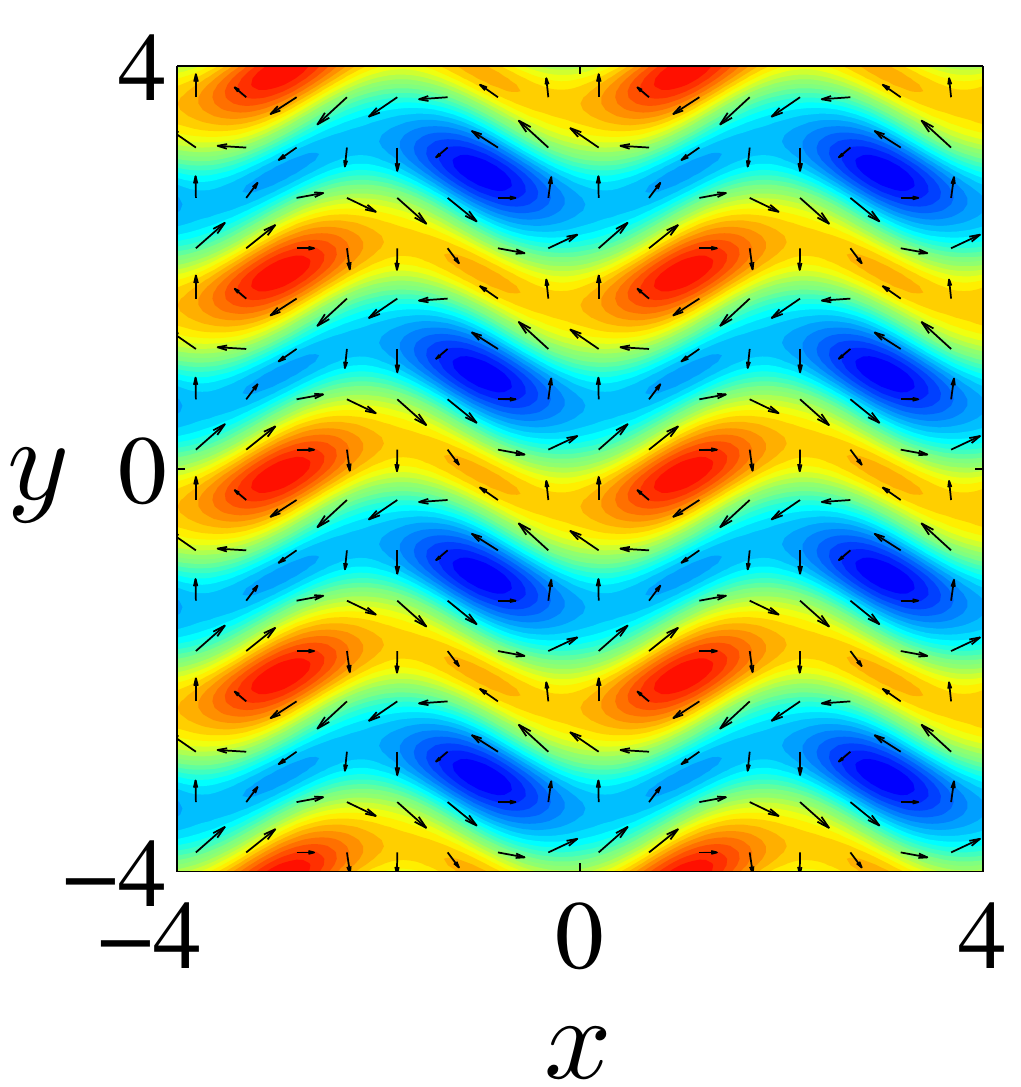}}\\
\subfloat[]{\includegraphics[height=1.6in]{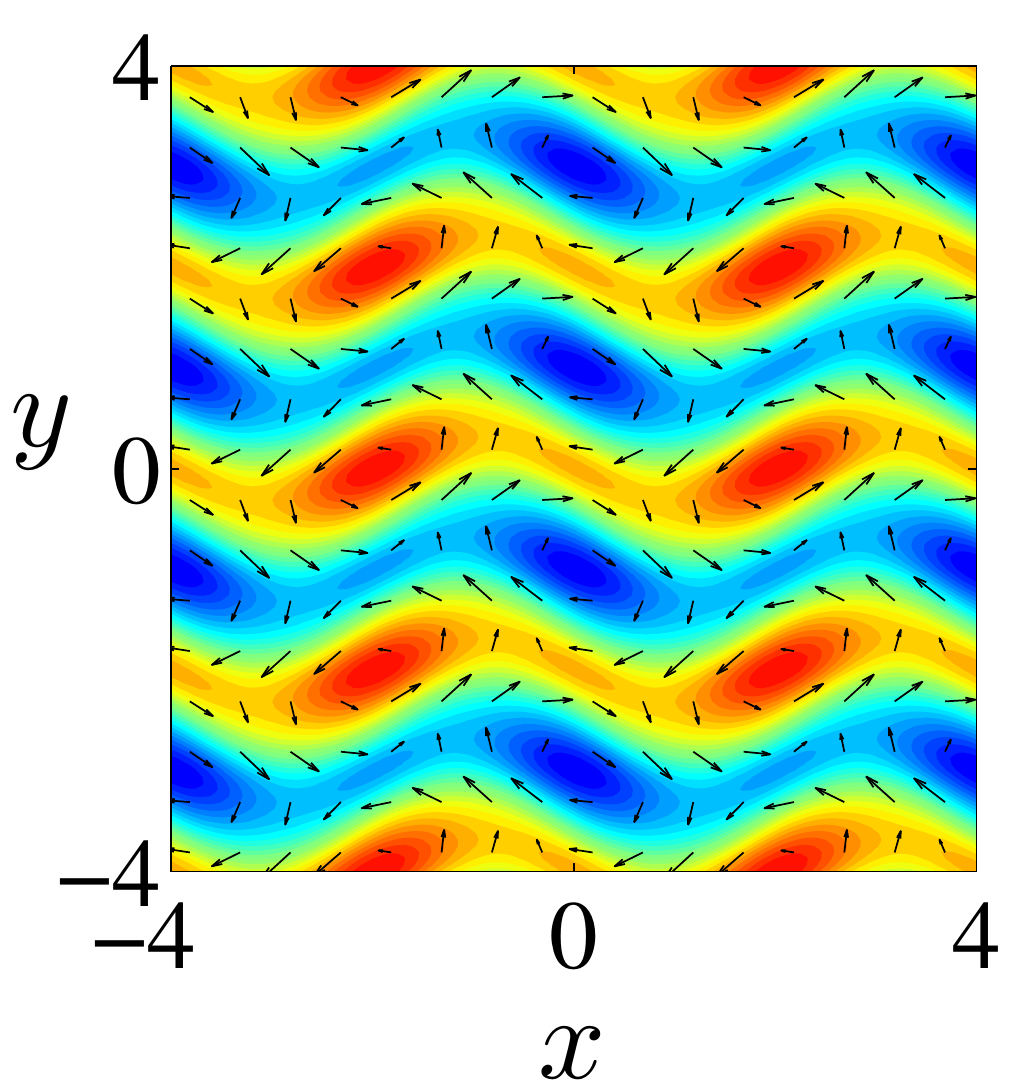}} \qquad
\subfloat[]{\includegraphics[height=1.6in]{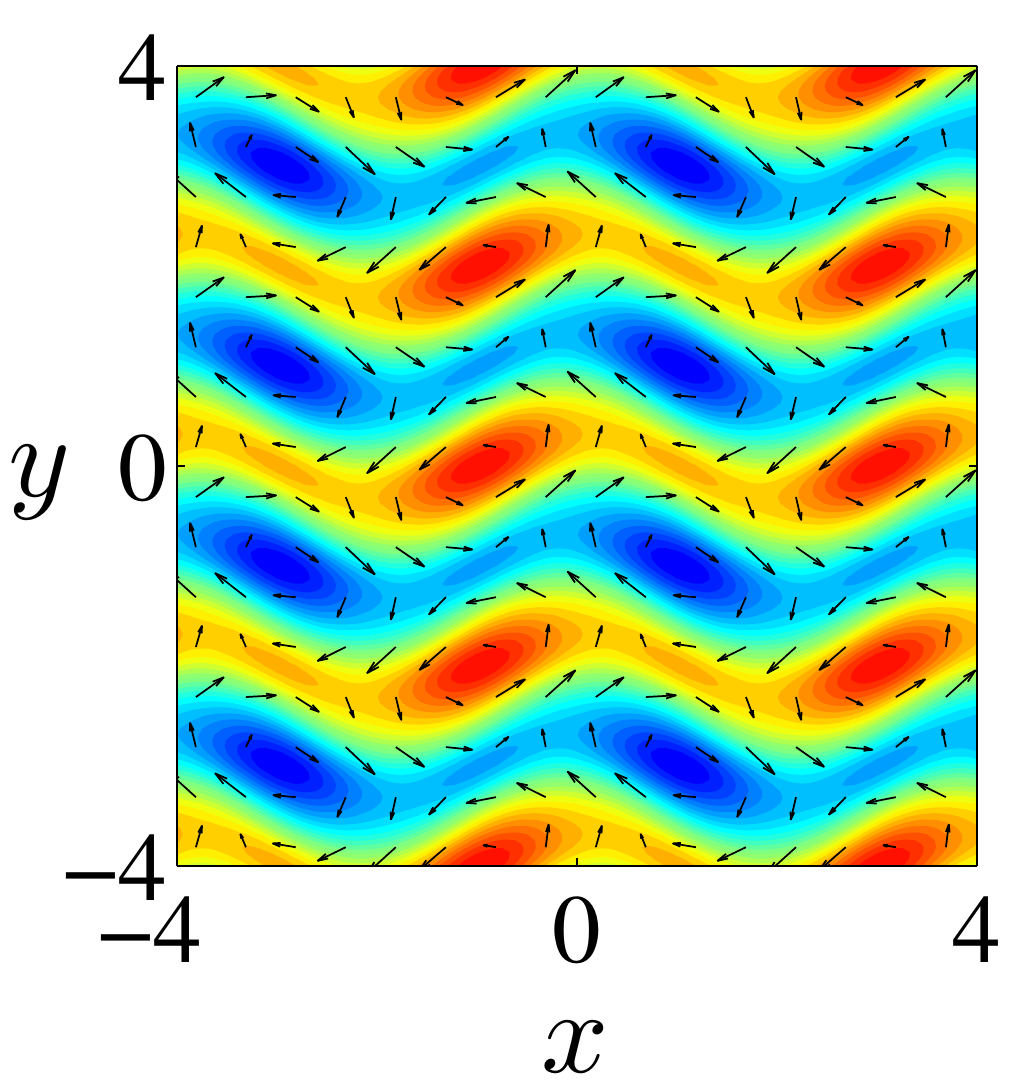}}\\
\caption{\label{fig:sym_copies_dps} 
Modulated flow fields (a) ${\bf u}^{{1}}_{{m}}$, (b) ${\bf u}^{{3}}_{{m}}$, (c) ${\bf u}^{{2}}_{{m}}$, and (d) ${\bf u}^{{4}}_{{m}}$ at $Re=14$ in the DPS. The vorticity colour scale is the same as that in figure \ref{fig:mod_state}.} 
\end{figure}

The equivariance of the governing equation under  $\mathcal{T}^{\delta x}_x$ with arbitrary $\delta x$ makes the choice of the coordinate origin $x=0$ for a modulated flow arbitrary. We fix it by requiring that ${\bf u}^{{1}}_{{m}}=\mathcal{R}_x\mathcal{R}_y{\bf u}^{{1}}_{{m}}$ for a particular modulated flow solution ${\bf u}^{{1}}_{{m}}$. 
Since both the discrete symmetries $\mathcal{R}_x\mathcal{R}_y$ and $\mathcal{R}_x\mathcal{T}^w_y$ include reflection of the flow about the line $x=0$, the choice of the origin determines whether a particular solution remains invariant under either of these discrete symmetries. 
Figure \ref{fig:sym_copies_dps} shows the four distinct solutions related by discrete translations $\mathcal{T}^{\delta x}_x$ with $\delta x=L_x/8$:
\begin{equation}\label{eq:shifts_dps}
 {\bf u}^{{3}}_{{m}}=\mathcal{T}^{\delta x}_x{\bf u}^{{1}}_{{m}},\quad {\bf u}^{{2}}_{{m}}=\mathcal{T}^{\delta x}_x{\bf u}^{{3}}_{{m}}, \quad  {\bf u}^{{4}}_{{m}}=\mathcal{T}^{\delta x}_x{\bf u}^{{2}}_{{m}}, \quad \quad {\bf u}^{{1}}_{{m}}=\mathcal{T}^{\delta x}_x{\bf u}^{{4}}_{{m}}.
\end{equation}
Each of these four solutions is invariant under $\mathcal{T}^{2w}_y$ and either $\mathcal{R}_x\mathcal{R}_y$ or $\mathcal{R}_x\mathcal{T}^w_y$. 
In particular,  ${\bf u}^{{1}}_{{m}}$ and ${\bf u}^{{2}}_{{m}}$ are invariant under $\mathcal{R}_x\mathcal{R}_y$, while ${\bf u}^{{3}}_{{m}}$ and ${\bf u}^{{4}}_{{m}}$ are invariant under $\mathcal{R}_x\mathcal{T}^w_y$.
Furthermore, the states ${\bf u}^{{1}}_{{m}}$ and ${\bf u}^{{2}}_{{m}}$ are related to each other via $\mathcal{R}_x\mathcal{T}^w_y$, i.e., ${\bf u}^{{1}}_{{m}}$  = $ \mathcal{R}_x\mathcal{T}^w_y {\bf u}^{{2}}_{{m}}$. Similarly, ${\bf u}^{{3}}_{{m}}$ and ${\bf u}^{{4}}_{{m}}$ are related via $\mathcal{R}_x\mathcal{R}_y$, i.e., ${\bf u}^{{3}}_{{m}}$ = $\mathcal{R}_x\mathcal{R}_y{\bf u}^{{4}}_{{m}}$. 
In summary,  by virtue of the continuous translational symmetry of the governing equation, the laminar flow in the DPS undergoes a circle pitchfork bifurcation with an infinite number of translation-related copies of a modulated flow. Only four of these copies, however, remain invariant under the discrete symmetries involving the reflection $\mathcal{R}_x$.

\subsection{SPS}

The no-slip boundary condition at $x=\pm L_x/2$ in the SPS destroys the equivariance under translation  $\mathcal{T}^{\delta x}_x$, reducing the symmetry group to $Z_2\times Z_n$. The governing equation, however, still remains equivariant under each of the discrete transformations $\mathcal{R}_x\mathcal{T}^w_y$ and $\mathcal{R}_x\mathcal{R}_y$. The loss of equivariance under $\mathcal{T}^{\delta x}_x$, which connected the states ${\bf u}_m^1$, ${\bf u}_m^2$  with ${\bf u}_m^3$, ${\bf u}_m^4$ in the DPS, implies either of $\mathcal{R}_x\mathcal{R}_y$ or $\mathcal{R}_x\mathcal{T}^w_y$ is broken in the straight to modulated transition in the SPS.  Breaking either of the discrete symmetries, $\mathcal{R}_x\mathcal{T}^w_y$ or $\mathcal{R}_x\mathcal{R}_y$ should generate (only) two branches, i.e, should result in a pitchfork bifurcation, since the modulated flow states in the SPS should remain symmetric with respect to $\mathcal{T}^{2w}_y$, which is not affected by confinement in $x$. Consequently, of the infinite number of modulated states in DPS only four, the counterparts of those shown in figure \ref{fig:sym_copies_dps}, will survive in the SPS, and should be formed via two distinct pitchforks.

\begin{figure}
\centering
\subfloat[]{\includegraphics[height=1.6in]{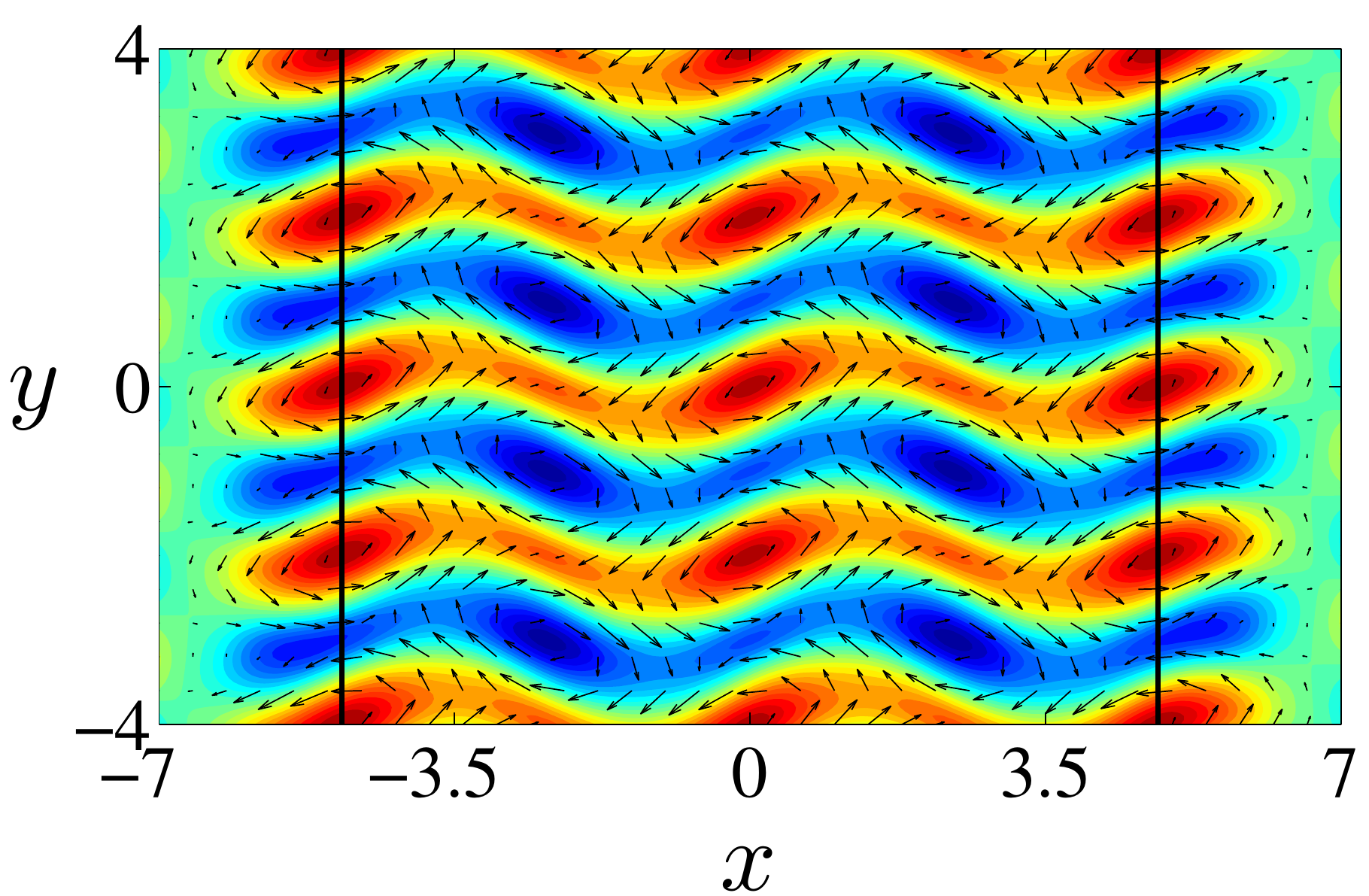}}\quad
\subfloat[]{\includegraphics[height=1.6in]{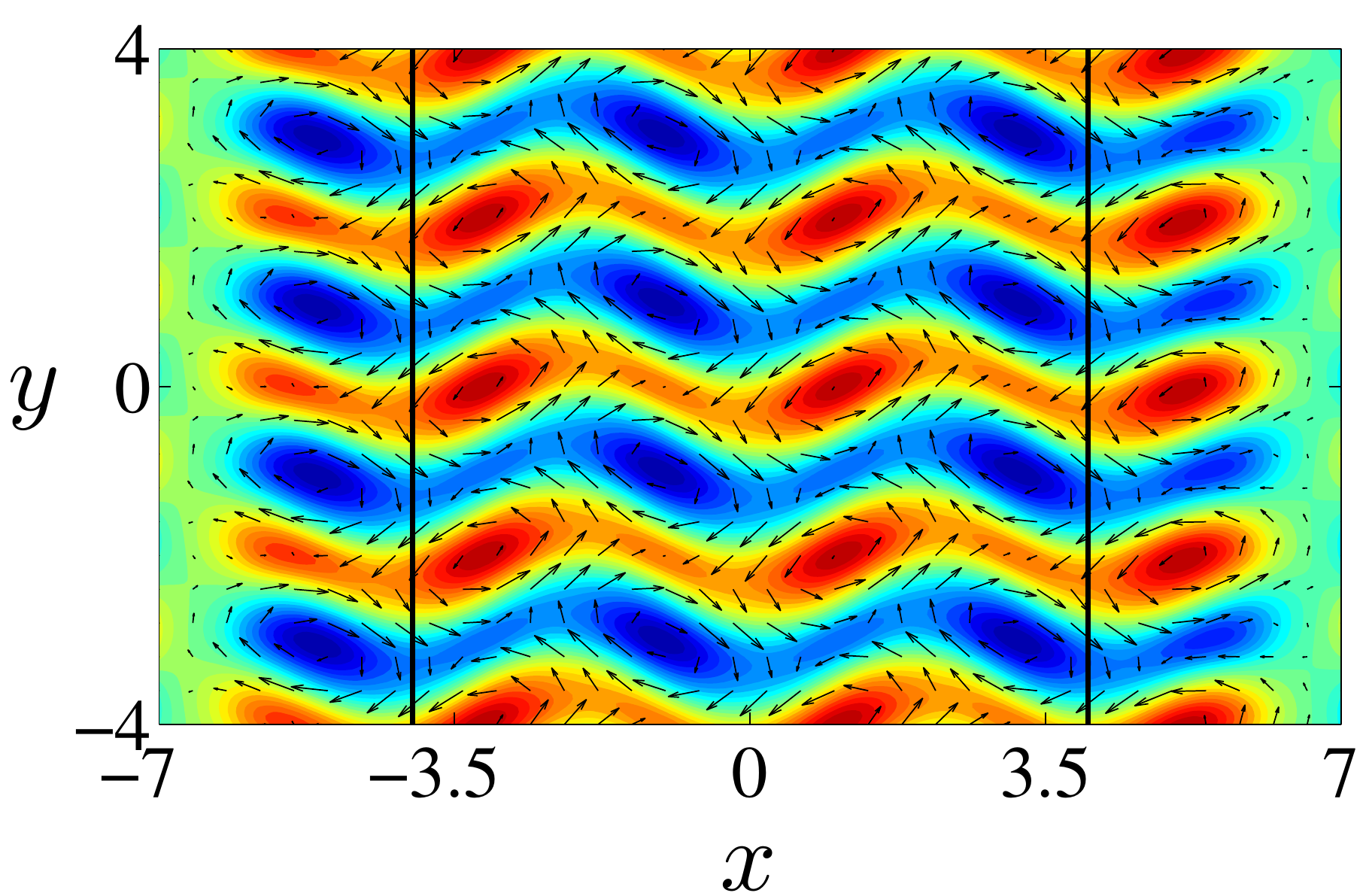}}\\
\subfloat[]{\includegraphics[height=1.6in]{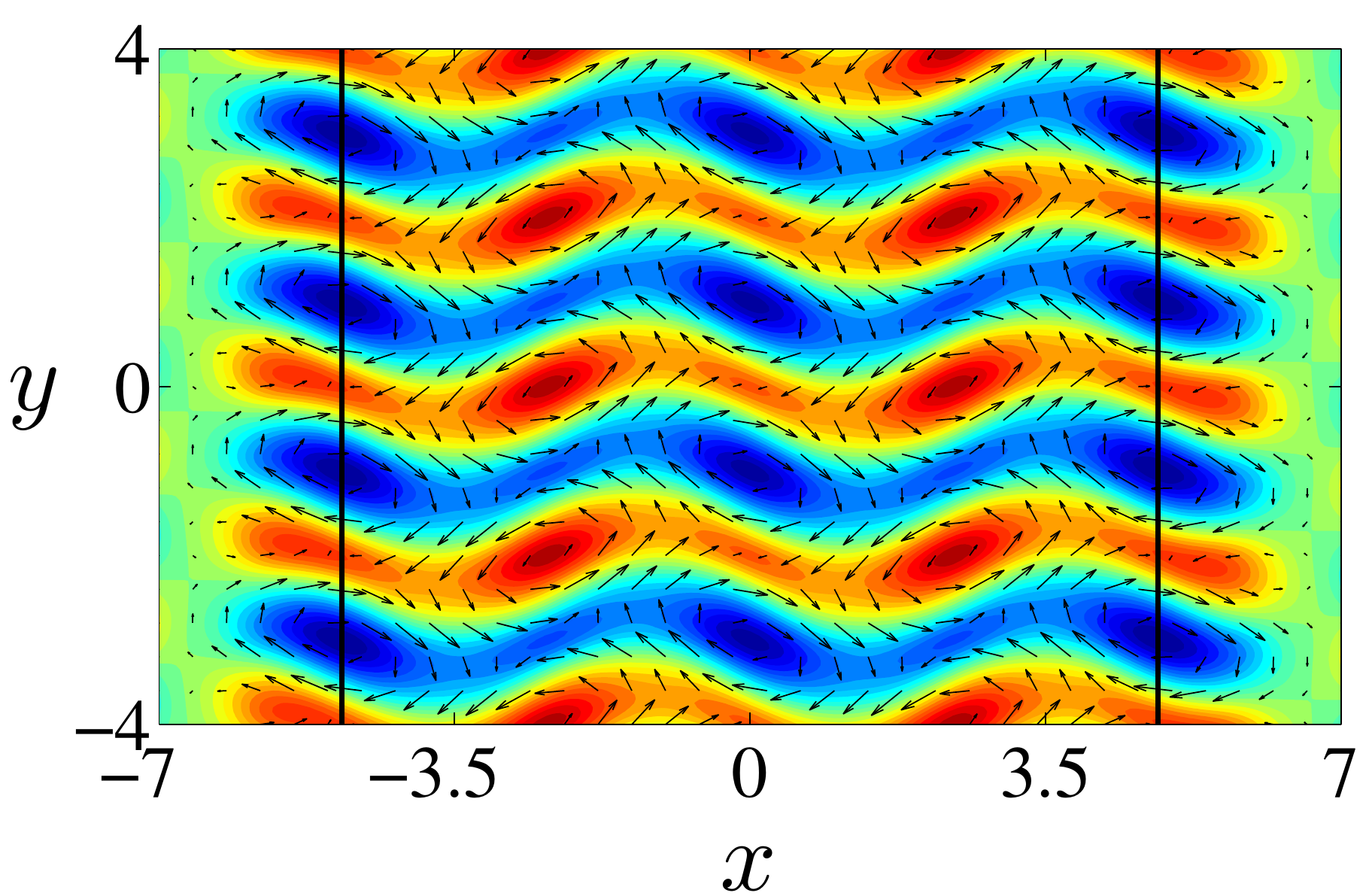}}\qquad
\subfloat[]{\includegraphics[height=1.6in]{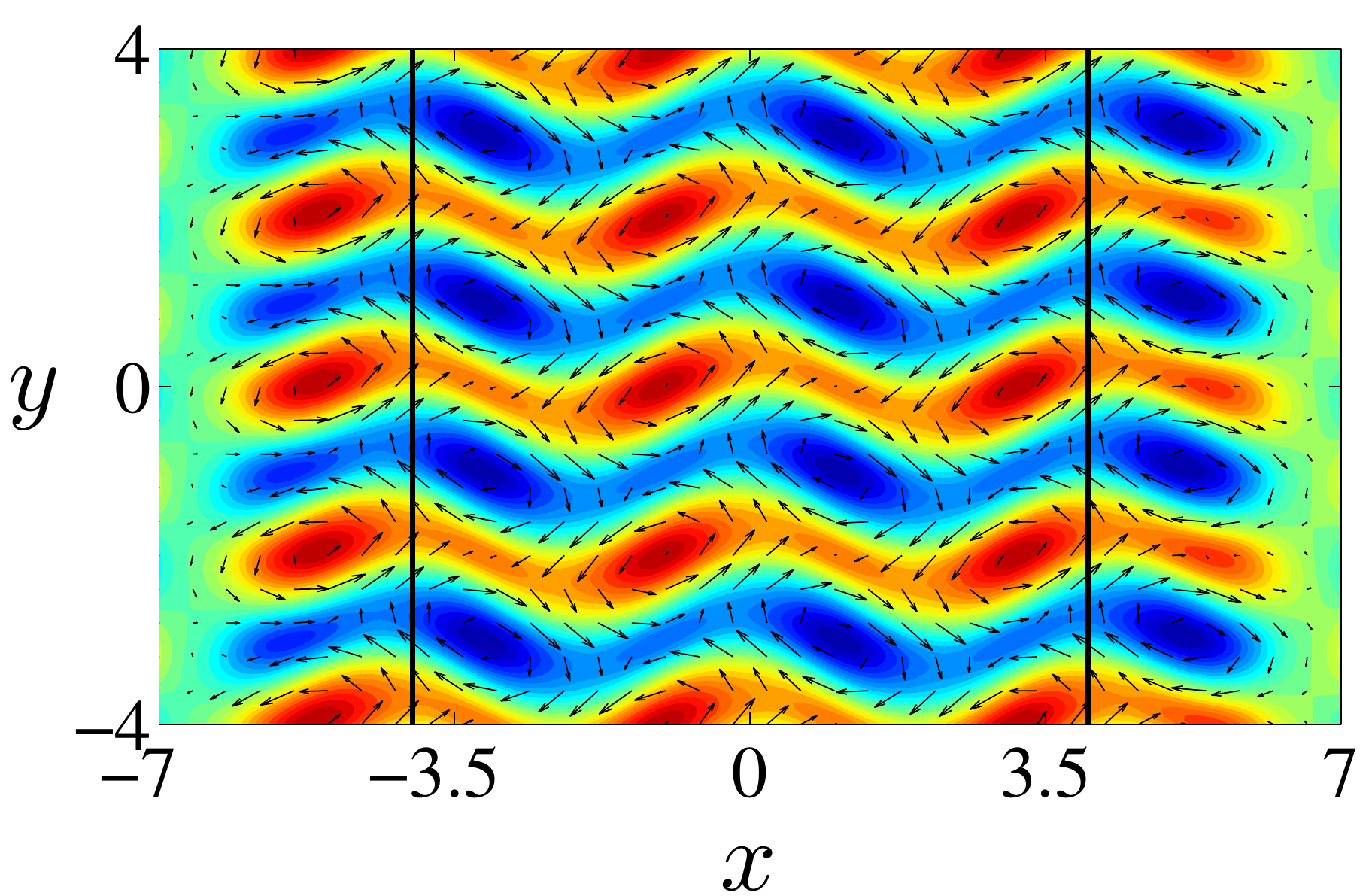}}
\caption{\label{fig:sym_copies_sps} Modulated flow fields (a) ${\bf u}^{{1}}_{{m}}$, (b) ${\bf u}^{{3}}_{{m}}$, (c) ${\bf u}^{{2}}_{{m}}$, and (d) ${\bf u}^{{4}}_{{m}}$ at $Re=14$ in the SPS. Vertical black lines indicate the central region which is analogous to the flow fields shown in figure \ref{fig:sym_copies_dps}. The vorticity colour scale is the same as that of figure \ref{fig:mod_state}.}
\end{figure}

This is indeed what we observe in the simulations, wherein two pairs of distinct solutions, shown in figure \ref{fig:sym_copies_sps}, are formed via two distinct pitchfork bifurcations of the straight flow. 
Like in the DPS, ${\bf u}^{{1}}_{{m}}$ and ${\bf u}^{{2}}_{{m}}$ are invariant under $\mathcal{R}_x\mathcal{R}_y$, while ${\bf u}^{{3}}_{{m}}$ and ${\bf u}^{{4}}_{{m}}$ are invariant under $\mathcal{R}_x\mathcal{T}^w_y$.  
In figure \ref{fig:pitchfork} (b) the  ${\bf u}^{{3}}_{{m}}$ and ${\bf u}^{{4}}_{{m}}$ branches are plotted to lie in a plane perpendicular to that containing ${\bf u}^{{1}}_{{m}}$ and ${\bf u}^{{2}}_{{m}}$. 
Since the bifurcations break either $\mathcal{R}_x\mathcal{R}_y$ or $\mathcal{R}_x\mathcal{T}^w_y$, each pair of branches is related via the broken symmetry, i.e., ${\bf u}^{{1}}_{{m}}=\mathcal{R}_x\mathcal{T}^w_y{\bf u}^{{2}}_{{m}}$ and ${\bf u}^{{3}}_{{m}}=\mathcal{R}_x\mathcal{R}_y{\bf u}^{{4}}_{{m}}$. 

Unlike the DPS where $\mathcal{T}^{\delta x}_x$ relates all the distinct solutions corresponding to the modulated flow (cf. equation (\ref{eq:shifts_dps})), there is no coordinate transformation that maps ${\bf u}^{{1}}_{{m}}$ and ${\bf u}^{{2}}_{{m}}$ to ${\bf u}^{{3}}_{{m}}$ and ${\bf u}^{{4}}_{{m}}$. 
On an infinite domain, all four branches of the modulated flow are created at exactly the same $Re$ (as in the DPS), however, on a finite domain, the pitchfork bifurcations that produce the two pairs of solutions would generally happen at different $Re$ (cf. figure \ref{fig:pitchfork} (b)) that depend on the confinement in the $x$-direction, i.e., on $L_x$. 
For  $L_x = 14$, chosen from experimental considerations, the bifurcation which gives rise to ${\bf u}^{{3}}_{{m}}$ and ${\bf u}^{{4}}_{{m}}$ occurs at a higher $Re$ than the bifurcation which gives rise to ${\bf u}^{{1}}_{{m}}$ and ${\bf u}^{{2}}_{{m}}$. For other choices of $L_x$, the sequence \textit{may} reverse.  
Note that the two modulated flow branches (${\bf u}^{{3}}_{{m}}$ and ${\bf u}^{{4}}_{{m}}$) that are formed from the second pitchfork are initially unstable,  because they bifurcate off of the unstable straight flow solution.

\subsection{NPS}

In the NPS, the additional no-slip boundary condition at $y=\pm L_y/2$ breaks the equivariance of the problem under $\mathcal{R}_x\mathcal{T}^w_y$, and hence $\mathcal{T}^{2w}_y$,  leaving the governing equation equivariant only under $\mathcal{R}_x\mathcal{R}_y$.
The pitchfork bifurcation that gives rise to the rotationally invariant solutions ${\bf u}^{{1}}_{{m}}$ and ${\bf u}^{{2}}_{{m}}$ in the SPS is associated with breaking of the $\mathcal{R}_xT^w_y$ symmetry. This symmetry is only approximate in the NPS for all $Re$, so one finds an imperfect pitchfork bifurcation instead, as shown in figure \ref{fig:pitchfork} (c). The straight flow ${\bf u}^{1}_{{s}}$ at lower $Re$  smoothly transitions to the modulated flow ${\bf u}^{{1}}_{{m}}$ at higher $Re$ without an instability taking place, i.e., the real part of the leading eigenvalue of the straight flow does not change sign as we increase $Re$ in the NPS. The shapes of the bifurcation curves close to $Re_c$ (figure \ref{fig:bif1} (a)) showcase the difference in the nature of the primary instability between the NPS and the two periodic simulations. 

\begin{figure}
\centering
\subfloat[]{\includegraphics[height=2.6in]{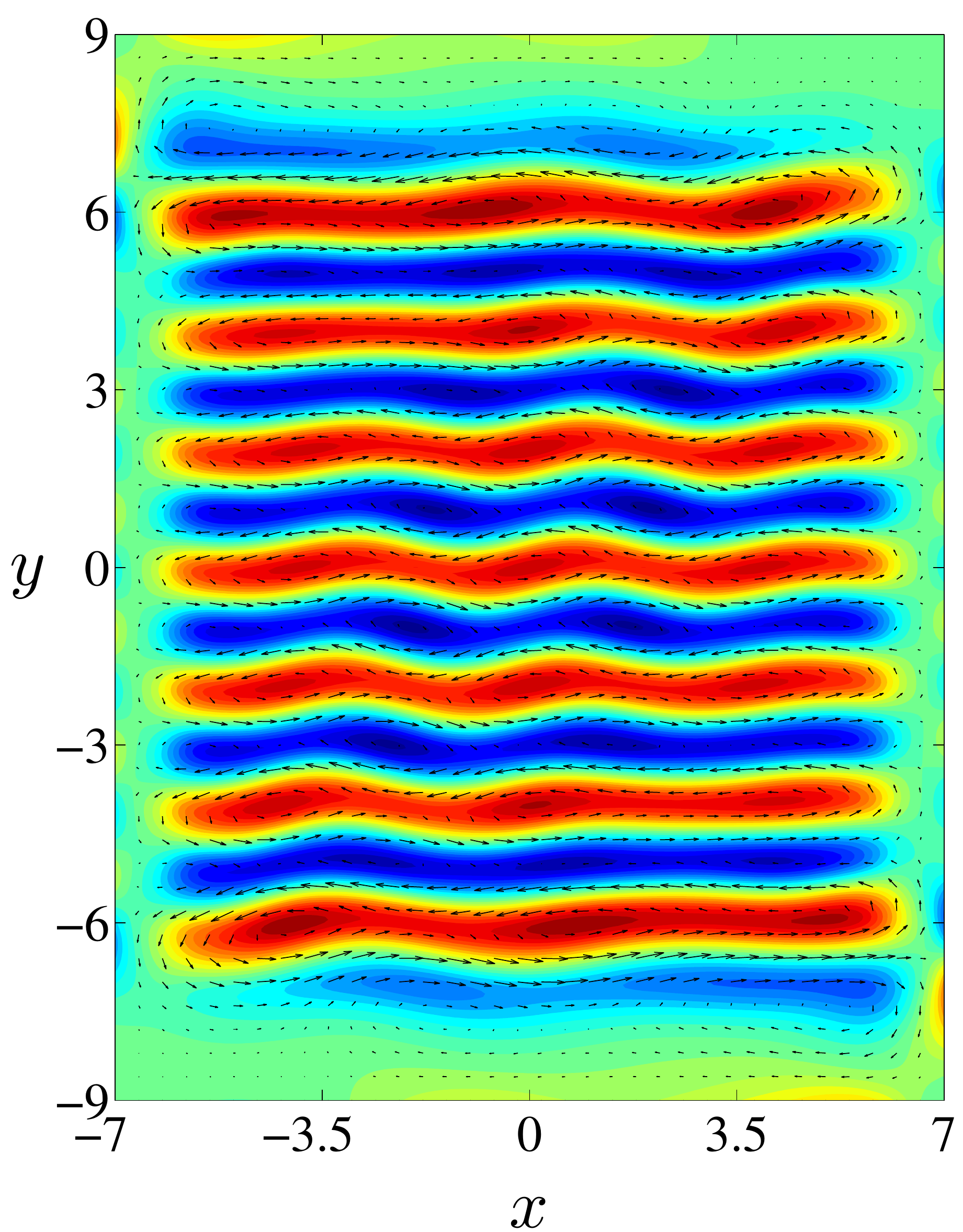}}\qquad
\subfloat[]{\includegraphics[height=2.6in]{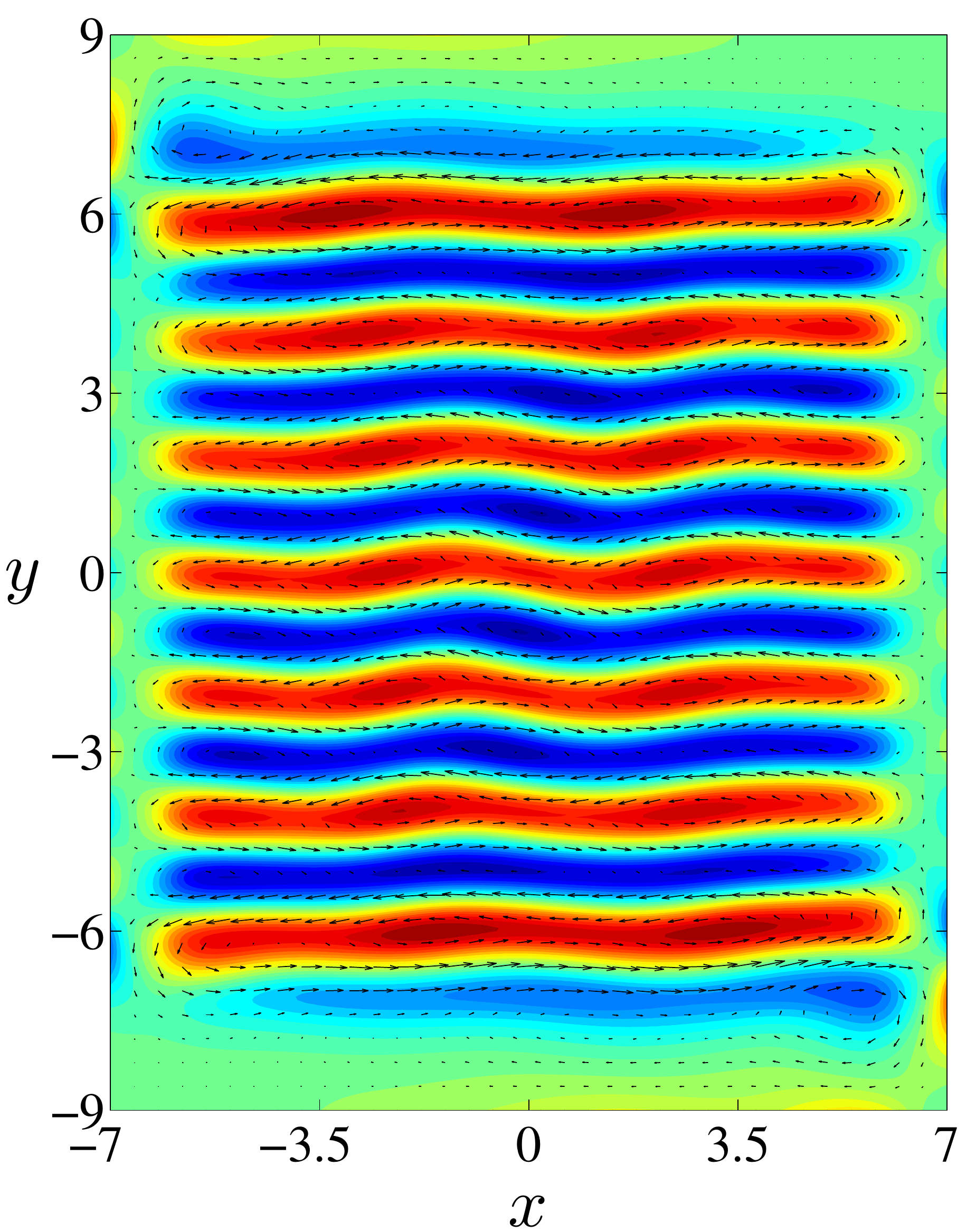}}
\caption{\label{fig:sym_copies_nps} Modulated flow fields at $Re=11.6$ in the NPS beyond the imperfect pitchfork bifurcation shown in figure \ref{fig:pitchfork} (c). The flow fields shown here are (a) ${\bf u}^{1}_{{m}}$ which emerges smoothly from the straight flow and (b) ${\bf u}^{2}_{{m}}$ which is formed through a saddle-node bifurcation.}
\end{figure}

In the NPS, the ${\bf u}^{{2}}_{{m}}$ branch and the higher-$Re$ branch of the straight flow ${\bf u}^{{2}}_{{s}}$ are created in a saddle-node bifurcation at $Re=10.72$.
The states ${\bf u}^{{1}}_{{m}}$ and ${\bf u}^{{2}}_{{m}}$, both of which are symmetric with respect to $\mathcal{R}_x\mathcal{R}_y$, are shown in figure \ref{fig:sym_copies_nps}. 
While $\mathcal{R}_xT^w_y$ is not an exact symmetry in the NPS, given the large transverse extent of the domain compared with the period of the forcing ($L_y/2w$ = 9), near the center of the domain this approximate symmetry holds and consequently ${\bf u}^{{2}}_{{m}} \approx \mathcal{R}_x\mathcal{T}^w_y{\bf u}^{{1}}_{{m}}$. 
However, unlike ${\bf u}^{{1}}_{{m}}$, which remains stable up to $Re=15.4$ in the NPS, ${\bf u}^{2}_{{m}}$ is unstable over the entire range of $Re$ where it exists ($Re\geq 10.72$).
This explains why our numerical simulations starting from randomized initial conditions have always converged to the modulated flow ${\bf u}^{{1}}_{{m}}$ and why ${\bf u}^{{2}}_{{m}}$ was never found in the numerical simulations or observed in the experiment. 

The pitchfork bifurcation which gave rise to the branches ${\bf u}^{{3}}_{{m}}$ and ${\bf u}^{{4}}_{{m}}$ in the SPS also does not carry over into the NPS. Instead ${\bf u}^{{2}}_{{s}}$ undergoes a Hopf bifurcation  at $Re = 12.6$. This change in the nature of the bifurcation is likely caused by transverse confinement, which has a more prominent effect on ${\bf u}^{{3}}_{{m}}$ and ${\bf u}^{{4}}_{{m}}$: these states are invariant under the $\mathcal{R}_x\mathcal{T}^w_y$ symmetry in the SPS, but this symmetry is broken in the NPS. While we fail to observe ${\bf u}^{{3}}_{{m}}$ and ${\bf u}^{{4}}_{{m}}$ in the NPS, the analogues of these solutions may appear for a different set of model parameters, forcing profile, and/or degree of confinement.  Details regarding the computation of the unstable branches associated with the various bifurcations are included in Appendix \ref{sec:methods_pitchfork}.

\section{Conclusions} \label{sec:conclusion}

In this article, we have presented a combined experimental and numerical study of bifurcations in a Q2D Kolmogorov-like flow.  
This flow is realized in the laboratory by electromagnetically driving a stratified layer of electrolyte above an immiscible layer of dielectric. 
This Q2D flow is described using a 2D model \eqref{eq:2dns_mod} derived from first principles by depth-averaging the 3D Navier-Stokes equation.  
In contrast, virtually all previous studies have modelled Q2D flows using equation (\ref{eq:2dns_wf}), a semi-empirical variation of the 2D Navier-Stokes equation with the addition of a linear friction.
Also unlike previous studies of Kolmogorov-like flows which have assumed a perfectly sinusoidal forcing profile, we have introduced a realistic model of the forcing which has been validated against 3D experimental measurements. 

To test the importance of lateral confinement, we have compared experimental measurements with numerical simulations using different boundary conditions. 
We have found that by incorporating realistic, no-slip boundary conditions at all lateral boundaries and a realistic forcing profile, quantitative agreement between the experiment and simulation can be achieved with no adjustable parameters. 
In particular, the Reynolds number $Re_c$ for the primary instability can be predicted to within about 5\% and the critical wavenumber $k_c$ can be predicted to an accuracy higher than the measurement accuracy.
These are significant improvements compared with previous studies, none of which were able to predict both $Re_c$ and $k_c$ with this level of accuracy, {\it despite using adjustable parameters}. 

We have also performed a systematic study of how lateral confinement affects the nature of the bifurcation describing the transition from the straight flow to the modulated flow. Previous studies have characterized this transition in the experiment as a pitchfork bifurcation, using analytical computations on a periodic domain. We have shown that because of confinement an imperfect pitchfork bifurcation is found instead. We have also numerically computed the two unstable branches of the imperfect pitchfork bifurcation describing flows that are not observed in either experiment or simulations under normal conditions.

Furthermore, we have demonstrated that the model reasonably accurately predicts the modulated flow pattern (both the wavenumber and the amplitude) beyond the onset of the primary instability. Moreover, this is the first study,  experimental or theoretical, to provide a quantitative analysis of the secondary instability of a Kolmogorov-like flow which generates a time-dependent pattern of vortices. Even for the secondary instability the numerical predictions of the critical Reynolds number $Re_p$ and the critical period $T_p$ are in general agreement with the experiment, although the accuracy of the numerical predictions decreases with increasing $Re$.

The discrepancy between the numerical predictions and experiments has been traced back to the variation of the forcing profile with height. This points to the limitations of a 2D model of what in reality is a 3D flow, albeit with a strongly suppressed vertical component of the velocity. Nonetheless, for a select range of $Re$ the experimental flow can be reproduced with extremely good quantitative accuracy by making fairly small adjustments to the model parameters, compared with their depth-averaged values computed for the simple straight flow. The ability of the model to closely reproduce the experimental flow is crucial for the utility of Q2D flows for testing the geometrical description of weakly turbulent flows and studying the dynamical role of exact coherent structures \citep{suri_2017}. Such tests will be the main focus of follow up studies.

\section*{Acknowledgements}

We thank Rich Kerswell for useful discussions and Daniel Borrero for his useful suggestions and insight. J.~T. is grateful to Samuel Raben for his help with the Prana PIV software package. This work was supported in part by the National Science Foundation under Grants No. CMMI-1234436 and DMS-1125302.

\appendix

\section{Scaling and Nondimensionalization\label{sec:nondim}} 

The governing equation (\ref{eq:2dns_mod}) was presented in dimensional form to highlight the dependence of parameters $\alpha$, $\beta$, and ${\nu}$ on the properties of the fluid layers.
The dimensional form also makes it easier to explore the sensitivity of the dynamics to changes in these parameters.
To simplify comparison of our results with other studies it is helpful to nondimensionalize this equation. 
Choosing the width of a magnet $w$ as the length scale, the rms velocity computed over the central region $|x|\le 4w$, $|y|\le 4w$ as the velocity scale $U$, and the ratio of these two scales as the time scale, one obtains the following nondimensional equation:
\begin{equation}\label{eq:q2dns_ndim}
\frac{\partial {\bf u}}{\partial t} + \beta {\bf u}\cdot\nabla{\bf u} = -\nabla p_0 + \frac{1}{Re}\left(\nabla^2 {\bf u}- \gamma{\bf u}\right) + {\bf f}_0,
\end{equation}
where $\gamma = \alpha w^2/{\nu}$ describes the relative strength of the Rayleigh friction and viscous terms in (\ref{eq:2dns_mod}). Finally, $p_0=wp/(U^2\bar{\rho})$ is the nondimensional pressure and ${\bf f}_0 = w{\bf f}/U^2$ is the nondimensional forcing profile. 

It is possible to eliminate the parameter $\beta$ from equation (\ref{eq:q2dns_ndim}) by making the length, time, and velocity scales independent. If we again choose the width of a magnet $w$ as the length scale, the rms velocity as the velocity scale $U$, and $w/(\beta U)$ as the time scale, we instead obtain the following nondimensional equation:
\begin{equation}\label{eq:2dns_scaled}
\frac{\partial {\bf u}}{\partial t} + {\bf u}\cdot\nabla{\bf u} = -\nabla p_1 + \frac{1}{Re'}\left(\nabla^2 {\bf u}- \gamma{\bf u}\right) + {\bf f}_1,
\end{equation}
where $Re'=Re/\beta$, $p_1=w\beta p/(U^2\bar{\rho})$ is the nondimensional pressure, and ${\bf f}_1 = w\beta{\bf f}/U^2$ is the nondimensional forcing profile. Although equation \eqref{eq:2dns_scaled} does not contain $\beta$ explicitly, the Reynolds number is rescaled by $\beta$.  Hence, the nondimensional equations \eqref{eq:q2dns_ndim} and \eqref{eq:2dns_scaled} as well as the dimensional equations \eqref{eq:2dns_wf} and \eqref{eq:2dns_mod} predict an identical sequence of bifurcations. However, the critical values of $Re$ scale as $1/\beta$, as we have found explicitly in equation \eqref{eq:Re_0}. 

\section{Numerical Methods} \label{sec:num_meth}

In this Appendix, we present the details of discretization methods and numerical integration schemes employed in the NPS, SPS, and DPS. Additionally, we also detail the computation of the unstable branches associated with the pitchfork bifurcation that cannot be obtained from simple numerical integration.

\subsection{Non-Periodic Simulation (NPS) and Singly-Periodic Simulation (SPS)} \label{sec:app_nps_sps}

Since the NPS, as well as the SPS, require prescribing no-slip (e.g., Dirichlet) boundary conditions on the velocity field ${\bf u}$, numerical simulations are performed using the primitive variable ($u_x$, $u_y$, and $p$) formulation by employing a semi-implicit fractional-step method detailed in \cite{armfield_1999}. Temporal discretization of equation (\ref{eq:2dns_mod}) is performed using the following difference scheme:
\begin{equation}\label{eq:2dns_mod_discrete}
\frac{{\bf u}_{\nt{n}+1} - {\bf u}_{\nt{n}}}{\Delta t} + \frac{3}{2}\mathcal{N}{\bf u}_{\nt{n}} -  \frac{1}{2}\mathcal{N}{\bf u}_{\nt{n}-1} = -\frac{1}{\bar \rho}\nabla p_{\nt{n}+1} + \frac{1}{2}\mathcal{L}({\bf u}_{\nt{n}+1} + {\bf u}_{\nt{n}}) + {\bf f}.
\end{equation}
In the above equation ${\bf u}_{\nt{n}}$ and $p_{\nt{n}+1}$ are the velocity and pressure fields, with the subscript $\nt{n}$ indicating a discrete time instant $t_{\nt{n}} = {\nt{n}}\Delta t$, where $\Delta t$ is the time step for the update. 
For purposes of brevity we have used the notation $\mathcal{N}{\bf u}_{\nt{n}} = \beta{\bf u}_{\nt{n}}\cdot\nabla{\bf u}_{\nt{n}}$ and $\mathcal{L}{\bf u}_{\nt{n}} = \nabla^2{\bf u}_{\nt{n}}-\alpha{\bf u}_{\nt{n}}$ to represent the nonlinear and linear terms, respectively. The above discretization is a semi-implicit approximation of equation (\ref{eq:2dns_mod}), where the linear terms in the update are treated implicitly using the Crank-Nicolson scheme, while the nonlinear term is handled explicitly using the Adams-Bashforth scheme. 
The velocity field ${\bf u}_{\nt{n}+1}$ at every instant satisfies the incompressibility condition:
\begin{equation} \label{eq:incomp_discrete}
\nabla\cdot{\bf u}_{\nt{n}+1}=0,
\end{equation}
which is enforced on each update through the three-fractional-step P2 (pressure correction) projection method discussed in \citet{armfield_1999}. 

Spatial discretization of the velocity and pressure fields is carried out using the standard marker and cell (MAC) staggered grid \citep{harlow_1965}. The spatial derivatives in equation (\ref{eq:2dns_mod_discrete}) are approximated using finite central differences; the 2D Laplacian operator ($\nabla^2$) uses a five-point stencil formula and the nonlinear term uses the three-point central difference formula. 

For both the NPS and the SPS, we have chosen 20 cells per magnet width $w$ to discretize the velocity and pressure fields. Since the dimensions of the NPS are identical to the lateral dimensions of the experiment, i.e., $14w\times 18w$, a total of $280\times360$ cells were used to sample the flow domain. 
The SPS, however, corresponds to a domain of dimensions $14w\times8w$, which maps to a region including the central eight magnets in the experiment. Hence, a total of $280\times160$ cells were used to discretize the SPS domain. For both the SPS and NPS, a time step of $\Delta t = 1/40$ s was used for all the numerical simulations.
 
To test the adequacy of the spatial resolution, the velocity field corresponding to the modulated flow at $Re\approx 15.5$ was recomputed by doubling the resolution, i.e., using 40 cells per magnet width. To compare this velocity field  (${\bf u}_{40}$) with the the one  computed on the 20-cell grid (${\bf u}_{20}$), we interpolated ${\bf u}_{40}$ onto the 20-cell grid to obtain ${\bf u}_{\mathrm{interp}}$ (interpolation was required due to the staggered nature of the grid). The difference between ${\bf u}_{\mathrm{interp}}$ and ${\bf u}_{20}$, computed as $\|({\bf u}_{\mathrm{interp}}-{\bf u}_{20})\|/\|{\bf u}_{20}\|$, was 1.2\%. Since the interpolation introduces error, it can be concluded that the actual error should be less than 1.2\%. We find that global measures, such as $Re$ or $\langle u_y^2 \rangle$ (used to characterize the primary instability), computed directly using ${\bf u}_{40}$ and ${\bf u}_{20}$ differed by less than 0.2\%. These tests confirm that a resolution of 20 cells per magnet width is sufficient to simulate the flow and characterize the bifurcations accurately.

\subsection{Doubly-Periodic Simulation (DPS)} \label{sec:app_dps}

Simulations on the doubly-periodic domain can be sped up significantly using a spectral method \citep{canuto_1988}. Since solving linear equations involving the Laplacian is very cheap in the spectral method, it is convenient to use the vorticity-stream function formulation instead of the velocity-pressure formulation. Taking the curl of equation (\ref{eq:2dns_mod}), we obtain the following equation for the $z$-component of vorticity  $\omega  = (\nabla\times{\bf u})\cdot{\bf\hat{z}}$:
\begin{equation}\label{eq:2dvor_modified}
\partial_t\omega + \beta{\bf u}\cdot{\bf{\nabla}}{\omega}=
{\nu} \nabla^2 \omega - \alpha\omega + W,
\end{equation}
where $W = (\nabla\times{\bf f})\cdot {\bf \hat z}$. The horizontal components of the velocity field $u_x = \partial\psi/\partial y$ and $u_y = -\partial\psi/\partial x$ can be computed using the stream function $\psi$, which satisfies the Poisson equation $\nabla^2\psi = -\omega$.

The vorticity field $\omega$ is discretized in the Fourier space using $128$ modes along each of the $x$- and $y$-directions. Since the lateral dimensions of the periodic domain are $8w\times8w$ units, the spatial resolution associated with the Fourier grid corresponds to 16 grid points per magnet width $w$. Taking the Fourier transform of equation (\ref{eq:2dvor_modified}), we obtain:
\begin{equation}\label{eq:2dvor_spectral}
\partial_t\Omega = - \beta{\mathcal F}[{\bf u}\cdot{\bf{\nabla}}{\omega}] + {\nu} \nabla^2 \Omega - \alpha\Omega + {\mathcal F}[W],
\end{equation} 
where ${\mathcal F}[\cdot]$ represents the Fourier transform and $\Omega={\mathcal F}[\omega]$.
 
Equation (\ref{eq:2dvor_spectral}) is stepped forward in time ($t \rightarrow t+\Delta t$) using a 3-substep semi-implicit Strang-Marchuk splitting algorithm \citep{ascher_1995,mitchell_2013} where the first and last substeps advance the vorticity field using the nonlinear term by means of a second-order explicit Runge-Kutta scheme (using a time step $\Delta t/2$), while the intermediate substep advances the vorticity field using the Crank-Nicolson scheme (using a time step $\Delta t$). We have used the time step $\Delta t=1/32$ s.

\subsection{Computing unstable branches in the pitchfork bifurcation}\label{sec:methods_pitchfork}

The schematic depicting the pitchfork bifurcation in figure \ref{fig:pitchfork} was constructed following the computation of all the stable and unstable states  using the matrix-free Newton-Krylov solver \citep{kelley_2003}. Guesses for the stable states, to initialize the Newton solver,  can be easily obtained using numerical integration. However, those for the unstable states should be constructed using continuation or using the eigenmode that goes unstable at the bifurcation. 

To begin with, initial guesses for the unstable straight flow branches in the SPS and NPS simulations were constructed by extrapolating the stable straight solutions in Reynolds number $Re$, i.e.,
\begin{equation}\label{eq:extrapolate_lam}
{\bf u}_s^2(Re_c + \epsilon) \approx {\bf u}_s^1(Re_c) + \epsilon \left(\frac{\partial {\bf u}_s^1}{\partial Re}\right)_{Re_c}, 
\end{equation}
where the derivative $\left({\partial {\bf u}_s^1}/{\partial Re}\right)$ was approximated using finite differences,
\begin{equation}
\left(\frac{\partial {\bf u}_s^1}{\partial Re}\right)_{Re} \approx \frac{{\bf u}_s^1(Re)- {\bf u}_s^1(Re-\Delta Re)} {\Delta Re}.
\end{equation}
This method proved particularly useful in obtaining a good initial guess for the unstable straight flow ${\bf u}_s^{2}$ in the NPS, since it is disconnected from ${\bf u}_s^{1}$, as shown in figure \ref{fig:pitchfork} (c). In the NPS, $Re_c \approx 10.5$ is not estimated by identifying the instability of ${\bf u}_s^1$, since there exists none.  Instead, it is computed using the intercept of a linear fit of the amplitude $\langle u_y^2 \rangle$ versus $Re$, shown in figure \ref{fig:bif1}, close to the onset of modulation. The  initial guess at  $Re \approx 10.75$ was constructed by extrapolating ${\bf u}_s^{1}$ from $Re \approx 10.25$ by  choosing $\epsilon = 0.5$ in equation (\ref{eq:extrapolate_lam}). 

For the unstable modulated branches emerging from the second pitchfork in the SPS, a good initial guess for ${\bf u}_m^{3}$ (${\bf u}_m^{4}$) is constructed  using  ${\bf u}_m^{3} \approx {\bf u}_s^{2} \pm p\,\hat{{\bf e}}_2$. Here $\hat{\bf e}_2$  is the second unstable eigenvector of the straight flow which has the symmetry $\mathcal{R}_x\mathcal{T}^w_y$. Since amplitude $p$ is not known \textit{a priori}, convergence to  ${\bf u}_m^{3}$ is tested by incrementing  $p$. In the NPS, the initial guess for the unstable branch ${\bf u}_m^{2}$ was similarly constructed, ${\bf u}_m^{2} \approx {\bf u}_s^{2} - p\,\hat{{\bf e}}_1$, using the eigenvector $\hat{\bf e}_1$ with the ${\mathcal R}_{x}{\mathcal R}_{y}$ symmetry. However, $p$ in the NPS case can be estimated using $p = \langle \hat{\bf e}_1|{\bf u}_m^{1}-{\bf u}_s^{2}\rangle$, since the stable modulated flow ${\bf u}_m^{1}$ is known from numerical integration and ${\bf u}_s^{2}$ is computed from extrapolation.

\section{Inherent Three-Dimensionality of the Forcing in the Experiment} \label{sec:sensitivity}

In \S \ref{sec:mod_flow_sec} we compared measurements of the average longitudinal wavelength $\bar{\lambda}_x$ of the modulated flow in the experiment and numerical simulations. 
The comparison between the experiment and the NPS (cf. figures \ref{fig:bif1} and \ref{fig:bif1_vary_params}) showed systematic differences in $\bar{\lambda}_x$ for both the depth-averaged and modified parameters, with the maximum difference being about 10\% at $Re - Re_c = 3.7$  (cf. figures \ref{fig:bif1}). Here we show that this deviation is likely due to the inherent three-dimensionality of the experiment, not captured by a strictly 2D model, rather than the choice of the model parameters.

As we discussed in \S \ref{sec:dipole}, the Lorentz force density due to the specific arrangement of magnets employed in the experiment is to a very good approximation given by ${\bf F} = JB_z{\bf \hat x}$, where $J$ is the magnitude of current density  and $B_z(x,y,z)$ is the $z$-component of the magnetic field at any given location within the electrolyte. 
In deriving equation (\ref{eq:2dns_mod}) it was assumed \citep{suri_2014} that $B_z$ can be decomposed as the product of a 2D horizontal profile $B_{2D}(x,y)$, which depends exclusively on the extended coordinates $(x,y)$, and a 1D vertical profile $D(z)$, which captures the variation of the magnetic field above the magnet array, i.e.,  ${B}_z(x,y,z) = D(z) B_{2D}(x,y)$. 
This implies that, when normalized, the planar magnetic field profiles at various heights $z$ within the electrolyte are identical. 
Such a magnetic field, which we call ``Q2D," facilitates the decomposition of the plane-parallel Q2D velocity field \eqref{eq:profile} which underpins the strictly 2D model (\ref{eq:2dns_mod}).

A magnetic field that is truly Q2D, however, cannot be created using a magnet array with finite dimensions, i.e., the \textit{shape} of the magnetic field profile generated by permanent magnets in the laboratory always changes with the vertical height $z$ to some extent. 
Experimental measurements of the magnetic field from previous studies have shown such changes in the shape of the field profile as a function of $z$ \citep{dovzhenko_1984, suri_2014}. This is very much the case in our experiment as well, as can be seen from the magnetic field profiles shown in figure \ref{fig:mag_prof} (a). 
For instance, if one rescales the transverse magnetic field profiles at heights $z=0.438$ and $z=0.265$ such that they match near the centre of the array, we see that these profiles would \emph{not} match near the end magnets. 
This is most apparent by comparing the relative heights of the peaks at $y=-6.5$ and $y=-4.5$ for the two profiles in figure \ref{fig:mag_prof} (a). 
A quantitative estimate of the deviation from quasi-two-dimensionality can be obtained by comparing the magnetic field profiles computed using the dipole summation at the bottom $B_{b} = B_z(x,y,z=0.236)$ and the top of the electrolyte layer $B_{t} = B_z(x,y,z=0.472)$. 
Normalizing $B_{b}$ and $B_{t}$ separately, using their respective spatial rms values computed over the entire lateral extent of the domain, we estimate the largest difference between the profiles to be approximately 12\%. This difference is fairly localised towards the ends of the magnet array and is likely the reason behind the larger discrepancy in longitundinal velocity measurements over the end magnets (cf. figure \ref{fig:vel_prof} (b)).

\begin{figure}
\subfloat[]{\includegraphics[width=2.5in]{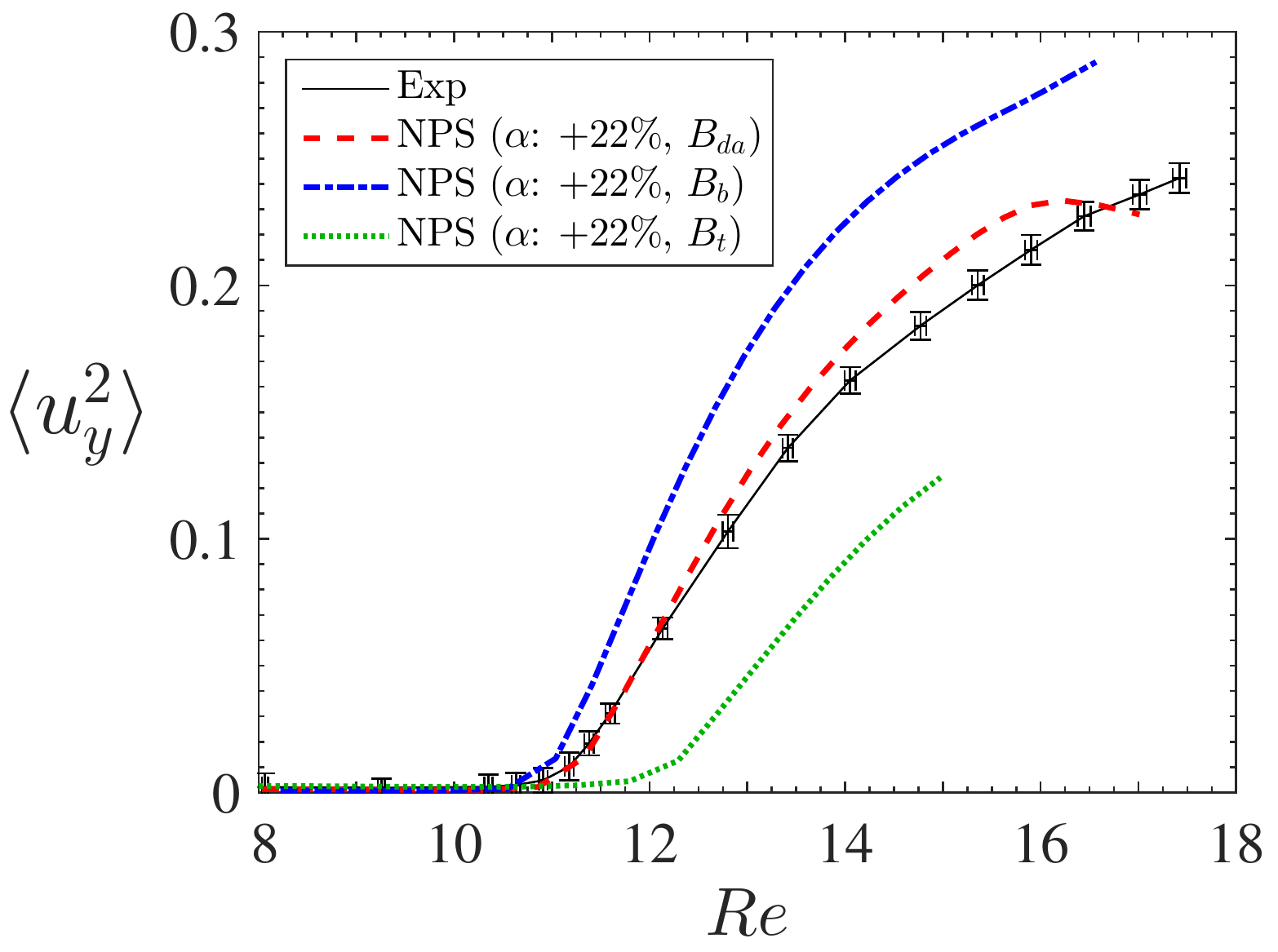}}\qquad
\subfloat[]{\includegraphics[width=2.45in]{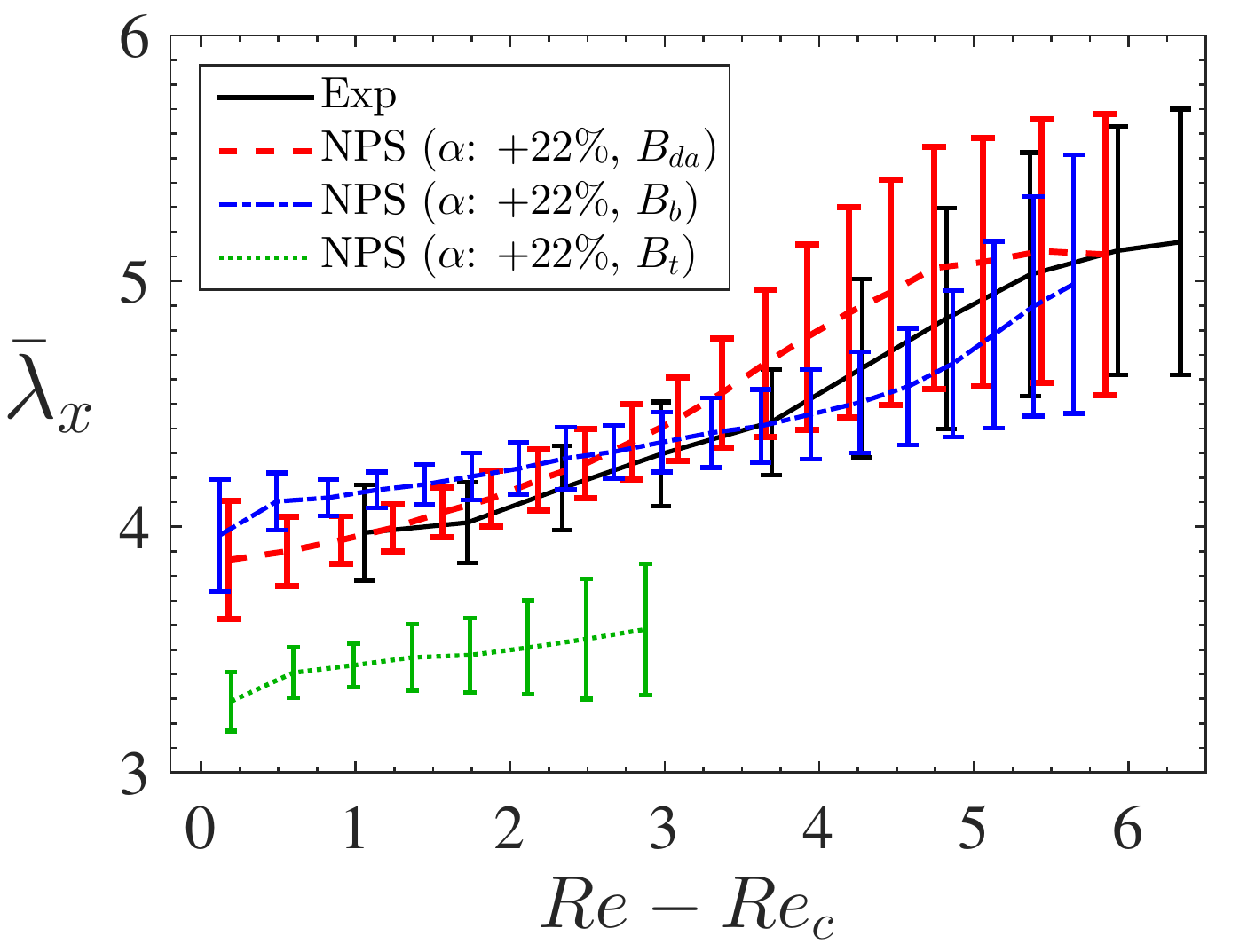}}
\caption{\label{fig:bif_4and7} Sensitivity to the magnetic field profile for $\alpha=0.078$ s$^{-1}$, $\beta=0.83$, and ${\nu}=3.26 \times 10^{-6}$ m$^2$/s. (a) A bifurcation diagram for the primary instability and (b) the average wavelength of the pattern in the modulated regime.  The simulations were performed with $\alpha$ increased by 22\% relative to the depth-averaged value for the straight flow and used either $B_{da}$, $B_{b}$, or $B_{t}$.}
\end{figure}

To demonstrate the impact of the $z$-dependence of the forcing profile on the flow, we have recomputed the straight and modulated flow fields in the NPS using $B_{2D}=B_{b}$ and $B_{2D}=B_{t}$, in addition to the depth-averaged magnetic field $B_{da}$.    
The value of $\alpha$ was increased by 22\% relative to the depth-averaged value to reduce the influence of the uncertainty in the model parameters on the flow pattern. 
This choice also yields the best agreement between the average wavelengths of the flow pattern in the simulation and experiment for $Re_c<Re<Re_p$ (see figure \ref{fig:bif1_vary_params} (b)). 
As figure \ref{fig:bif_4and7} (a) shows, the forcing profile strongly affects both $Re_c$ and the amplitude of the modulation of the flow for $Re>Re_c$. It also shows that $B_{da}$ produces substantially better agreement with experiment that either $B_b$ or $B_t$.
Similarly, we find that the forcing profile strongly influences the modulation wavelength. As figure \ref{fig:bif_4and7} (b) shows, for both $B_b$ or $B_{da}$ the average wavelength agrees reasonable well with experiment, while $B_t$ produces a very poor agreement. Similar results (not shown) are obtained if, instead of $\alpha$, either $\beta$ or ${\nu}$ is modified to match $Re_c$.

The above analysis shows that, although the NPS with the depth-averaged magnetic field profile captures the salient features of the dynamics fairly well, the flow pattern depends fairly sensitively on the details of the forcing. 
Hence, one should expect systematic deviations between the 2D model derived for a Q2D flow and the experiment where quasi-two-dimensionality is broken by the forcing. 
It should be mentioned that the wavelength of the modulated flow measured by either seeding the dielectric-electrolyte interface or the top surface of the electrolyte are virtually identical. 
This implies that viscous coupling across the fluid layers produces a Q2D flow despite the fact that the forcing profile driving the flow is not perfectly Q2D.

\bibliographystyle{jfm}

\bibliography{bifurcations}

\begin{thebibliography}{59}
\expandafter\ifx\csname natexlab\endcsname\relax\def\natexlab#1{#1}\fi

\bibitem[Akkermans {\em et~al.\/}(2010)Akkermans, Kamp, Clercx \& van
  Heijst]{akkermans_2010}
{\sc Akkermans, R. A.~D., Kamp, L. P.~J., Clercx, H. J.~H. \& van Heijst, G.
  J.~F.} 2010 Three-dimensional flow in electromagnetically driven shallow
  two-layer fluids. {\em Phys. Rev. E\/} {\bf 82}, 026314.

\bibitem[Akkermans {\em et~al.\/}(2008)Akkermans, Kamp, Clercx \&
  Van~Heijst]{akkermans_2008a}
{\sc Akkermans, R. A.~D., Kamp, L. P.~J., Clercx, H. J.~H. \& Van~Heijst, G.
  J.~F.} 2008 Intrinsic three-dimensionality in electromagnetically driven
  shallow flows. {\em Europhys. Lett.\/} {\bf 83}~(2), 24001.

\bibitem[Armbruster {\em et~al.\/}(1992)Armbruster, Heiland, Kostelich \&
  Nicolaenko]{armbruster_1992}
{\sc Armbruster, D., Heiland, R., Kostelich, E.~J. \& Nicolaenko, B.} 1992
  Phase-space analysis of bursting behavior in {K}olmogorov flow. {\em Physica
  D\/} {\bf 58}~(1), 392--401.

\bibitem[Armfield \& Street(1999)]{armfield_1999}
{\sc Armfield, S. \& Street, R.} 1999 The fractional-step method for the
  {N}avier-{S}tokes equations on staggered grids: The accuracy of three
  variations. {\em J. Comput. Phys.\/} {\bf 153}~(2), 660 -- 665.

\bibitem[Arnold \& Meshalkin(1960)]{arnold_1960}
{\sc Arnold, V.~I. \& Meshalkin, L.~D.} 1960 Seminar led by {A}. {N}.
  {Kolmogorov} on selected problems of analysis (1958-1959). {\em Usp. Mat.
  Nauk\/} {\bf 15}~(247), 20--24.

\bibitem[Ascher {\em et~al.\/}(1995)Ascher, Ruuth \& Wetton]{ascher_1995}
{\sc Ascher, U.~M., Ruuth, S.~J. \& Wetton, B. T.~R.} 1995 Implicit-explicit
  methods for time-dependent partial differential equations. {\em SIAM J.
  Numer. Anal.\/} {\bf 32}~(3), 797--823.

\bibitem[Batchaev \& Dowzhenko(1983)]{batchaev_1983}
{\sc Batchaev, A.~M. \& Dowzhenko, V.~A.} 1983 Experimental modeling of
  stability loss in periodic zonal flows. In {\em Dokl. Akad. Nauk\/}, , vol.
  273, p. 582.

\bibitem[Batchaev \& Ponomarev(1989)]{batchaev_1989}
{\sc Batchaev, A.~M. \& Ponomarev, V.~M.} 1989 Experimental and theoretical
  investigation of {K}olmogorov flow on a cylindrical surface. {\em Fluid
  Dyn.\/} {\bf 24}~(5), 675--680.

\bibitem[Boffetta \& Ecke(2012)]{boffetta_2012}
{\sc Boffetta, G. \& Ecke, R.~E.} 2012 Two-dimensional turbulence. {\em Annu.
  Rev. Fluid Mech.\/} {\bf 44}, 427--451.

\bibitem[Bondarenko {\em et~al.\/}(1979)Bondarenko, Gak \&
  Dolzhanskiy]{bondarenko_1979}
{\sc Bondarenko, N.~F., Gak, M.~Z. \& Dolzhanskiy, F.~V.} 1979 Laboratory and
  theoretical models of plane periodic flows. {\em Izv. Akad. Nauk SSSR, Fiz.
  Atmos. Okeana\/} {\bf 15}~(10), 711--716.

\bibitem[Burgess {\em et~al.\/}(1999)Burgess, Bizon, McCormick, Swift \&
  Swinney]{burgess_1999}
{\sc Burgess, J.~M., Bizon, C., McCormick, W.~D., Swift, J.~B. \& Swinney,
  Harry~L.} 1999 Instability of the {K}olmogorov flow in a soap film. {\em
  Phys. Rev. E\/} {\bf 60}, 715--721.

\bibitem[Canuto {\em et~al.\/}(1988)Canuto, Hussaini, Quarteroni \&
  Zang]{canuto_1988}
{\sc Canuto, C., Hussaini, M.~Y., Quarteroni, A. \& Zang, T.~A.} 1988 {\em
  Spectral methods in fluid dynamics\/}. Springer.

\bibitem[Chandler \& Kerswell(2013)]{chandler_2013}
{\sc Chandler, G.~J. \& Kerswell, R.~R.} 2013 Invariant recurrent solutions
  embedded in a turbulent two-dimensional {K}olmogorov flow. {\em J. Fluid
  Mech.\/} {\bf 722}, 554--595.

\bibitem[Couder(1984)]{couder_1984}
{\sc Couder, Y.} 1984 Two-dimensional grid turbulence in a thin liquid film.
  {\em J. Phys. Lett.\/} {\bf 45}~(8), 353--360.

\bibitem[Couder {\em et~al.\/}(1989)Couder, Chomaz \& Rabaud]{couder_1989}
{\sc Couder, Y., Chomaz, J.~M. \& Rabaud, M.} 1989 On the hydrodynamics of soap
  films. {\em Physica D\/} {\bf 37}~(1), 384--405.

\bibitem[Dennis \& Sogaro(2014)]{dennis_2014}
{\sc Dennis, D. J.~C. \& Sogaro, F.~M} 2014 Distinct organizational states of
  fully developed turbulent pipe flow. {\em Phys. Rev. Lett.\/} {\bf 113}~(23),
  234501.

\bibitem[Dolzhanskii {\em et~al.\/}(1992)Dolzhanskii, Krymov \&
  Manin]{dolzhanskii_1992}
{\sc Dolzhanskii, F.~V., Krymov, V.~A. \& Manin, D.~Yu.} 1992 An advanced
  experimental investigation of quasi-two-dimensional shear flows. {\em J.
  Fluid Mech.\/} {\bf 241}, 705--722.

\bibitem[Dolzhansky(2013)]{dolzhansky_2012}
{\sc Dolzhansky, F.~V.} 2013 {\em Fundamentals of {G}eophysical
  {H}ydrodynamics\/}, {\em {E}ncyclopaedia of {M}athematical {S}ciences\/},
  vol. 103. Springer, translated by B.~A. Khesin.

\bibitem[Dovzhenko {\em et~al.\/}(1984)Dovzhenko, Krymov \&
  Ponomarev]{dovzhenko_1984}
{\sc Dovzhenko, V.~A., Krymov, V.~Ao \& Ponomarev, V.~M.} 1984 Experimental and
  theoretical investigation of the shear flow generated by an axially symmetric
  force. {\em Izv. Akad. Nauk SSSR, Fiz. Atmos. Okeana\/} {\bf 20}, 693.

\bibitem[Dovzhenko {\em et~al.\/}(1981)Dovzhenko, Obukhov \&
  Ponomarev]{dovzhenko_1981}
{\sc Dovzhenko, V.~A., Obukhov, A.~M. \& Ponomarev, V.~M.} 1981 Generation of
  vortices in an axisymmetric shear flow. {\em Fluid Dyn.\/} {\bf 16}~(4),
  510--518.

\bibitem[Drew {\em et~al.\/}(2013)Drew, Charonko \& Vlachos]{prana}
{\sc Drew, B., Charonko, J. \& Vlachos, P.~P.} 2013 {QI} -- {Q}uantitative
  {I}maging ({PIV} and more).

\bibitem[Eckhardt {\em et~al.\/}(2007)Eckhardt, Schneider, Hof \&
  Westerweel]{eckhardt_2007}
{\sc Eckhardt, B., Schneider, T.~M., Hof, B. \& Westerweel, J.} 2007 Turbulence
  transition in pipe flow. {\em Annu. Rev. Fluid Mech.\/} {\bf 39}~(1),
  447--468.

\bibitem[Eckstein \& Vlachos(2009)]{eckstein_2009}
{\sc Eckstein, A. \& Vlachos, P.~P.} 2009 Digital particle image velocimetry
  {(DPIV)} robust phase correlation. {\em Meas. Sci. Technol.\/} {\bf 20}~(5),
  055401.

\bibitem[Gallet \& Young(2013)]{gallet_2013}
{\sc Gallet, B. \& Young, W.~R.} 2013 A two-dimensional vortex condensate at
  high {R}eynolds number. {\em J. Fluid Mech\/} {\bf 715}, 359--388.

\bibitem[Gibson {\em et~al.\/}(2009)Gibson, Halcrow \&
  Cvitanovi{\'c}]{gibson_2009}
{\sc Gibson, J.~F., Halcrow, J. \& Cvitanovi{\'c}, P.} 2009 Equilibrium and
  travelling-wave solutions of plane {C}ouette flow. {\em J. Fluid Mech.\/}
  {\bf 638}, 243--266.

\bibitem[Green(1974)]{green_1974}
{\sc Green, J. S.~A.} 1974 Two-dimensional turbulence near the viscous limit.
  {\em J. Fluid Mech.\/} {\bf 62}~(02), 273--287.

\bibitem[Haller(2015)]{haller_2015}
{\sc Haller, G.} 2015 Lagrangian coherent structures. {\em Annu. Rev. Fluid
  Mech.\/} {\bf 47}, 137--162.

\bibitem[Haller \& Yuan(2000)]{haller_2000}
{\sc Haller, G. \& Yuan, G.} 2000 Lagrangian coherent structures and mixing in
  two-dimensional turbulence. {\em Physica D\/} {\bf 147}~(3), 352--370.

\bibitem[Harlow \& Welch(1965)]{harlow_1965}
{\sc Harlow, F.~H. \& Welch, J.~E.} 1965 Numerical calculation of time
  dependent viscous incompressible flow of fluid with free surface. {\em Phys.
  Fluids\/} {\bf 8}~(12), 2182--2189.

\bibitem[Hof {\em et~al.\/}(2004)Hof, van Doorne, Westerweel, Nieuwstadt,
  Faisst, Eckhardt, Wedin, Kerswell \& Waleffe]{hof_2004}
{\sc Hof, B., van Doorne, C. W.~H., Westerweel, J., Nieuwstadt, F. T.~M.,
  Faisst, H., Eckhardt, B., Wedin, H., Kerswell, R.~R. \& Waleffe, F.} 2004
  Experimental observation of nonlinear traveling waves in turbulent pipe flow.
  {\em Science\/} {\bf 305}~(5690), 1594--1598.

\bibitem[Hussain(1986)]{hussain_1986}
{\sc Hussain, A. K. M.~F.} 1986 Coherent structures and turbulence. {\em J.
  Fluid Mech.\/} {\bf 173}, 303--356.

\bibitem[Iudovich(1965)]{iudovich_1965}
{\sc Iudovich, V.~I.} 1965 Example of the generation of a secondary stationary
  or periodic flow when there is loss of stability of the laminar flow of a
  viscous incompressible fluid. {\em J. Appl. Math. Mech.\/} {\bf 29}~(3),
  527--544.

\bibitem[J\"uttner {\em et~al.\/}(1997)J\"uttner, Marteau, Tabeling \&
  Thess]{juttner_1997}
{\sc J\"uttner, B., Marteau, D., Tabeling, P. \& Thess, A.} 1997 Numerical
  simulations of experiments on quasi-two-dimensional turbulence. {\em Phys.
  Rev. E\/} {\bf 55}, 5479--5488.

\bibitem[Kawahara {\em et~al.\/}(2012)Kawahara, Uhlmann \& van
  Veen]{kawahara_2012}
{\sc Kawahara, G., Uhlmann, M. \& van Veen, L.} 2012 The significance of simple
  invariant solutions in turbulent flows. {\em Annu. Rev. Fluid Mech.\/} {\bf
  44}, 203--225.

\bibitem[Kelley(2003)]{kelley_2003}
{\sc Kelley, C.} 2003 {\em Solving Nonlinear Equations with Newton's Method\/}.
  SIAM.

\bibitem[Kelley \& Ouellette(2011)]{kelley_2011a}
{\sc Kelley, D.~H. \& Ouellette, N.~T.} 2011 Onset of three-dimensionality in
  electromagnetically driven thin-layer flows. {\em Phys. Fluids\/} {\bf
  23}~(4), 045103.

\bibitem[Kerswell(2005)]{kerswell_2005}
{\sc Kerswell, R.~R.} 2005 Recent progress in understanding the transition to
  turbulence in a pipe. {\em Nonlinearity\/} {\bf 18}~(6), R17.

\bibitem[Kliatskin(1972)]{kliatskin_1972}
{\sc Kliatskin, V.~I.} 1972 On the nonlinear theory of stability of periodic
  flows. {\em J. Appl. Math. Mech.\/} {\bf 36}~(2), 243--250.

\bibitem[Krymov(1989)]{krymov_1989}
{\sc Krymov, V.~A.} 1989 Stability and supercritical regimes of
  quasi-two-dimensional shear flow in the presence of external friction
  (experiment). {\em Fluid Dynamics\/} {\bf 24}~(2), 170--176.

\bibitem[de~Lozar {\em et~al.\/}(2012)de~Lozar, Mellibovsky, Avila \&
  Hof]{lozar_2012}
{\sc de~Lozar, A., Mellibovsky, F., Avila, M. \& Hof, B.} 2012 Edge state in
  pipe flow experiments. {\em Phys. Rev. Lett.\/} {\bf 108}, 214502.

\bibitem[Lucas \& Kerswell(2014)]{lucas_2014}
{\sc Lucas, D. \& Kerswell, R.~R.} 2014 Spatiotemporal dynamics in
  two-dimensional {K}olmogorov flow over large domains. {\em J. Fluid Mech.\/}
  {\bf 750}, 518--554.

\bibitem[Lucas \& Kerswell(2015)]{lucas_2015}
{\sc Lucas, D. \& Kerswell, R.~R.} 2015 Recurrent flow analysis in
  spatiotemporally chaotic 2-dimensional {K}olmogorov flow. {\em Phys.
  Fluids\/} {\bf 27}~(4), 045106.

\bibitem[Marteau {\em et~al.\/}(1995)Marteau, Cardoso \&
  Tabeling]{marteau_1995}
{\sc Marteau, D., Cardoso, O. \& Tabeling, P.} 1995 Equilibrium states of
  two-dimensional turbulence: {A}n experimental study. {\em Phys. Rev. E\/}
  {\bf 51}, 5124--5127.

\bibitem[Meshalkin \& Sinai(1961)]{meshalkin_1961}
{\sc Meshalkin, L.~D. \& Sinai, Ia~G.} 1961 Investigation of the stability of a
  stationary solution of a system of equations for the plane movement of an
  incompressible viscous liquid. {\em J. Appl. Math. Mech.\/} {\bf 25}~(6),
  1700--1705.

\bibitem[Mitchell(2013)]{mitchell_2013}
{\sc Mitchell, R.} 2013 Transition to turbulence and mixing in a
  quasi-two-dimensional {L}orentz force-driven {K}olmogorov flow. PhD thesis,
  Georgia Institute of Technology.

\bibitem[Nagata(1997)]{nagata_1997}
{\sc Nagata, M.} 1997 Three-dimensional traveling-wave solutions in plane
  {C}ouette flow. {\em Phys. Rev. E\/} {\bf 55}, 2023--2025.

\bibitem[Nepomniashchii(1976)]{nepomniashchii_1976}
{\sc Nepomniashchii, A.~A.} 1976 On stability of secondary flows of a viscous
  fluid in unbounded space. {\em J. Appl. Math. Mech.\/} {\bf 40}~(5),
  886--891.

\bibitem[Obukhov(1983)]{obukhov_1983}
{\sc Obukhov, A.~M.} 1983 Kolmogorov flow and laboratory simulation of it. {\em
  Russ. Math. Surv.\/} {\bf 38}~(4), 113.

\bibitem[Paret \& Tabeling(1997)]{paret_1997a}
{\sc Paret, J. \& Tabeling, P.} 1997 Experimental observation of the
  two-dimensional inverse energy cascade. {\em Phys. Rev. Lett.\/} {\bf 79},
  4162--4165.

\bibitem[Rivera \& Ecke(2005)]{rivera_2005}
{\sc Rivera, M.~K. \& Ecke, R.~E.} 2005 Pair dispersion and doubling time
  statistics in two-dimensional turbulence. {\em Phys. Rev. Lett.\/} {\bf 95},
  194503.

\bibitem[Smaoui(2001)]{smaoui_2001}
{\sc Smaoui, N.} 2001 A model for the unstable manifold of the bursting
  behavior in the 2{D} {N}avier--{S}tokes flow. {\em SIAM J. Sci. Comp.\/} {\bf
  23}~(3), 824--839.

\bibitem[Sommeria(1986)]{sommeria_1986}
{\sc Sommeria, J.} 1986 Experimental study of the two-dimensional inverse
  energy cascade in a square box. {\em J. Fluid Mech.\/} {\bf 170}, 139--168.

\bibitem[Sommeria {\em et~al.\/}(1988)Sommeria, Meyers \&
  Swinney]{sommeria_1988}
{\sc Sommeria, J., Meyers, S.~D. \& Swinney, H.~L.} 1988 Laboratory simulation
  of {J}upiter's great red spot. {\em Nature\/} {\bf 331}~(6158), 689--693.

\bibitem[Sommeria \& Moreau(1982)]{sommeria_1982}
{\sc Sommeria, J. \& Moreau, R.} 1982 Why, how, and when, {MHD} turbulence
  becomes two-dimensional. {\em J. Fluid Mech.\/} {\bf 118}, 507--518.

\bibitem[Suri {\em et~al.\/}(2017)Suri, Tithof, Grigoriev \& Schatz]{suri_2017}
{\sc Suri, B., Tithof, J., Grigoriev, R.~O. \& Schatz, M.~F.} 2017 Forecasting
  fluid flows using the geometry of turbulence. {\em Phys. Rev. Lett.\/} {\bf
  118}, 114501.

\bibitem[Suri {\em et~al.\/}(2014)Suri, Tithof, Mitchell, Grigoriev \&
  Schatz]{suri_2014}
{\sc Suri, B., Tithof, J., Mitchell, R., Grigoriev, R.~O. \& Schatz, M.~F.}
  2014 Velocity profile in a two-layer {K}olmogorov-like flow. {\em Phys.
  Fluids\/} {\bf 26}~(5), 053601.

\bibitem[Tabeling {\em et~al.\/}(1991)Tabeling, Burkhart, Cardoso \&
  Willaime]{tabeling_1991}
{\sc Tabeling, P., Burkhart, S., Cardoso, O. \& Willaime, H.} 1991 Experimental
  study of freely decaying two-dimensional turbulence. {\em Phys. Rev. Lett.\/}
  {\bf 67}, 3772--3775.

\bibitem[Thess(1992)]{thess_1992a}
{\sc Thess, A.} 1992 Instabilities in two-dimensional spatially periodic flows.
  {P}art {I}: {K}olmogorov flow. {\em Phys. Fluids A\/} {\bf 4}~(7),
  1385--1395.

\bibitem[Waleffe(1998)]{waleffe_1998}
{\sc Waleffe, F.} 1998 Three-dimensional coherent states in plane shear flows.
  {\em Phys. Rev. Lett.\/} {\bf 81}~(19), 4140.

\end{thebibliography}

\end{document}